\documentclass[floats,floatfix,amssymb,prd,twocolumn,superscriptaddress,nofootinbib,nolongbibliography,reprint]{revtex4-2}

\usepackage{amssymb,amsmath,verbatim,mathtools,needspace,enumitem,etoolbox,graphicx,physics,microtype,afterpage,xspace,tabularx,lmodern,multirow}
\usepackage{gensymb}
\usepackage[dvipsnames, usenames]{xcolor}
\definecolor{linkcolor}{rgb}{0.0,0.3,0.5}
\usepackage[unicode, colorlinks=true, linkcolor=linkcolor, citecolor=linkcolor, filecolor=linkcolor, urlcolor=linkcolor, linktocpage, breaklinks]{hyperref}
\usepackage[all]{hypcap}
\usepackage[T1]{fontenc}
\usepackage[utf8]{inputenc}
\usepackage[usenames,dvipsnames]{xcolor}
\hypersetup{colorlinks=true,citecolor=romared,linkcolor=romared,urlcolor=romared}

\definecolor{romared}{RGB}{142,0,28}

\newcommand{\be}{\begin{equation}}
\newcommand{\ee}{\end{equation}}

\def\be{\begin{equation}}
\def\ee{\end{equation}}
\newcommand{\beq}{\begin{eqnarray}}
\newcommand{\eeq}{\end{eqnarray}}

\usepackage{aas_macros}
\usepackage{makecell}
\usepackage{soul}
\usepackage[nolist,nohyperlinks]{acronym}
\usepackage{longtable}
\usepackage{booktabs}
\usepackage{ulem} 

\newcommand{\orcid}[1]{\href{https://orcid.org/#1}{\includegraphics[width=10pt]{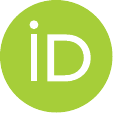}}}

\def\i{{\rm i}}
\def\e{{\rm e}}

\usepackage[nolist,nohyperlinks]{acronym}

\allowdisplaybreaks

\begin{document}

\pagenumbering{arabic}

\title{Unveiling the merger structure of black hole binaries in generic planar orbits}

\author{Gregorio Carullo \orcid{0000-0001-9090-1862}}
\email{gregorio.carullo@nbi.ku.dk}
\affiliation{Niels Bohr International Academy, Niels Bohr Institute, Blegdamsvej 17, 2100 Copenhagen, Denmark}
\author{Simone Albanesi \orcid{0000-0001-7345-4415}}
\affiliation{INFN sezione di Torino, Torino, 10125, Italy}
\affiliation{Dipartimento di Fisica, Universit\`a di Torino, Torino, 10125, Italy}
\author{Alessandro Nagar \orcid{}}
\affiliation{INFN sezione di Torino, Torino, 10125, Italy}
\affiliation{Institut des Hautes Etudes Scientifiques, 35 Route de Chartres, Bures-sur-Yvette, 91440, France}
\author{Rossella Gamba \orcid{0000-0001-7239-0659}}
\affiliation{Theoretisch-Physikalisches Institut, Friedrich-Schiller-Universit{\"a}t Jena, 07743, Jena, Germany}
\author{Sebastiano Bernuzzi \orcid{0000-0002-2334-0935}}
\affiliation{Theoretisch-Physikalisches Institut, Friedrich-Schiller-Universit{\"a}t Jena, 07743, Jena, Germany}
\author{Tomas Andrade \orcid{}}
\affiliation{Departament de F{\'\i}sica Qu\`antica i Astrof\'{\i}sica, Institut de Ci\`encies del Cosmos, Universitat de Barcelona, Mart\'{\i} i Franqu\`es 1, E-08028 Barcelona, Spain}
\author{Juan Trenado \orcid{}}
\affiliation{Departament de F{\'\i}sica Qu\`antica i Astrof\'{\i}sica, Institut de Ci\`encies del Cosmos, Universitat de Barcelona, Mart\'{\i} i Franqu\`es 1, E-08028 Barcelona, Spain}

\begin{abstract}
  The precise modeling of binary black hole coalescences in generic planar orbits is a crucial step to disentangle dynamical and isolated binary formation channels through gravitational-wave observations.
  The merger regime of such coalescences exhibits a significantly higher complexity compared to the quasicircular case, and cannot be readily described through standard parameterizations in terms of eccentricity and anomaly.
  In the spirit of the Effective One Body formalism, we build on the study of the test-mass limit, and introduce a new modelling strategy to describe the general-relativistic dynamics of two-body systems in generic orbits.
  This is achieved through gauge-invariant combinations of the binary energy and angular momentum, such as a dynamical ``impact parameter'' at merger.
  These variables reveal simple ``quasi-universal'' structures of the pivotal merger parameters, allowing to build an accurate analytical representation of generic (bounded and dynamically-bounded) orbital configurations.
  We demonstrate the validity of these analytical relations using 311 numerical simulations of bounded noncircular binaries with progenitors from the RIT and SXS catalogs, together with a custom dataset of dynamical captures generated using the Einstein Toolkit, and test-mass data in bound orbits.
  Our modeling strategy lays the foundations of accurate and complete waveform models for systems in arbitrary orbits, bolstering observational explorations of dynamical formation scenarios and the discovery of new classes of gravitational wave sources.
\end{abstract}

\maketitle


\acrodef{LSC}[LSC]{LIGO Scientific Collaboration}
\acrodef{BH}{black hole}
\acrodef{NS}{neutron star}
\acrodef{PN}{Post-Newtonian}
\acrodef{BBH}{binary black-hole}
\acrodef{BNS}{binary neutron-star}
\acrodef{NSBH}{neutron-star black-hole}
\acrodef{EOB}{effective-one-body}
\acrodef{NR}{numerical relativity}
\acrodef{GW}{gravitational wave}
\acrodef{PSD}{power spectral density}
\acrodef{aLIGO}{Advanced Laser interferometer Gravitational-Wave Observatory}
\acrodef{AZDHP}{aLIGO zero detuned high power density}
\acrodef{GR}{general relativity}
\acrodef{PE}{parameter estimation}
\acrodef{LAL}{LIGO algorithm library}
\acrodef{TPI}{tensor-product interpolant}
\acrodef{SVD}{singular value decomposition}
\acrodef{SNR}{signal-to-noise ratio}
\acrodef{ODE}{ordinary differential equation}
\acrodef{PDE}{partial differential equation}
\acrodef{ROM}{reduced order model}
\acrodef{QNM}{quasi-normal mode}
\acrodef{IMR}{inspiral-merger-ringdown}
\acrodef{LVK}{LIGO-Virgo-KAGRA}
\acrodef{SXS}{Simulating eXtreme Spacetimes}

\newcommand{\hlm}{h_{\ell m}}
\newcommand{\ylm}{{}_{-2}Y_{\ell m}}
\newcommand{\Eeffmrg}{\hat{E}_{\mathrm{eff}}^{\mathrm{mrg}}}
\newcommand{\Eeff}{\hat{E}_{\mathrm{eff}}}

\noindent {\textbf{\textit{Introduction}}.}
%
\begin{figure}[thbp]
         \includegraphics[width=0.98\columnwidth]{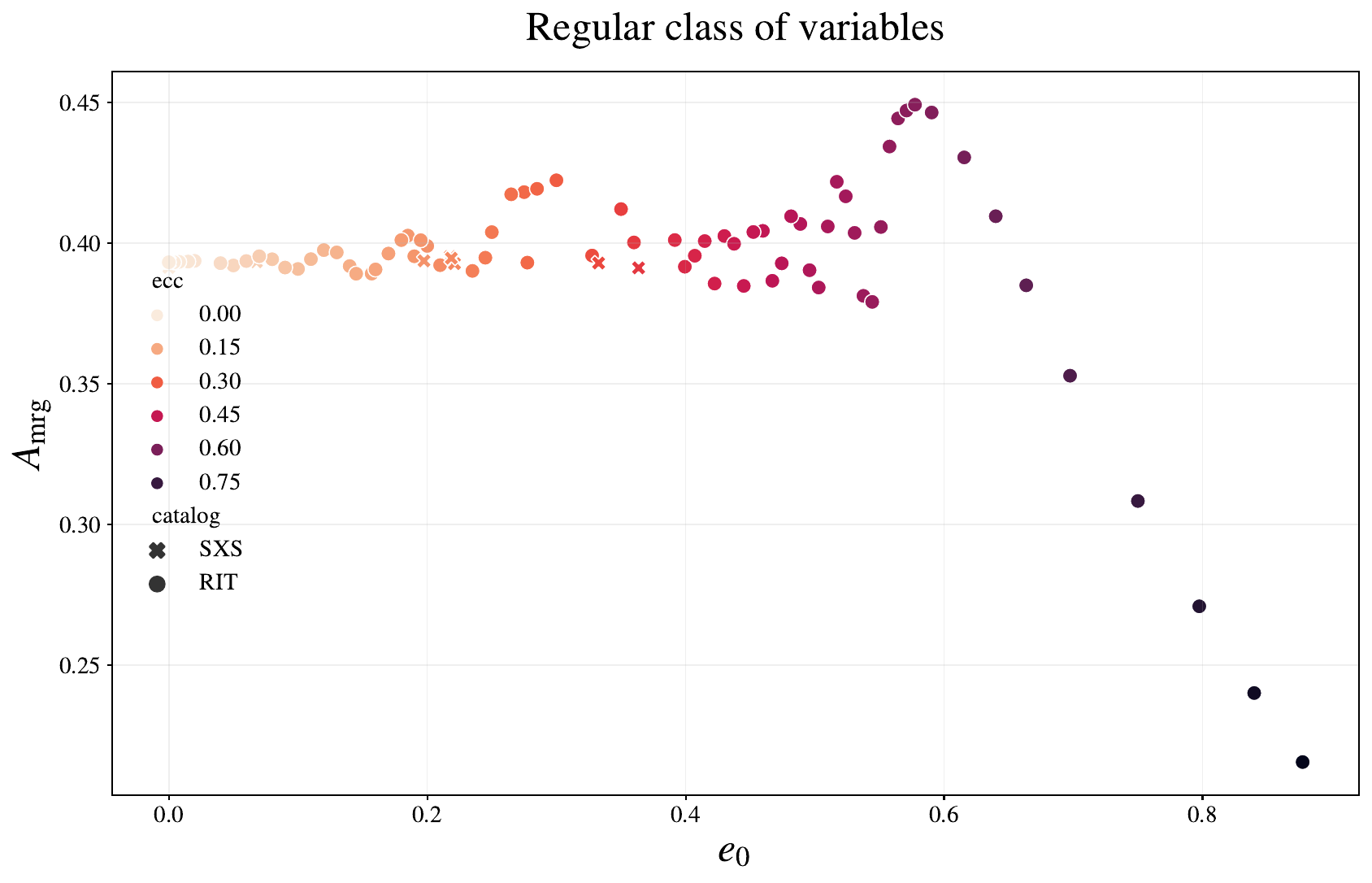}
         \includegraphics[width=0.98\columnwidth]{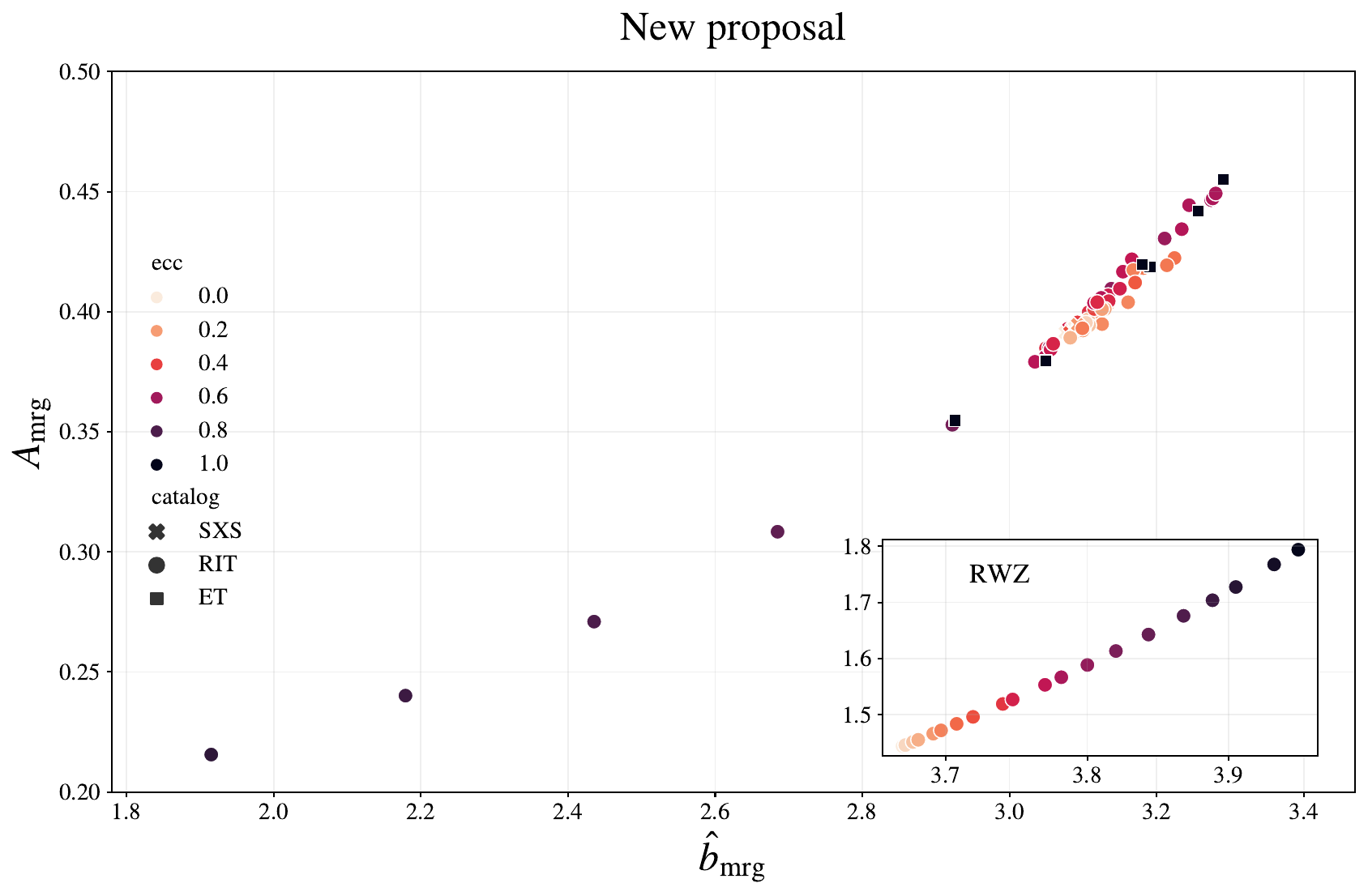}
   \caption{
   Top panel: merger amplitude as a function of the initial nominal eccentricity for bounded equal mass nonspinnning binaries. Their relationship is oscillatory and multi-valued.
   Bottom panel: same quantity as above, but as a function of a suitably defined dynamical impact parameter evaluated at merger.
   The amplitude now displays a simple monotonic dependence, even for hyperbolic-like systems (square markers).
   The inset highlights the same relationship for bounded test-mass data.}
   \label{fig:Apeak_complex_par}
\end{figure}
Black hole (BH) binary mergers are unique probes of dynamical formation channels in dense environments~\cite{Mandel:2018hfr,Mapelli:2021taw}, and allow to push searches of new physics~\cite{LIGOScientific:2021sio,Narayan:2023vhm} into a stronger-field regime.
Gravitational-wave (GW) signals emitted by these systems can be detected by interferometric observatories both on the ground~\cite{TheLIGOScientific:2014jea,TheVirgo:2014hva} and in space~\cite{LISA:2017pwj}, or by Pulsar Timing Arrays~\cite{Antoniadis:2023rey, NANOGrav:2023gor, Reardon:2023gzh, Xu:2023wog}.
Recently, significant effort has been placed in their modelling and search~\cite{LIGOScientific:2019dag, Romero-Shaw:2019itr, Nitz:2019spj, Gayathri:2020coq, Ramos-Buades:2020eju, Veske:2020idq, Romero-Shaw:2020thy, Veske:2020zch, Romero-Shaw:2021ual, OShea:2021ugg, LIGOScientific:2021tfm, Gayathri:2021xwb, Iglesias:2022xfc, Romero-Shaw:2022xko, Ebersold:2022zvz, Dandapat:2023zzn, Garg:2023lfg, Ramos-Buades:2023yhy,Gayathri:2019kop, LIGOScientific:2020ufj, Tagawa:2020qll, Tagawa:2021ofj, OBrien:2021sua, Barrera:2022yfj, Gayathri:2023bha,Samsing:2017xmd, Zevin:2018kzq, Tagawa:2020jnc, Samsing:2020tda, Zevin:2021rtf, 2021arXiv210704639F, Chattopadhyay:2023pil}, with one signal being already consistent with a noncircular scenario~\cite{Romero-Shaw:2020thy,Gayathri:2020coq,Gamba:2021gap,CalderonBustillo:2020xms}.
Crucially, a significant fraction of dynamically-formed sources are expected to lie in the high-end of the mass distribution, due to hierarchical mergers~\cite{Gerosa:2021mno}, pushing their inspiral outside the sensitive band of ground-based detectors, and giving rise to a signal dominated by the merger-ringdown portion.
While several noncircular inspiral models have been developed, both in bounded and dynamically-bounded~\cite{Arun:2009mc, Huerta:2014eca, Moore:2016qxz, Chiaramello:2020ehz, Nagar:2020xsk,Placidi:2021rkh, Paul:2022xfy, Albanesi:2022xge} orbits, merger-ringdown waveforms instead still rely on a quasicircular description~\cite{Huerta:2017kez, Cao:2017ndf, Hinder:2017sxy, Nagar:2021gss, Ramos-Buades:2021adz, Liu:2021pkr, Chattaraj:2022tay, Liu:2023dgl}, with the exception of a numerical surrogate valid for small eccentricity~\cite{Islam:2021mha}.
Further, analytical models have shown good accuracy also for scattering orbits~\cite{Damour:2014afa, Hopper:2022rwo}, a set of configurations which is crucial in the ongoing effort of connecting quantum scattering amplitude calculations with classical gravity~\cite{Damour:2016gwp, Damour:2017zjx, Damour:2019lcq, Bern:2019nnu, Bern:2021dqo, Dlapa:2022lmu, Dlapa:2021npj, Dlapa:2021vgp, Bern:2021yeh, Damour:2022ybd, Khalil:2022ylj, Rettegno:2023ghr}.
Aside from theoretical considerations, complete models are urgently needed to extend standard template searches based on quasicircular waveforms~\cite{LIGOScientific:2021djp}, which exhibit dramatic sensitivity loss to systems in arbitrary orbits~\cite{East:2012xq}.
Laying the foundations to go beyond quasicircular merger-ringdown models is the main goal of this paper.

To this end, appropriate modeling variables capable of capturing the noncircular merger structure are required.
For the bounded case, proposed modeling choices in the literature are generalizations of the Newtonian definitions of eccentricity and anomaly parameters.
The most recent proposals~\cite{Mora:2002gf, Islam:2021mha, Bonino:2022hkj, Shaikh:2023ypz} are based on waveform-constructed quantities, ensuring gauge-invariance.
However, these definitions rely on ``pericenter'' and ``apocenter'' frequencies constructed from interpolation of the waveform frequency minima and maxima.
As a consequence, this method does not apply to situations where only a modest number of waveform cycles is available, which include the vast majority of numerical simulations with intermediate to high eccentricity, and dynamical-capture systems.
Moreover, an eccentricity defined from frequency minima and maxima is intrinsically ill-defined at merger, nor can be readily extended to the merger resulting from hyperbolic initial conditions.
It is thus intuitive to expect that these parameterisations do not allow for a simple plunge-merger-ringdown modelling.

This is showcased in the top panel of Fig.~\ref{fig:Apeak_complex_par}, displaying the behavior of the merger GW amplitude $A_{\mathrm{mrg}}$ (defined below) in the nonspinning equal-mass binary case as a function of the initial orbital eccentricity.
The displayed one-dimensional relationship presents a highly complex structure and wide bifurcations: the initial eccentricity parameter does not allow for a smooth merger modeling and, already for intermediate eccentricities, it does not uniquely map the noncircular amplitude value.
In the Supplemental Material, which includes Refs.~\cite{Kalin:2019inp, Santamaria:2010yb, Lousto:1997ge, Martel:2001yf, Reisswig:2009rx, Damour:2007xr}, we show the same result to hold even when using the time-evolved and gauge-invariant eccentricity parameter of Refs.~\cite{Mora:2002gf, Bonino:2022hkj, Shaikh:2023ypz} for all the reference times available. Additionally, the required interpolation fails for the vast majority of the simulations shown, leaving a much smaller dataset to be considered.
The latter point also prevents to obtain simple relationships when performing a two-dimensional fit including a gauge-invariant anomaly parameter~\cite{Shaikh:2023ypz}: the relationship becomes single-valued, but it presents a highly complicated oscillatory structure difficult to resolve with the very few points for which such parameter can be computed.  
An equivalent behavior has been observed for the late-time ringdown amplitudes, see Fig.10 of Ref.~\cite{Forteza:2022tgq}.

\begin{figure*}[thbp]
    \includegraphics[width=0.48\textwidth]{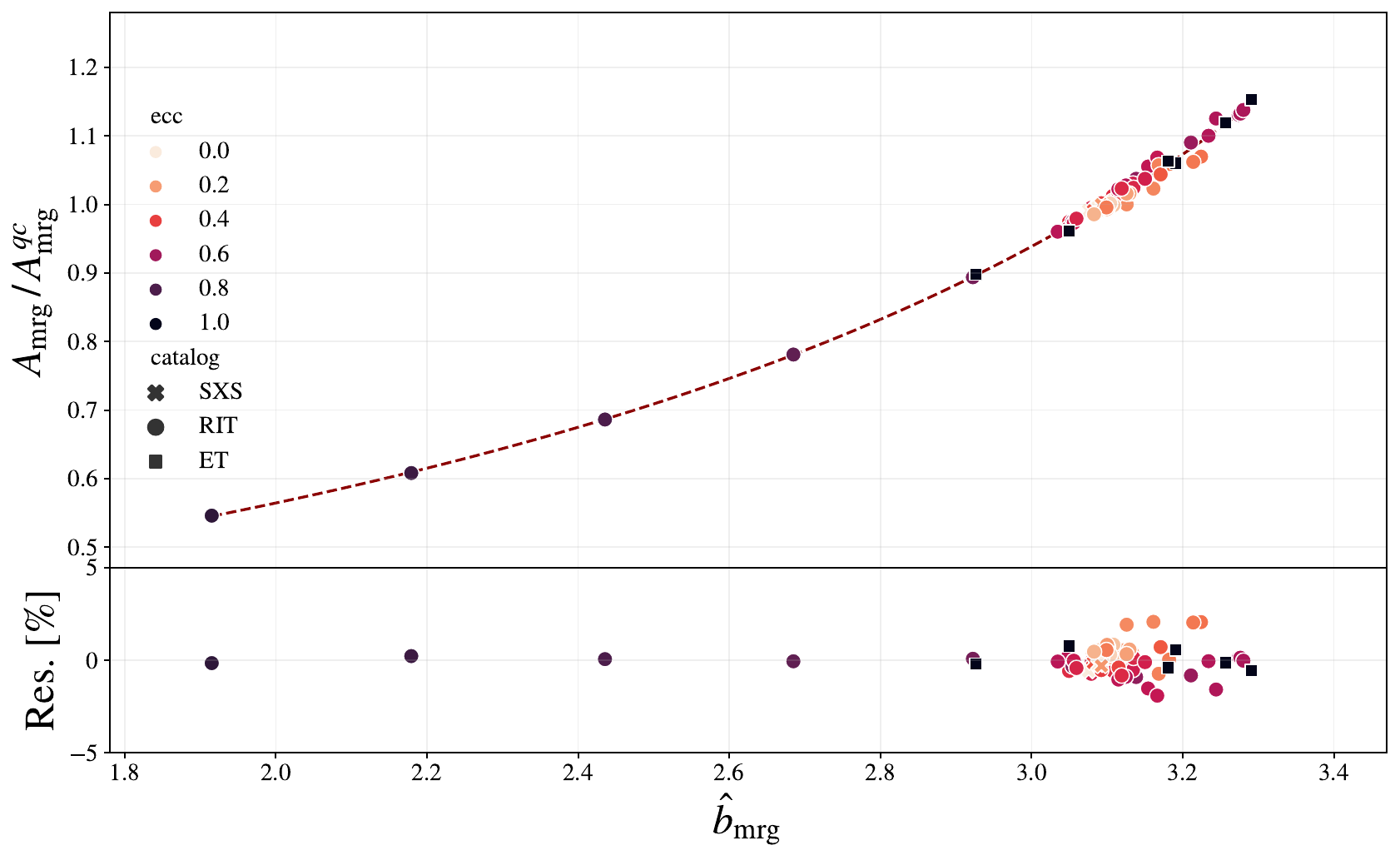}   
    \includegraphics[width=0.48\textwidth]{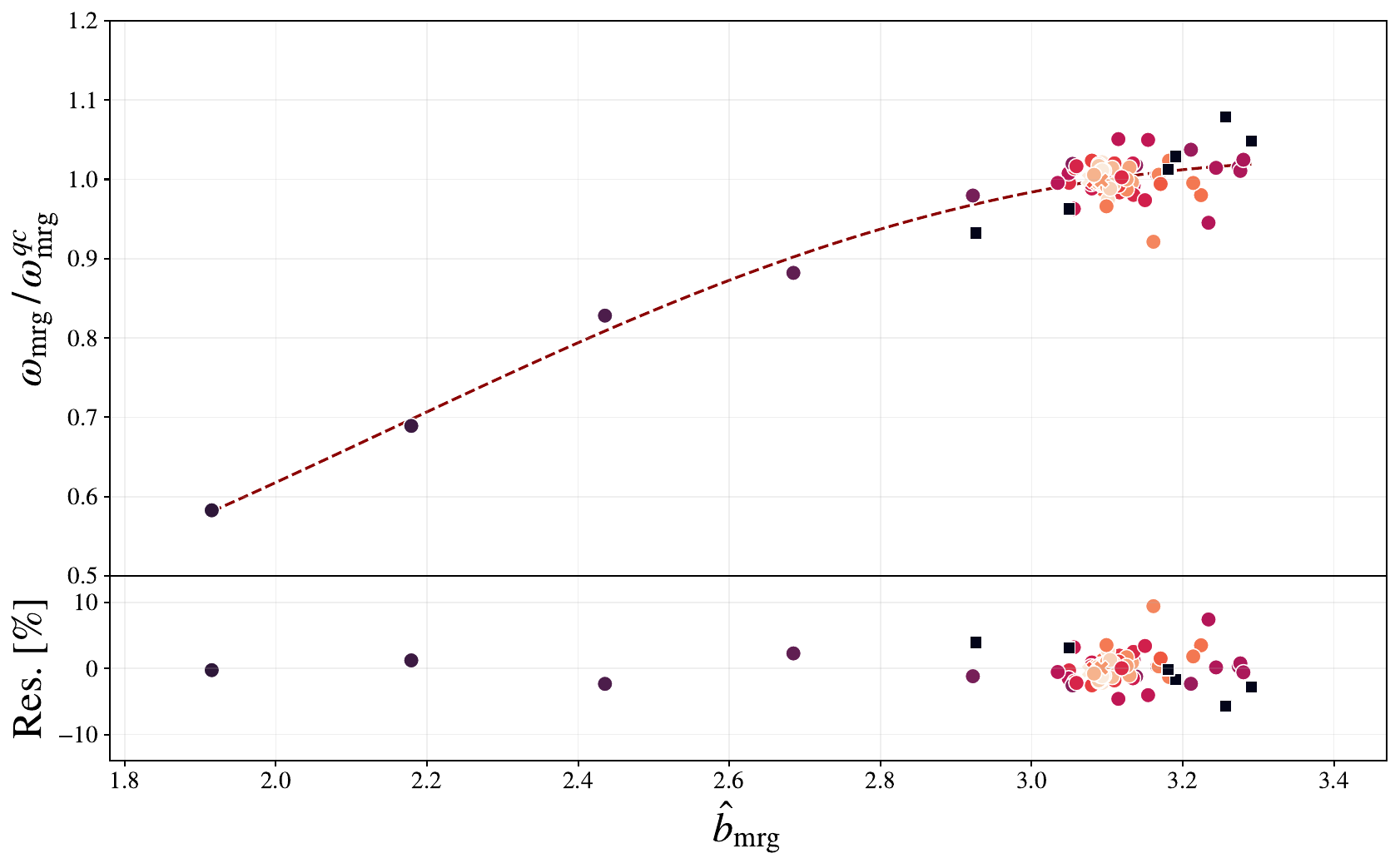}
    \includegraphics[width=0.48\textwidth]{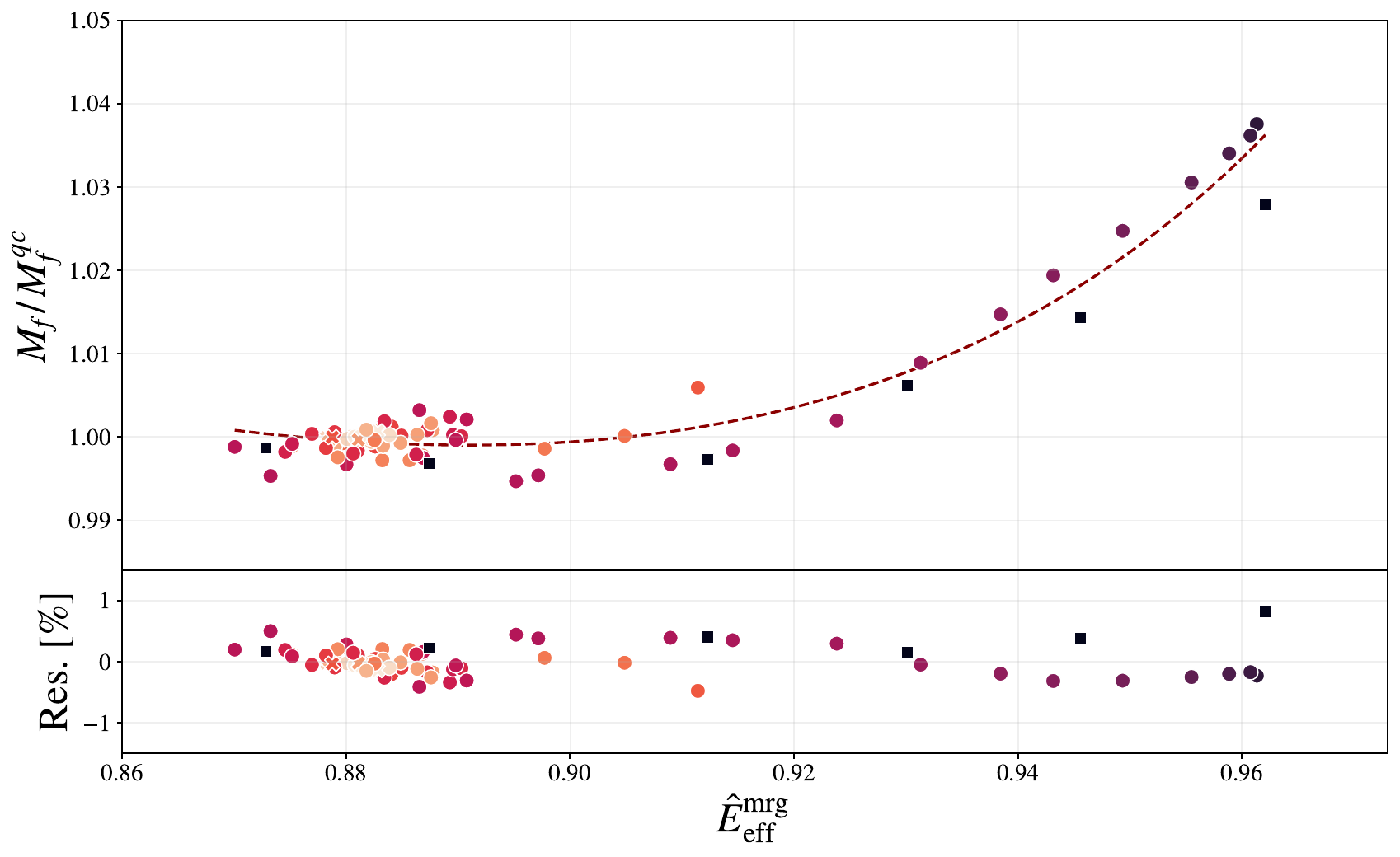}
    \includegraphics[width=0.48\textwidth]{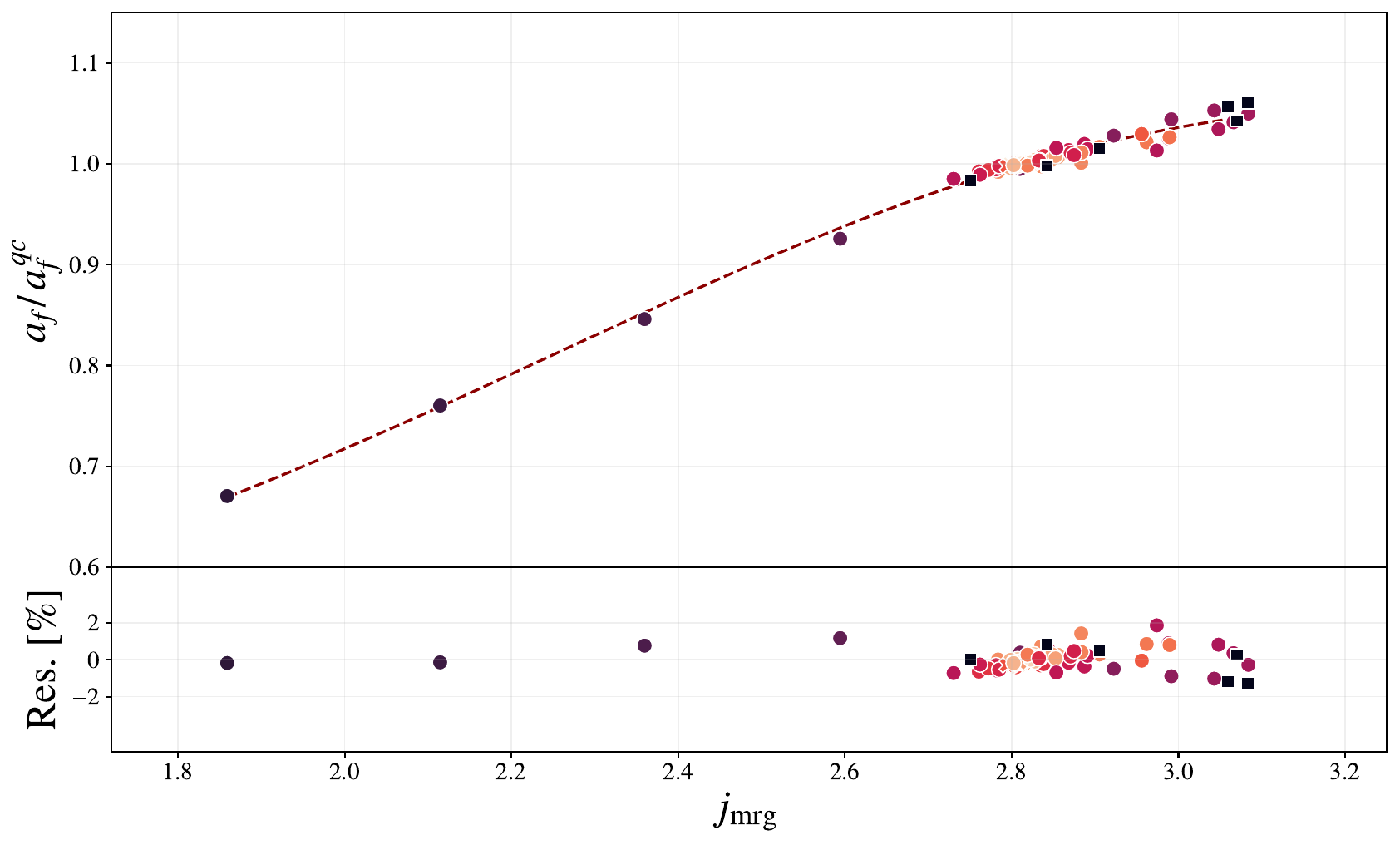}    
   \caption{
   Quasi-universal relationships of the merger quantities in terms of dimensionless evolved variables, for nonspinning configurations.
   For visual purposes, ET data are assigned eccentricity value 1.
   Note the smaller variation range of $M_f$ compared to the other quantities.
   Smaller panels indicate residuals as defined in the text.
   }
   \label{fig:1D_equal_mass}
\end{figure*}

Here, we improve on the above parameterizations and introduce a new modelling strategy valid for arbitrary binary planar orbits, relying on results obtained in the test mass limit~\cite{Albanesi:2023bgi}.
The merger amplitude of a particle in bounded eccentric orbits is a smooth function of a suitably defined dynamical ``impact parameter'', as shown in the inset of the bottom panel in Fig.~\ref{fig:Apeak_complex_par}.
Now, no bifurcations arise and a simple structure emerges.
One of the main results of the paper is showing how such parametrization can be generalized to the comparable-mass case, previewed in the main bottom panel of Fig.~\ref{fig:Apeak_complex_par}.
We also consider dynamically-bounded systems, showcasing the applicability of our parameterization to comparable-mass binaries in \textit{generic} orbits.
This is achieved through a \textit{single} variable representing the two-dimensional space of initial conditions in the noncircular case, a highly non-trivial feature attesting the effectiveness of our modeling strategy.
Such variable thus uncovers a new and non-trivial ``quasi-universality'' in the merger-remnant structure, deepening the understanding of the two-body dynamics.
Our methodology, which leverages the natural variables (energy and angular momentum) describing a generic binary dynamic, readily incorporates BH spins, as we show by including in our study binaries with spins aligned to the orbital angular momentum.
Below, we detail the construction and validation of our modeling variables, and of the displayed relationships, similarly built for all key merger parameters, shown in Fig.~\ref{fig:1D_equal_mass} in the nonspinning case.
The constructed relationships provide for the first time the required corner stone for the completion of semi-analytical aligned-spins models, significantly increasing their agreement to numerical solutions (at the $\sim 99 \%$ match level even in the challenging late stages of dynamical captures)~\cite{Andrade:2023trh}.
Such models will significantly extend the discovery horizon of GW searches for compact binary coalescences.\\

\noindent {\textbf{\textit{Conventions.}}}
%
We use geometric units $c=G=1$.
The gravitational-wave strain is decomposed in spin-weighted spherical harmonics modes, $h_{\ell m}(t)$, split in amplitude and phase as $h_{\ell m}(t) = A_{\ell m}(t)\,\e^{\i \phi_{\ell m}(t)}$, with GW frequency, $\omega_{\ell m}(t) \equiv 2\pi f_{\ell m}(t) = \dot{\phi}_{\ell m}(t) $.
Individual ADM masses of the two BHs are denoted as $m_{1,2}$, with $M$ = $m_1 + m_2$, the mass ratio $q \equiv m_1 /m_2 \geq 1$, the symmetric mass ratio $\nu = m_1 m_2 / M^2$,  individual BH spin components aligned to the orbital angular momentum are denoted as $\chi_{1,2}$, and the effective binary spin $\chi_{\rm eff} = (m_1 \chi_1 + m_2 \chi_2)/M  $.
Since we will be focusing on the dominant $(\ell,m)=(2, \pm 2)$ mode, we will drop the mode subscript.
All available modes are however used in the fluxes computations.\\

\noindent {\textbf{\textit{Variables construction.}}}
%
To model systems in arbitrary orbits, we start by considering the initial ADM energy and angular momentum $(E^{\rm ADM}_0, J_0^{\rm ADM})$. 
Their values at time $t$, $(E(t), J(t))$ are obtained by subtracting from $(E^{\rm ADM}_0, J_0^{\rm ADM})$ the gravitational wave losses (see the Supplemental Material for details), $E(t) = E_0^{\rm ADM} - \int_{t_0}^{t}{\dot{E}(t^\prime) dt^\prime}, \, J(t) = J_0^{\rm ADM} - \int_{t_0}^{t}{\dot{J} (t^\prime) dt^\prime}$,
where $t_0$ is the simulation starting time and $(\dot{E}(t), \dot{J}(t))$ are the radiated fluxes of energy and angular momentum.
We define the merger quantities $E_{\mathrm{mrg}} \equiv E(t_{\mathrm{mrg}}), J_{\mathrm{mrg}} \equiv J(t_{\mathrm{mrg}})$,
where the merger time $t_{\mathrm{mrg}}$ is the merger time.
In the generic parameter space under study, a robust and physically meaningful definition of $t_{\mathrm{mrg}}$ can be obtained selecting the time corresponding to the peak of the emission immediately before a quasinormal-driven regime (ringdown) begins.
For all the configurations under consideration here, this time simply corresponds to the \textit{last} peak of the \textit{quadrupolar} waveform amplitude $A_{22}$. 
It is important to note that for generic initial data, this last peak does not coincide with the largest maximum of $A_{22}$, which can occur at the periastron passage previous to the plunge-merger, after which the system still displays a two-body orbital dynamics incompatible with a merger definition.
For more extreme eccentricities (very close to the head-on limit) and larger mass ratios the above definition can be generalized by considering the peak of the total amplitude, including higher harmonics, or by simultaneously fitting for quasinormal-modes, still guaranteeing the applicability of the above definition.

Even when considering merger-evolved quantities, a key point to obtain accurate fits is to use dimensionless variables 
factoring out the appropriate mass-scaling. 
In particular, we use the mass-normalized energy $h_{\mathrm{mrg}} \equiv E_{\mathrm{mrg}}/M$
and angular momentum $j_{\mathrm{mrg}} \equiv J_{\mathrm{mrg}}/(\nu M^2)$. 
The effective-one-body (EOB) approach to the two-body general relativistic dynamics~\cite{Buonanno:1998gg} is a powerful analytical method that maps
the dynamics of a two-body system into the dynamics of an effective body of mass $\mu=m_1 m_2/M$ moving into an effective external potential. 
The map between the real energy $h$ and the effective energy (per unit mass) $\Eeff\equiv E_{\rm eff}/\mu$ 
is given by $\Eeff(h, \nu) = 1 + (h^2-1) / (2\nu)$. 
For scattering configurations, an impact parameter of the form $b_{\rm EOB}/M=j h/\sqrt{\Eeff^2-1}$ can be defined, see Eq.~(2.29) in Ref.~\cite{Damour:2019lcq} or Eq.~(2.5) in Ref.~\cite{Kalin:2019rwq}. 
In the test-mass limit, i.e. the case of a particle moving on a Schwarzschild BH, we have $h\to 1$, while
$\Eeff$ becomes the real energy of the particle. 
The parameter $b_{\rm EOB}/M$ is well defined only when $\Eeff>1$ and thus it cannot be straightforwardly applied, 
as it is our intention, to characterize the dynamics of bound configurations, where $\Eeff<1$. 
Thus, inspired by recent results in the test-mass limit~\cite{Albanesi:2023bgi}, we use as dynamical impact parameter at merger the quantity $\hat{b}_{\rm mrg}= b_{\rm mrg}/M \equiv j_{\rm mrg} \, h(\Eeffmrg, \nu)/\Eeffmrg$, that is the one used in the right panel of Fig.~\ref{fig:Apeak_complex_par}.
As fitting variables, we will employ functions of $\{ \Eeffmrg, j_{\mathrm{mrg}}, \hat{b}_{\mathrm{mrg}}, \nu \chi_{\rm eff} \}$, 
depending on whether we discuss quasi-universal relations, or consider the full dimensionality of the parameter space.
This choice of parameters is key to obtain simple and accurate relationships in arbitrary orbits, shifting the focus from orbit-based quantities, to evolved dynamical ones, more naturally incorporating radiation-reaction effects.

The key quantities determining the merger emission are $\{ A_{\mathrm{mrg}}, \omega_{\mathrm{mrg}}, M_f, a_f \}$, where 
$a_f \equiv J_f / M_f^2$, $(M_f,J_f)$ denote the mass and spin of the remnant BH, $A_{\rm mrg} \equiv A(t=t_{\mathrm{mrg}})$
and $\omega_{\mathrm{mrg}} \equiv \omega(t=t_{\rm mrg})$.
We decide to focus on modeling these quantities ratio compared to the quasicircular case.
This allows to scale leading order dependences (e.g. on $\nu$) observed independently of the orbital configuration.
We thus obtain a factorized fit, straightforwardly applicable on top of quasicircular values, maintaining the accuracy of the quasicircular limit, where more simulations are available. 
Both astrophysical inference and searches for new physics rely on merger-remnant predictions in the quasicircular case, well-studied both for remnant mass and spin~\cite{Healy:2014yta,Healy:2016lce,Hofmann:2016yih,Jimenez-Forteza:2016oae,NathanDCC,Varma:2019csw,Boschini:2023ryi, Ferguson:2019slp,LIGOScientific:2021djp,LIGOScientific:2021sio}, and for merger quantities~\cite{Nagar:2018zoe,Nagar:2019wds, Carullo:2018gah}.
As fitting template, we consider $Y = Y_0 (1 + p_1 \, Q + p_2 \, Q^2) \cdot (1 + p_3 \, Q + p_4 \, Q^2)^{-1}$ where $Q$ is the dimensionless fitting variable, $Y$ the ratio of the modeled quantity with its quasicircular value, and $p_k = b_k \, (1 +c_k X)$, with $X\equiv 1 - 4\nu$, $b_k, c_k \in \mathbb{R}$.
We found this function sufficiently simple and flexible to capture the structure of our dataset~\cite{Bernuzzi:2014kca}. 
Residuals are defined by: $\Delta Y \equiv (Y - Y_{\mathrm{NR}})/Y_{\mathrm{NR}}$.
Technical details of the fits are provided in the Supplemental Material.
There, we also discuss how to incorporate these relationships in an waveform template.
It is important to note that the above relationships only enter the merger and post-merger portions of the waveform, while fluxes are derived only through the pre-merger waveform.\\

\begin{figure*}[thbp]

         \includegraphics[width=0.463\textwidth]{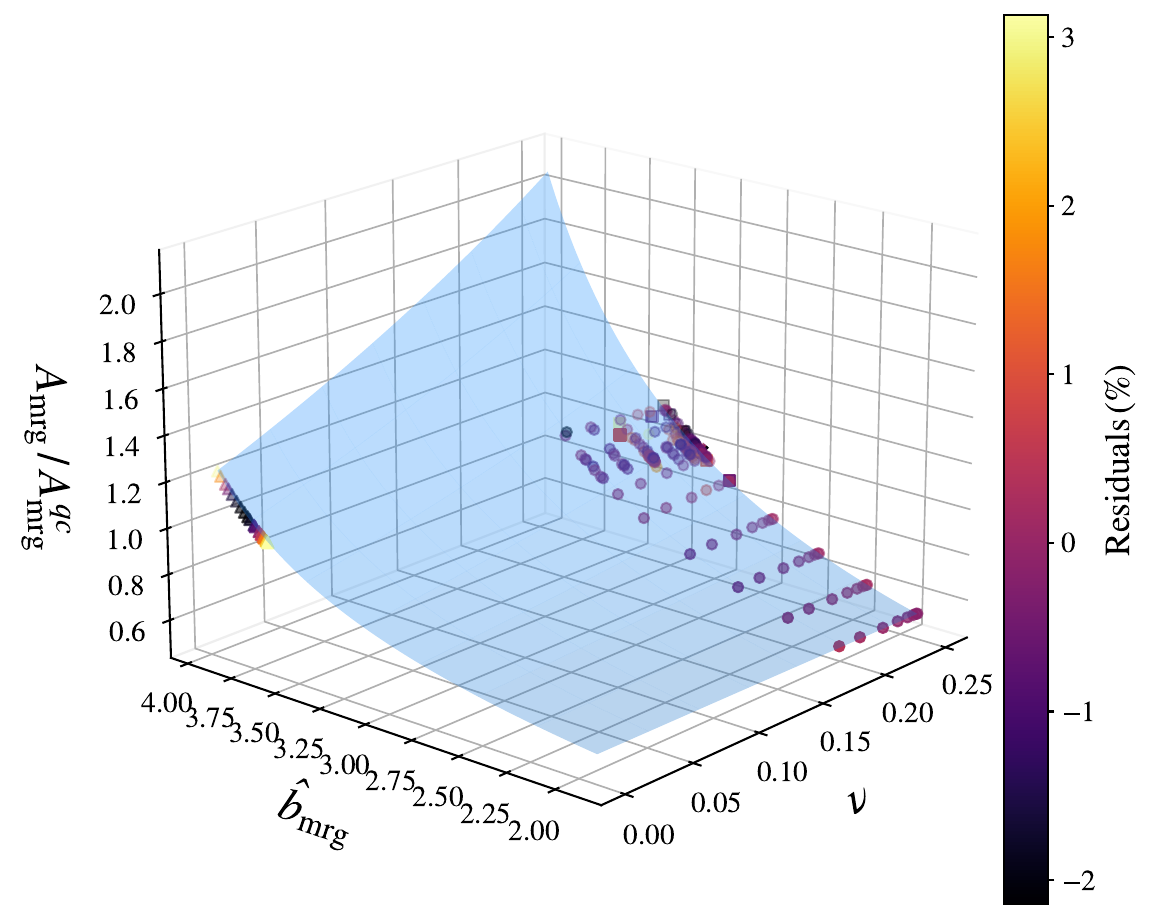}
         \includegraphics[width=0.47\textwidth]{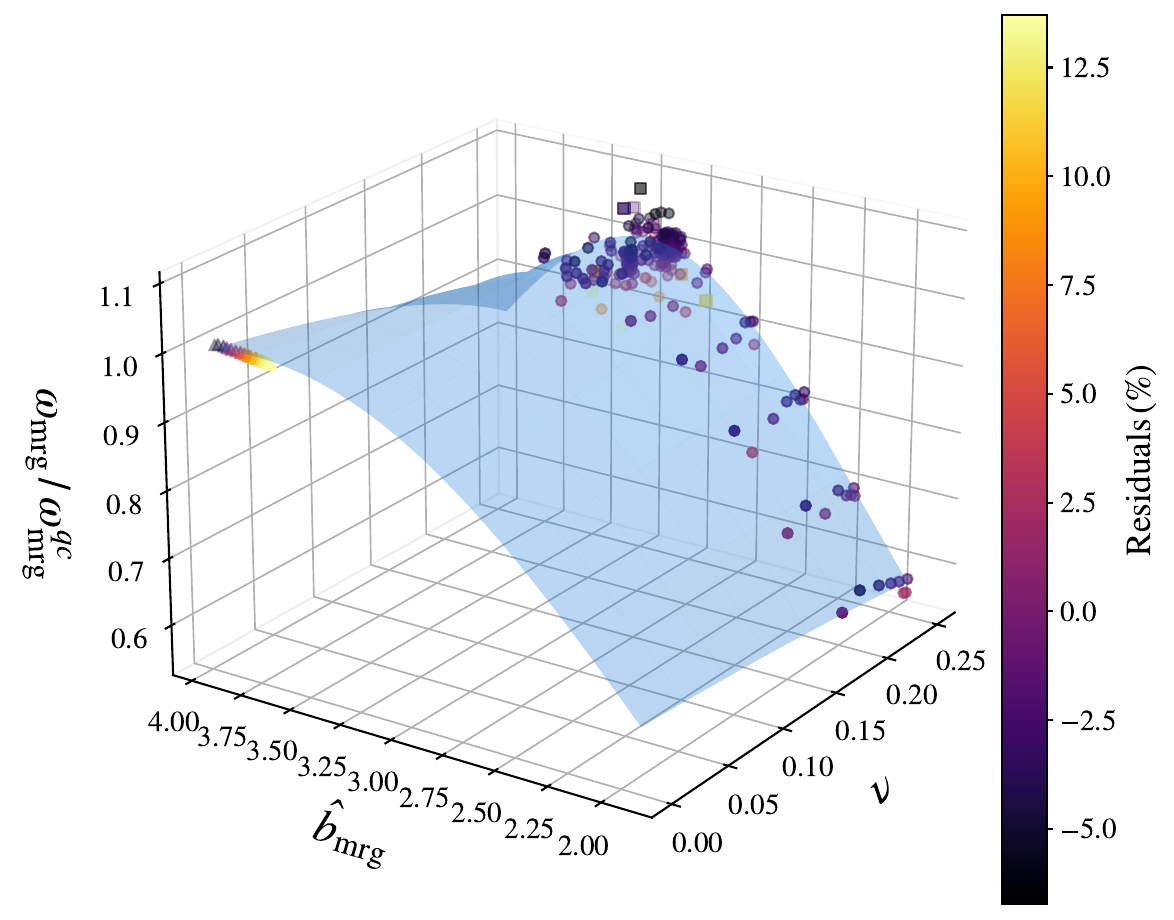}
         \includegraphics[width=0.48\textwidth]{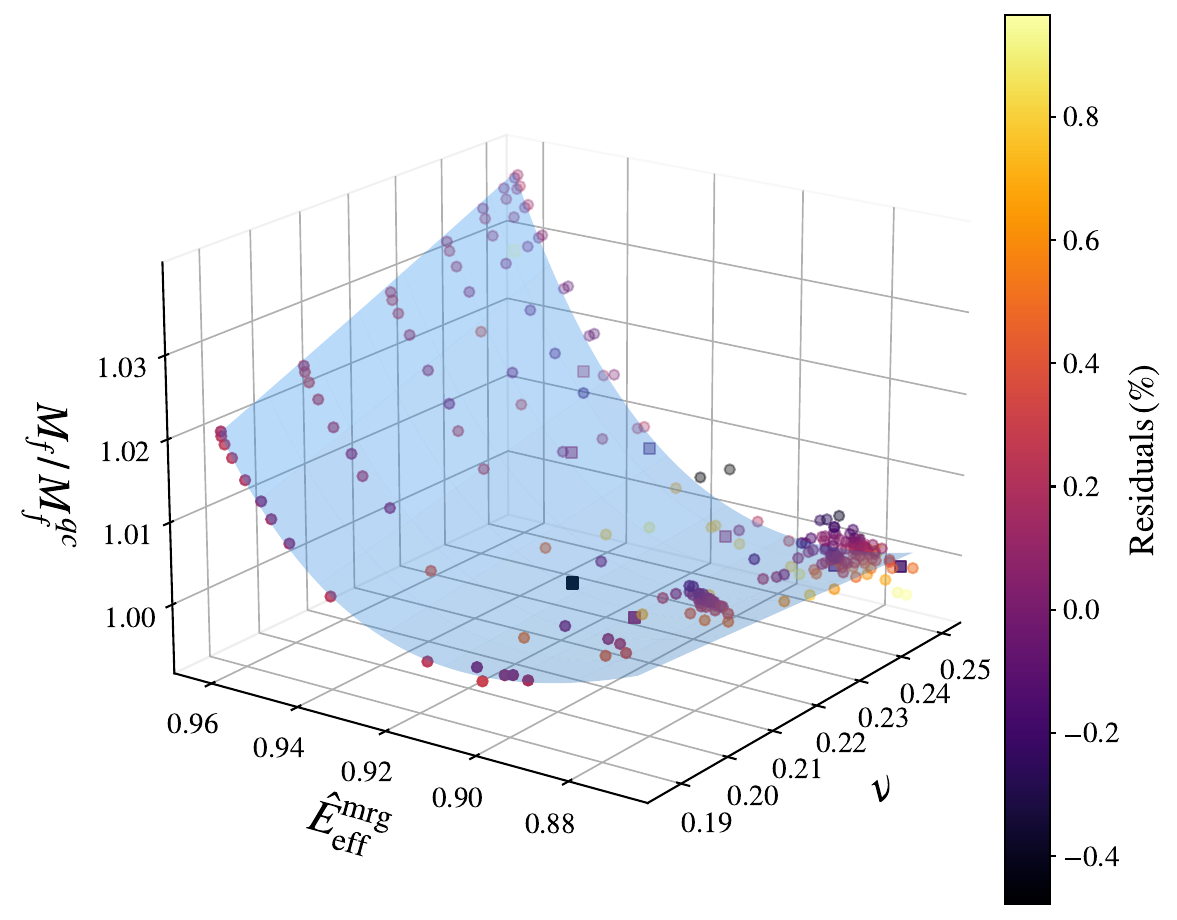}
         \includegraphics[width=0.463\textwidth]{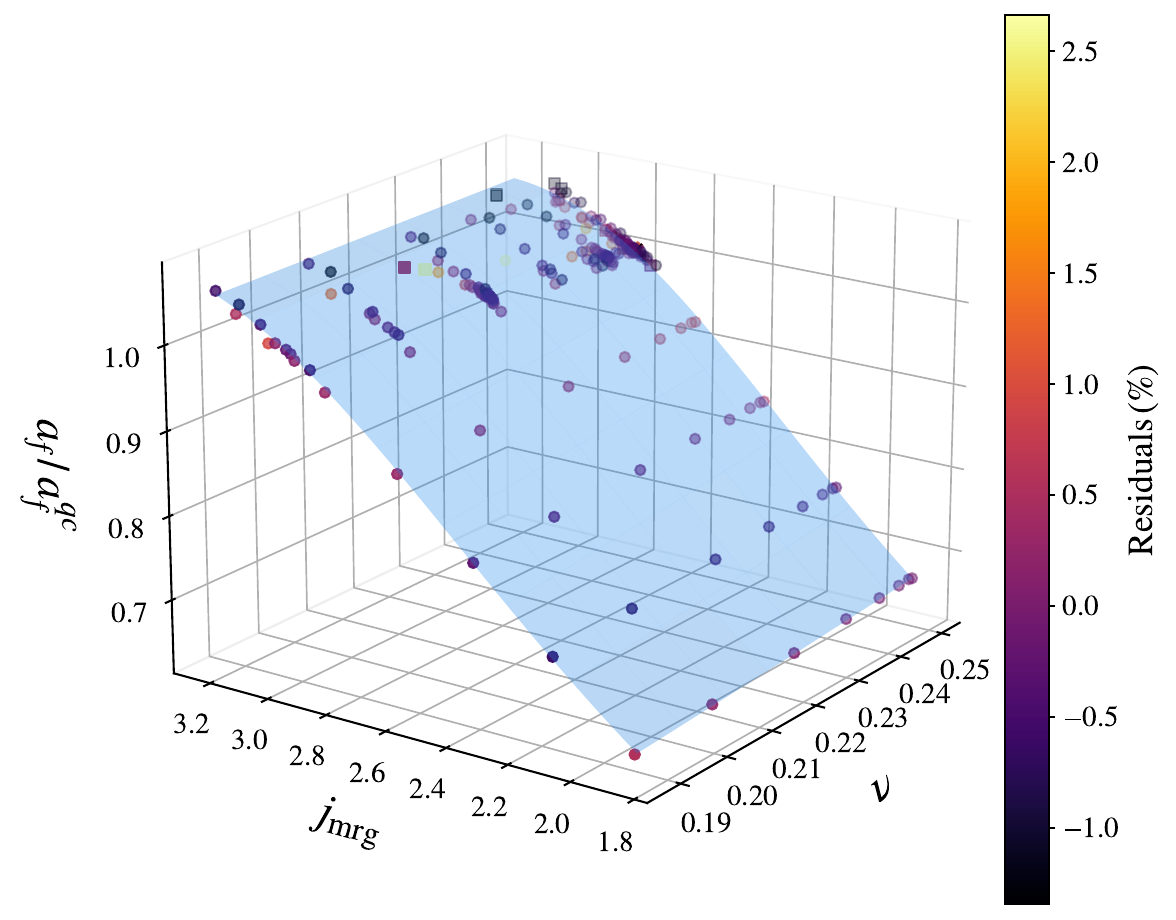}
         
   \caption{
   Quasi-universal relationships for the nonspinning unequal mass dataset (SXS, ET, RIT catalogs, with markers matching the previous figures, and triangles indicating RWZ data).
   The same structure observed in the equal mass case holds for arbitrary mass ratios, including the test mass limit.
   }
   \label{fig:unequal_mass_single_par}
\end{figure*}

\noindent {\textbf{\textit{Dataset}}.}
%
We restrict to binaries with progenitors spins aligned to the orbital angular momentum and employ 311 publicly available noncircular bounded simulations. 
The vast majority of them is contained in the RIT catalog~\cite{Healy:2022wdn}, complemented by available simulations from the SXS collaboration~\cite{Chu:2009md,Lovelace:2010ne,Lovelace:2011nu,Buchman:2012dw,Hemberger:2013hsa,Scheel:2014ina,Blackman:2015pia,Lovelace:2014twa,Mroue:2013xna,Kumar:2015tha,Chu:2015kft,Boyle:2019kee,SXS:catalog}.
The RIT catalog spans the ranges $q=[1,32]$, $e_0 = [0,1]$, $\chi_{1,2} = [-0.8, 0.8]$, while the SXS one has ranges $q=[1, 3]$ $e_0 = [0,0.2]$, $\chi_{1,2} = [-0.5, 0.85]$.
Here, $e_0$ is the nominal gauge-dependent eccentricity of the simulation.
Additionally, we employ a custom dataset of the latest stages of nonspinning dynamical captures simulations generated with the Einstein Toolkit (ET) package~\cite{Loffler:2011ay, Brandt:1997tf, Ansorg:2004ds, Baumgarte:1998te, Shibata:1995we}, with a range $q=[1.0, 2.15]$, see Ref.~\cite{Andrade:2023trh}.
These additional simulations allow us to show that the quasi-universal behavior under investigation is not restricted to bound orbits, but is a generic feature in planar orbits.
Finally, we consider also the test-mass data of Ref.~\cite{Albanesi:2023bgi}. 
The latter are generated by eccentric inspirals of a non-spinning test-particle around a Schwarzschild BH, driven by a EOB-based radiation reaction.
They span a large range of eccentricities, and have been computed by numerically solving the Regge-Wheeler 
and Zerilli (RWZ) inhomogeneous equations~\cite{Regge:1957td,Zerilli:1970se,Nagar:2005ea,Martel:2005ir} 
with \texttt{RWZHyp}~\cite{Bernuzzi:2010ty,Bernuzzi:2011aj,Bernuzzi:2012ku}.

Details of simulations used and our selection criterion based on the quality of the numerical data (quantified by balance laws), are provided in the Supplemental Material.

\noindent {\textbf{\textit{Results}}.}
%
We start by considering the equal mass nonspinning case, with five free parameters in our template.
The results are shown in Fig.~\ref{fig:1D_equal_mass}.
The merger parameters are well-described by $Q=\hat{b}_{\mathrm{mrg}}$ in our template, as expected from perturbation theory, 
while the remnant BH properties $M_f / M_f^{\rm qc}$ ($a_f / a_f^{\rm qc}$) are naturally expressed in terms of $Q=\hat{E}^{\, \mathrm{eff}}_{\mathrm{mrg}}$ ($j_{\mathrm{mrg}}$).
This level of discrepancy in $A_{\rm mrg} / A_{\mathrm{mrg}}^{\rm qc}$ is fully compatible with the expected numerical error of the simulations ($1\%$) for the vast majority of the cases~\cite{Healy:2022wdn}. 
Note the non-monotonic behaviour of the amplitude as a function of eccentricity: a stronger merger emission is observed for moderate eccentricities, while larger eccentricities display a highly suppressed merger amplitude.
The non-monotonic behaviour of the amplitude as a function of eccentricity is smoothly accounted for by our variable.
In the case of $\omega_{\rm mrg} / \omega_{\rm mrg}^{\rm qc}$, the residuals stay below $3\%$ for the majority of the cases (83/91 simulations), with a few outliers that reach up to $10\%$.
As discussed in Ref.~\cite{Albanesi:2023bgi}, the $\omega_{\mathrm{mrg}}$ variation with eccentricity is suppressed compared to the amplitude for intermediate eccentricities, hence the impact of numerical noise on this quantity is larger (compare the  patterns regularity between top left and right panels of Fig.~\ref{fig:unequal_mass_single_par}).
The impact on noncircular corrections on $M_f / M_f^{\rm qc}$ is the smallest among all considered quantities.
Our parameterization captures its variation with an accuracy better than $0.5 \%$ for almost all simulations (89/91 cases). 
The visual spread of the datapoints with respect to our model is apparently larger compared to the one encountered for other quantities, but fully within numerical accuracy $O(0.1\%)$, Ref.~\cite{Healy:2022wdn}, and consistent with our flux-based analysis discussed above.
The noncircular behavior of $a_f$ is very well captured by the evolved variable $j_{\mathrm{mrg}}$, with better than $1\%$ accuracy for almost all simulations considered (88/91 cases). Note the large variation (up to  $35\%$) compared to the quasicircular case.

Although our strategy allows for an accurate representation with a \textit{single} effective parameter (quasi-universality), residuals show a subdominant trend.
This is expected, since initial conditions in the noncircular case are determined by \textit{two} parameters, not one.
The generalization of the above relationship beyond quasi-universality, including two parameters, is discussed in the Supplemental Material.
Such generalizations naturally provide yet smaller residuals (given the larger dimensionality of the parameter space employed), allowing to construct even more accurate models.

Our description maintains a comparable accuracy also in the nonspinning unequal mass case, where $\nu-$dependent corrections are folded-in through the $p_k$ factors in our template.
The number of free parameters is now nine, and results are shown in Fig.~\ref{fig:unequal_mass_single_par}.
Even in this extended parameter space, the datapoints lie on a surface, and display smooth changes in terms of the fitting variables all the way to the test mass case.
In the Supplemental Material we discuss the extension of these fits beyond quasi-universality.
The inclusion of multiple variables has a more pronounced impact on the unequal-mass dataset, bringing the accuracy to the remarkable level already achieved in the equal mass case, with the bulk of the residuals of $O(1\%)$  on $(A_{\rm mrg}, a_f)$, $O(3\%)$ on $\omega_{\rm mrg}$, and $O(0.1\%)$ on $M_f$.

Finally, our generic strategy can be straightforwardly applied to the case of progenitors with spin.
Since this application is conceptually identical to the nonspinning case, we defer these results to the Supplemental Material.
There, we show how the same relationships are naturally extended to this case, with an accuracy at the same level of the nonspinning one.

\noindent {\textbf{\textit{Discussion}}.}
%
We have identified the existence of nontrivial order and a simple structure in all public NR simulations of noncircular binaries with aligned-spins progenitors.
This structure relies on evolved dynamical variables, such as the dynamical impact parameter $\hat{b}_{\mathrm{mrg}}$, uncovering quasi-universal 
relationships yielding highly accurate models of the merger properties.
Such construction, inspired from the behavior observed in the test-mass limit, avoids the issues arising from eccentricity-based parameterizations and frequency peaks interpolations.
Our results apply equally well to bounded and hyperbolic-like orbits from multiple NR catalogs (including test-mass data), unifying these a priori different regimes into a unique framework.
The presented relationships yield the necessary building blocks required to construct highly accurate, semi-analytical full-waveform models, 
valid for generic planar orbits with mismatches smaller than $1 \%$~\cite{Andrade:2023trh}.
This strategy will allow to include numerical information into phenomenological waveform models (see Suppl. Mat.) even beyond what has been considered here. 
For example, a natural application will be to generalize the templates of Refs.~\cite{Damour:2014yha,London:2018gaq}, incorporating their coefficients dependence on $\hat{b}_{\mathrm{mrg}}$.
These results enable new exciting discoveries of binaries in non-standard configurations, with dramatic implications on our understanding of binary formation in chaotic astrophysical settings.

Our parameterization will also aid parameter estimation of GW signals emitted by noncircular binaries.
The correlations shown in Fig.~2 of Ref.~\cite{Gamba:2021gap} can now be seen to indicate the impact parameter as leading-order measurable quantity.
The same reasoning applies to the band-like structure observed in the number of encounters as a function of the binary initial conditions~\cite{Nagar:2020xsk, Gamba:2021gap}.
We thus expect our new variables to allow for a much easier sampling convergence (due to the higher degree of smoothness of our parameterization), unlocking new observational investigations of GW signals sourced by binaries in generic orbits, previously hindered by computational cost.\\

\noindent {\textbf{\textit{Acknowledgments}}.}
%
We are grateful to Carlos Lousto, James Healy, the Einstein Toolkit community and the SXS collaboration for maintaining the respective public waveform databases and accompanying software.
We thank Piero Rettegno, and all the participants to the workshop ``EOB@Work 2023'' for helpful interactions.
G.C. thanks Shilpa Kastha for stimulating interest in this problem and collaboration in the initial attempts to tackle it; David Pereñiguez, Jaime Redondo-Yuste, Maarten van de Meent for fruitful discussions and comments; Luis Lehner and Perimeter Institute for Theoretical Physics for generous hospitality during the last stages of this work.
A.~N.~thanks the Niels Bohr Institute for hospitality during the development of this work.
G.C. acknowledges funding from the European Union’s Horizon 2020 research and innovation program under the Marie Sklodowska-Curie grant agreement No. 847523 ‘INTERACTIONS’, from the Villum Investigator program supported by VILLUM FONDEN (grant no. 37766) and the DNRF Chair, by the Danish Research Foundation.
R.G. acknowledges support from the Deutsche Forschungsgemeinschaft (DFG) under Grant No. 406116891 within the Research Training Group RTG 2522/1.
S.B. knowledges funding from the EU Horizon under ERC Consolidator Grant, no. InspiReM-101043372.
The work of T.A. is supported in part by the ERC Advanced Grant GravBHs-692951 and 
by Grant CEX2019-000918-M funded by Ministerio de Ciencia e Innovaci\'on (MCIN)/Agencia 
Estatal de Investigaci\'on (AEI)/10.13039/501100011033.
This research was supported in part by Perimeter Institute for Theoretical Physics. Research at Perimeter Institute is supported in part by the Government of Canada through the Department of Innovation, Science and Economic Development Canada and by the Province of Ontario through the Ministry of Colleges and Universities.\\

We release the dataset behind the figures and a \texttt{python} implementation of our fits in the repository: \href{https://github.com/GCArullo/noncircular_BBH_fits}{github.com/GCArullo/noncircular\_BBH\_fits}.
This study made use of the open-software \texttt{python} packages: \texttt{core-watpy, gw\_eccentricity, h5py, json, lal, matplotlib, numpy, pandas, pyRing, scipy, seaborn, sxs}~\cite{core, Shaikh:2023ypz, hdf5, json, lalsuite, matplotlib, numpy, pandas, pandas_zenodo, pyRing_2p3p0, scipy, seaborn, Boyle_The_sxs_package_2023}.

\clearpage


\renewcommand{\thesubsection}{{S.\arabic{subsection}}}
\setcounter{section}{0}

\section*{Supplemental material}

\begin{figure}[thbp]
         \includegraphics[width=0.98\columnwidth]{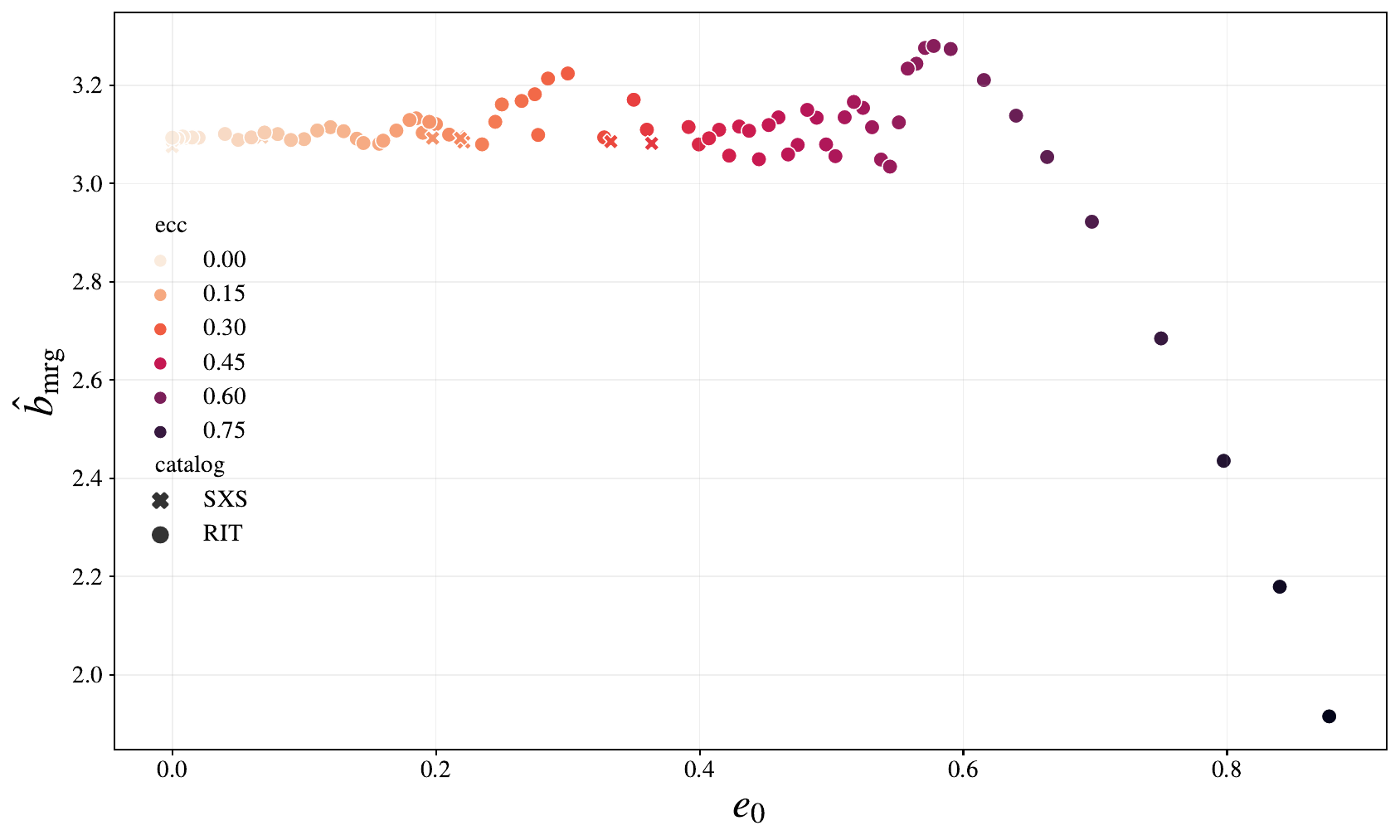}
   \caption{Impact parameter at merger as a function of the initial nominal eccentricity for bounded equal mass binaries. }
   \label{fig:bmrg_ecc}
\end{figure}

\noindent {\textbf{\textit{Alternative parameterizations}}.}
%
In Fig.~\ref{fig:bmrg_ecc} we show the relationship between $\hat{b}_{\rm mrg}$ and $e_0$. 
As expected from Fig.~1 in the main text, the relationship is oscillatory and multi-valued. 
It can be appreciated how even for very large eccentricities, the impact parameter still attains values far from the head-on case ($\hat{b}_{\rm mrg}=0$).\\

\begin{figure}[thbp]
         \includegraphics[width=0.98\columnwidth]{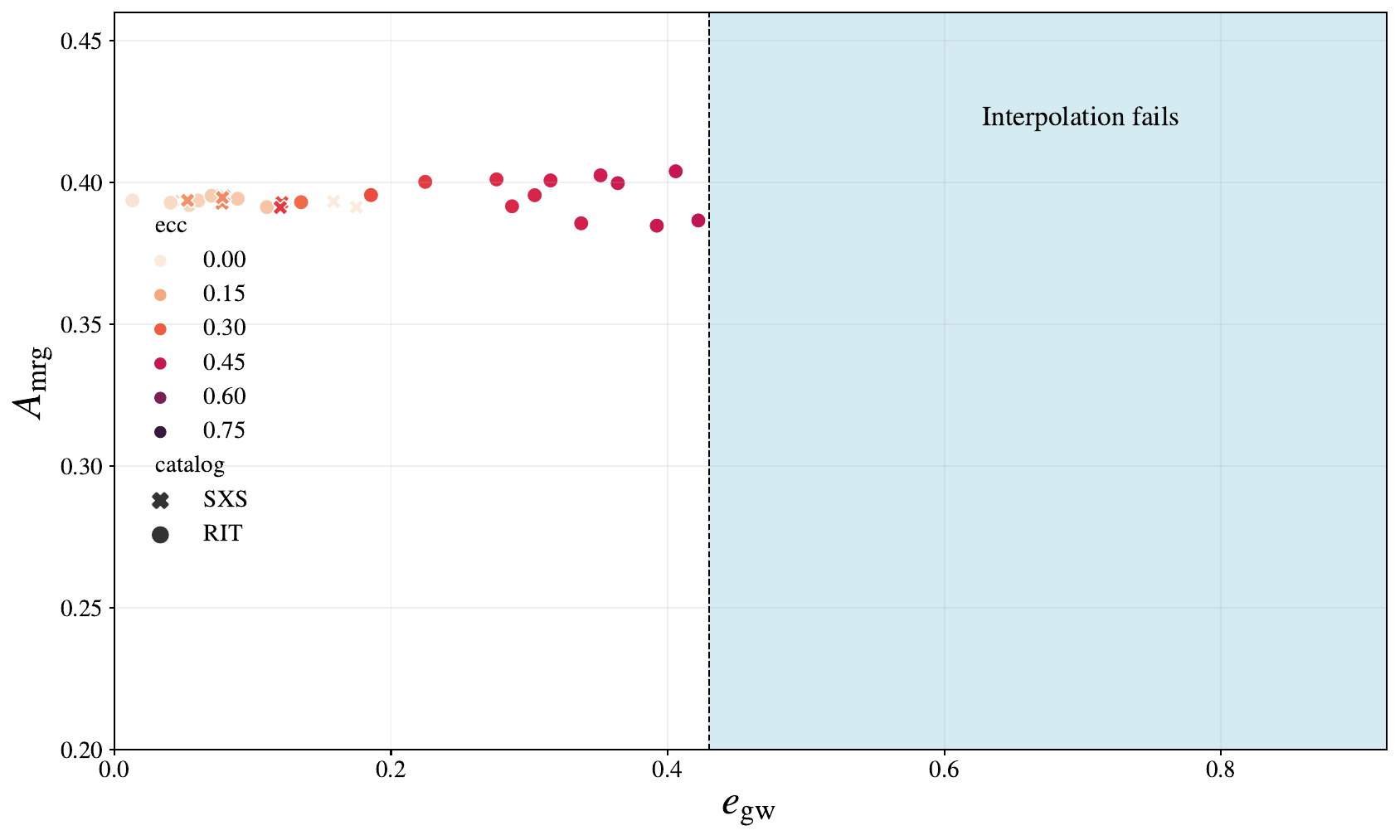}
   \caption{Merger amplitude as a function of the gauge-invariant eccentricity computed through the \texttt{gw\_eccentricity} package, for bounded equal mass binaries.}
   \label{fig:Amrg_ecc_gi}
\end{figure}

Instead, Fig.~\ref{fig:Amrg_ecc_gi} displays the relationship between $A_{\rm mrg}$ and $e_{\rm gw}$, the gauge-invariant eccentricity parameter introduced in Ref.~\cite{Shaikh:2023ypz}. 
The latter was computed using the \texttt{gw\_eccentricity-v1.0.4} package~\cite{Shaikh:2023ypz}, with the ``Amplitude'' method applied to $h_{22}$, at a reference time $t_{\rm ref}=-500M$ with respect to the waveform peak.
We have also experimented with all other methods available in the package and $t_{\rm ref}$ ranging within $[-1500,0] M$, obtaining a smaller number of successes in measuring the eccentricity and anomaly parameters.
This shows how the results shown in Fig.1 of the main text are not dependent on the specific eccentricity parameter employed, since the relationship remains multi-valued even when using a gauge-invariant eccentricity parameter.
At the same time, a much smaller number of points remain available to perform the fits, with no data above $e_{\rm gw}=0.4$, implying that this method cannot be employed for our scopes.\\

\noindent {\textbf{\textit{Fluxes computation}}.}
%
The energy and angular momentum fluxes are computed according to:

\begin{align}\label{eq:fluxes}
\dot{E} &= \frac{1}{16 \pi} \,  \sum_{\ell=2}^{\ell_{\mathrm{max}}} \sum_{m=-\ell}^{\ell} |\dot{h}_{\ell m}|^2\\
\dot{J} &= \frac{1}{16 \pi} \, \sum_{\ell=2}^{\ell_{\mathrm{max}}} \sum_{m=-\ell}^\ell m \cdot \Im(h_{\ell m} \cdot \dot{h}^{*}_{\ell m})\\
\end{align}

with all the available modes: $\ell_{\mathrm{max}} = 5$ for RIT, $\ell_{\mathrm{max}} = 8$ for SXS, and $\ell_{\mathrm{max}} = 8$ for ET data.
For ET data we do not include the $m=0$ modes as they are not well-resolved in the simulation. These modes are however completely negligible compared to the total flux for all the configurations considered in this study.
Derivatives are computed using centered second order accurate finite-differencing.
For data which are equally spaced (such as in the RIT catalog), integrals are computed using second order accurate discrete anti-derivatives, while for unequally spaced data (such as the ones available from the SXS catalog) we use the trapezoidal rule accessed through \texttt{numpy}.
We verified that for all the data considered, the time step is sufficiently small that numerical inaccuracies introduce negligible errors.
An implementation of the second order accurate formulas we use is publicly available in the \texttt{core-watpy} waveform analysis package~\cite{core}.


\noindent {\textbf{\textit{Template implementation}}.}
%
In our study, we construct appropriate variables evaluated at $t_{\rm mrg}$ to model the key merger parameters.
Waveform models are instead generated by specifying a set of \textit{initial} conditions.
Hence, when applying our variables to the construction of an inspiral-plunge-merger-ringdown model, the following practical procedure would be followed:

\begin{enumerate}

	\item Initial conditions are set by ($E_0^{\rm ADM}, J_0^{\rm ADM}$). If working within an EOB framework, the appropriate transformation between EOB and ADM coordinates should be applied. 
	An implementation of this step was already achieved in Ref.~\cite{Gamba:2021gap};
	
	\item The pre-merger waveform is used to compute the GW fluxes, yielding ($E (t), J (t)$). 
	A tabulation or analytic representation of the fluxes could also be pre-implemented to reduce the ``online'' computational cost.
	Inaccuracies due to semi-analytical approximations in the waveform model would enter this step, eventually deteriorating the accuracy of the predicted merger/remnant quantities.
	Given the smooth functional form of the relationships found, we expect our fits to be little sensitive against small errors in the fluxes.
	The propagation of such errors on merger/remnant fits would be less than linear, hence suppressed, when using evolved energy/angular momentum variables.
	This is because approximate semi-analitic fluxes control the \textit{variation} of $E(t),J(t)$ with respect to initial conditions, not their absolute values;
	
	\item The modelling variables $\{ \Eeffmrg, j_{\mathrm{mrg}}, \hat{b}_{\mathrm{mrg}} \}$ are constructed using evolved quantities ($E (t), J (t)$), and $\{ A_{\mathrm{mrg}}, \omega_{\mathrm{mrg}}, M_f, a_f \}$ computed through Eq.~(1) in the main text or Eq.~\eqref{eq:template_2D} below;
	
	\item The post-peak model in which $\{ A_{\mathrm{mrg}}, \omega_{\mathrm{mrg}}, M_f, a_f \}$ are used is constructed.
	
\end{enumerate}

The key point is that quantities discussed in this paper directly enter only the post-merger portion, while fluxes are computed through the pre-merger part of the waveform only.
We stress that this strategy is not restricted to EOB models, but can instead be applied to a generic IMR waveform template.\\

\noindent {\textbf{\textit{Details of the fits}}.} 
%
Lacking a simulation error estimate for our full dataset, we simply use a non-linear least squares algorithm to minimise the difference between our target quantities $\{ A_{\mathrm{mrg}}, \omega_{\mathrm{mrg}}, M_f, a_f \}$ and our template, Eq.~(1) in the main text.
We adopt the \texttt{scipy.optimize.least\_squares} function, bounding all the coefficients within the interval $[-10,10]$.
To avoid missing the minimum of the parameter space the algorithm is exploring, we repeat the fit with 10 different seed values for the coefficients, selecting the seed yielding the coefficients with the smallest value of the cost function.
The seed values are extracted from a normal distribution $\mathcal{N}(1, 1)$.
We verified that the different seeds deliver compatible residuals.
This points to a good convergence of the procedure, as expected in a low-dimensional problem when employing a set of variables that allow for a smooth parameterization, 
such as the ones described in the main text.
It is worth stressing that in Fig.~(2) of the main text we are quoting residuals relative to the quasicircular case. 
Since the quantities under consideration have absolute values below unity, for the vast majority of our dataset the error on the \textit{absolute values} of the noncircular quantities will be smaller than the ones reported.
The resulting coefficients and a \texttt{python} implementation of our factorised fits, which can be straightforwardly applied to augment existing quasicircular ones~\cite{Healy:2014yta, Jimenez-Forteza:2016oae, Nagar:2018zoe, Nagar:2019wds}, are publicly available in the repository: \href{https://github.com/GCArullo/noncircular_BBH_fits}{github.com/GCArullo/noncircular\_BBH\_fits}.\\

\begin{figure*}[thbp]
         
    \includegraphics[width=0.4\textwidth]{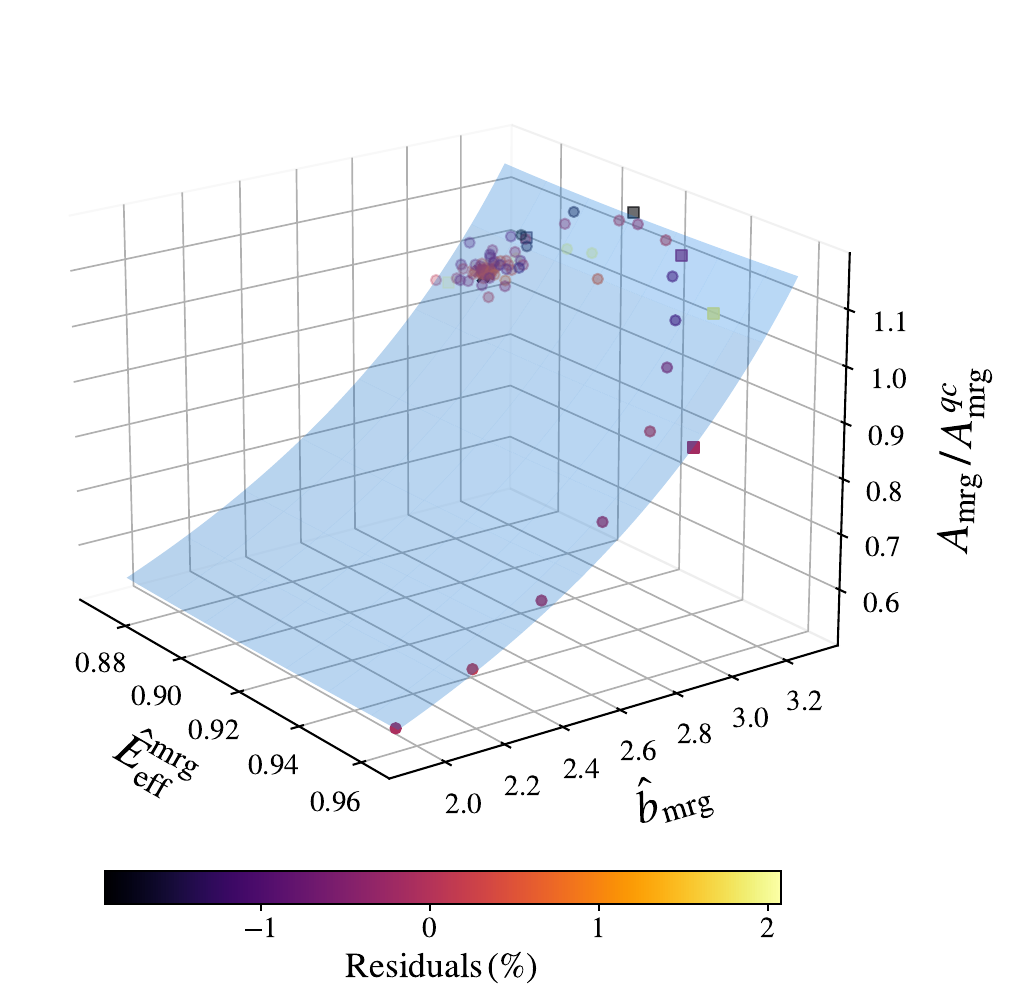}   
    \includegraphics[width=0.4\textwidth]{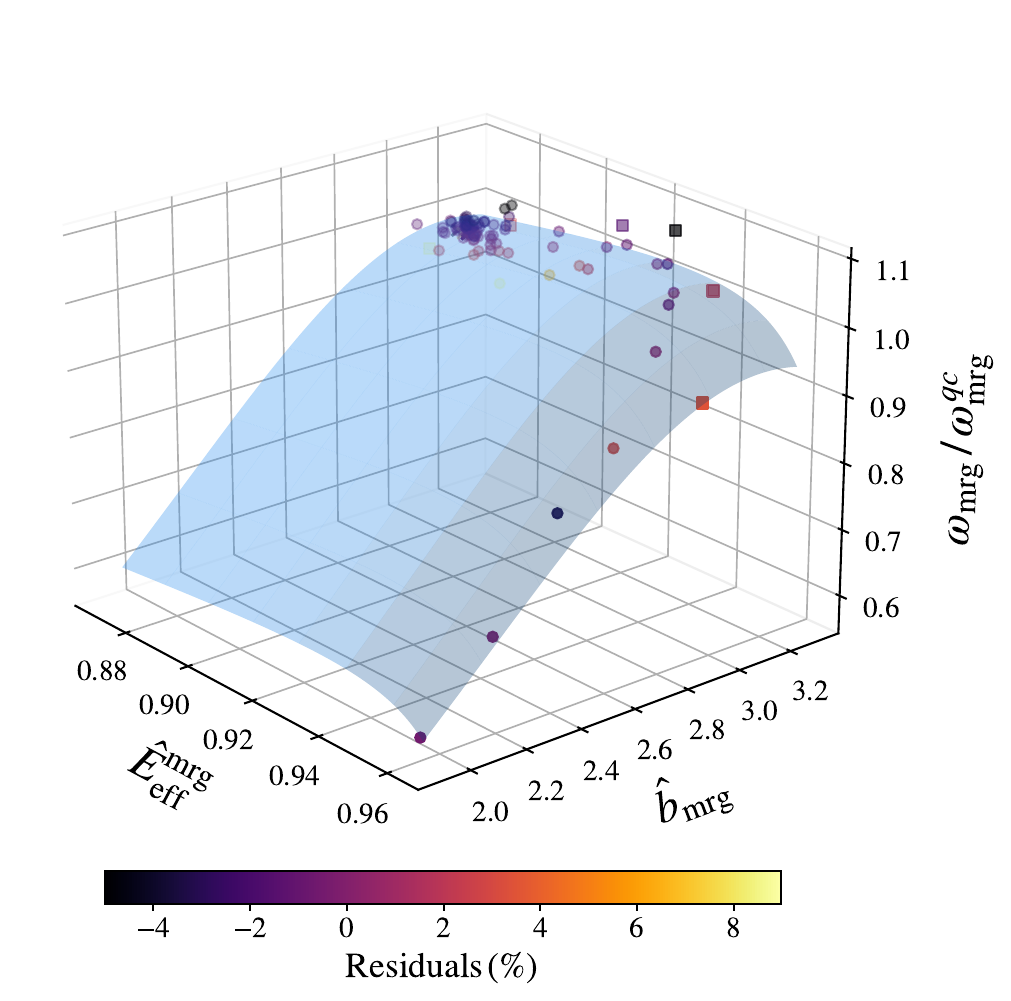}   
    \includegraphics[width=0.4\textwidth]{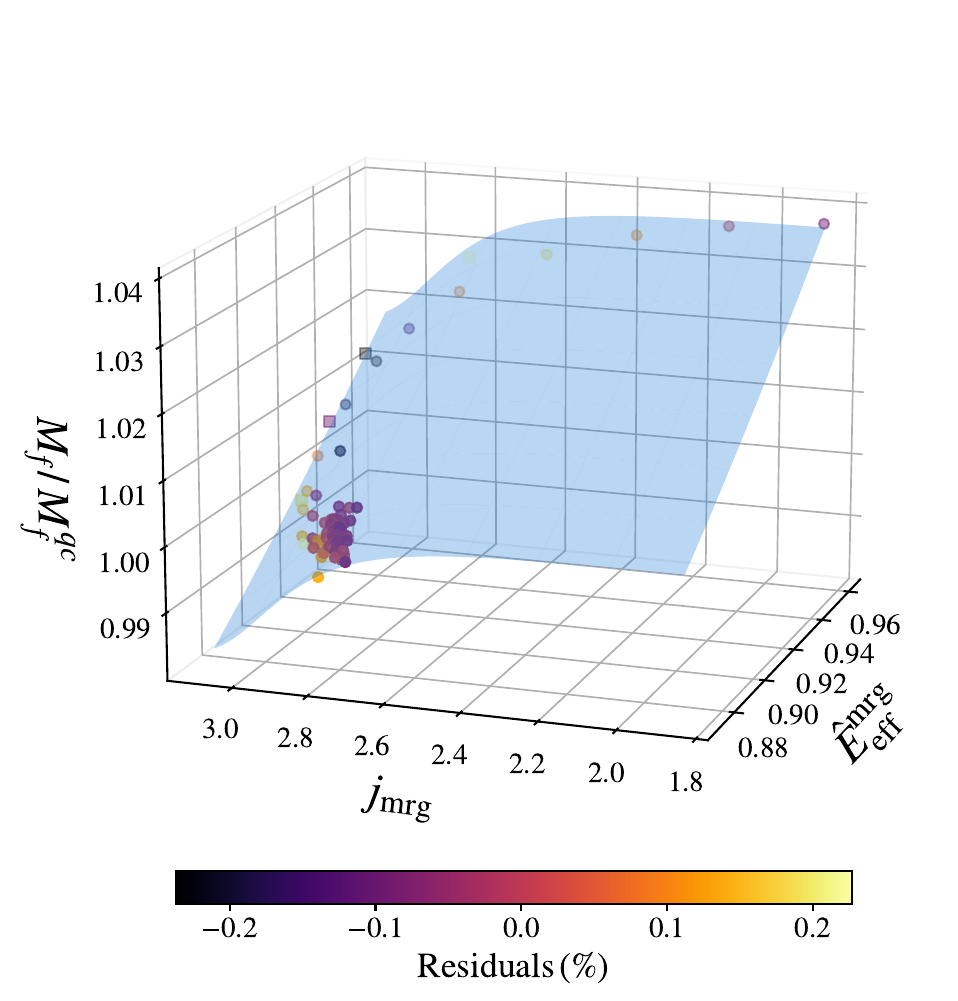}   
    \includegraphics[width=0.4\textwidth]{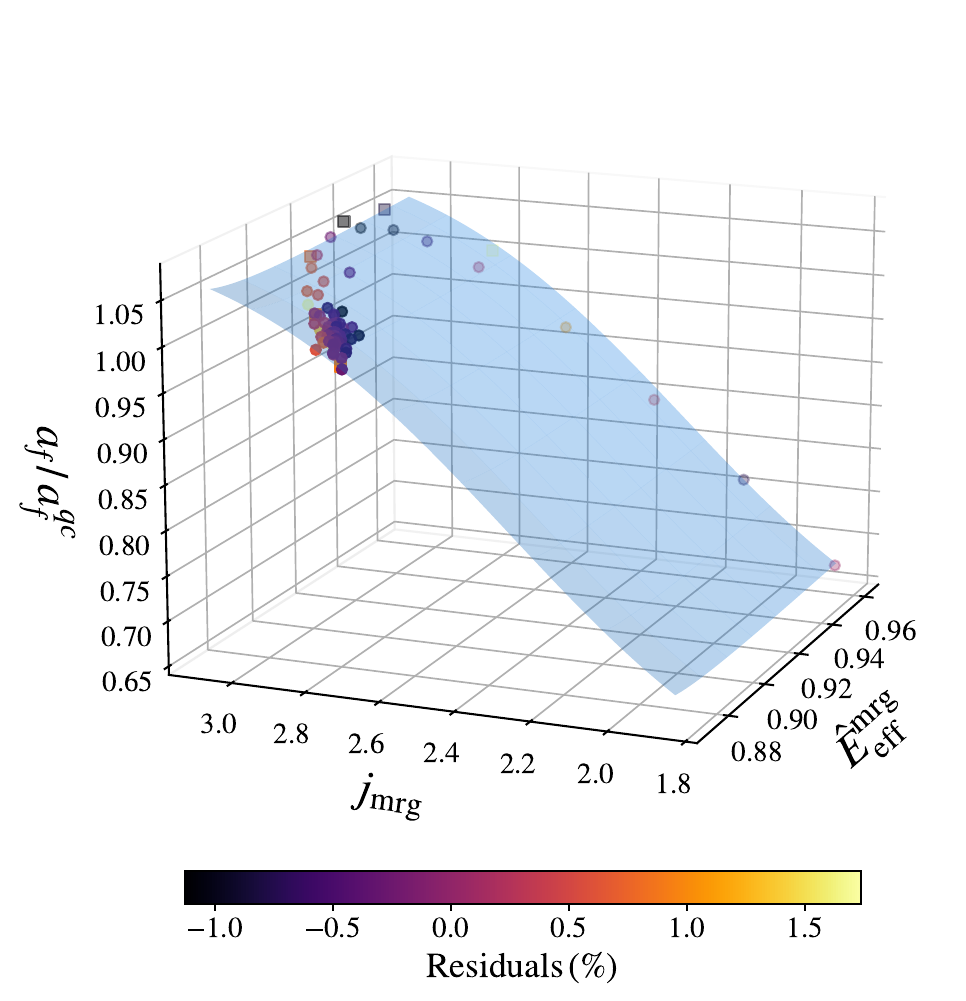}   
   \caption{Two-dimensional relationships of the merger quantities in terms of the dimensionless evolved variables for the comparable mass nonspinning dataset (SXS, ET, RIT, with markers matching the previous figures).}\label{fig:2D_equal_mass}
   
\end{figure*}

\noindent {\textbf{\textit{Beyond quasi-universality}}.}
%
In the main text we showed how the merger quantities dependence on the initial conditions can be very well-captured by a single effective parameter.
However, at a given initial frequency, the full binary dynamics in the noncircular case is determined by two parameters, typically chosen to be an eccentricity and anomaly parameter in the bounded case.
In the scattering case, to go beyond a description based on the impact parameter, one would need to include the asymptotic impulse~\cite{Kalin:2019rwq, Kalin:2019inp}.
To complement $\hat{b}_{\mathrm{mrg}}$, we instead use $\Eeffmrg$ as an additional parameter.
In the case of the remnant mass, which was already modeled through $\Eeffmrg$, we chose instead $j_{\mathrm{mrg}}$.

When using more than one fitting variable, we generalise Eq.~(1) in the main text to include a product of rational functions, one per fitting quantity:

\be\label{eq:template_2D}
    \tilde{Y} = \prod_{i=1}^{2} \, \tilde{Y}_0 \, \left( \frac{1 + p_{1,i} \, Q_i + p_{2,i} \, Q_i^2}{1 + p_{3,i} \, Q_i + p_{4,i} \, Q_i^2} \right) \,,
\ee
where $p_{k,i} = b_{k,i} \, (1 +c_{k,i} X)$ and $b_{k,i}, c_{k,i} \in \mathbb{R}$.

\begin{figure*}[thbp]

         \includegraphics[width=0.4\textwidth]{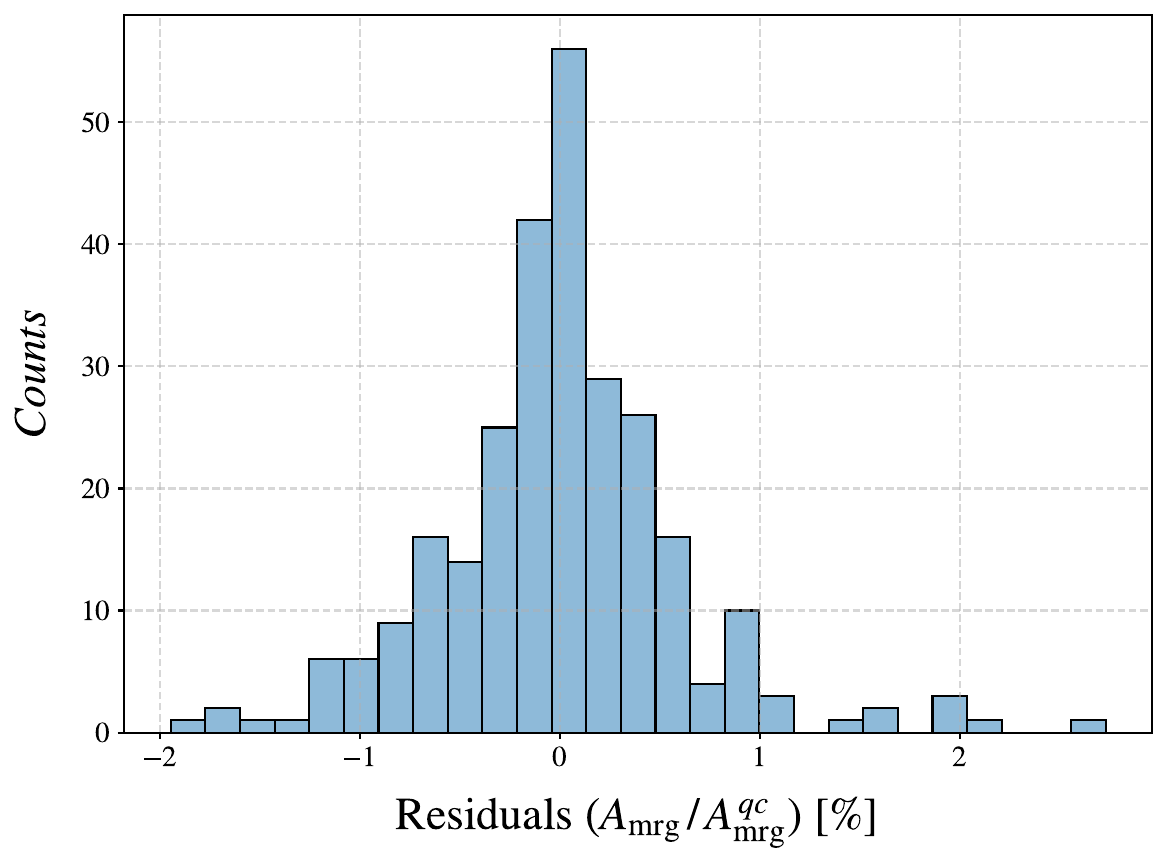}
         \includegraphics[width=0.4\textwidth]{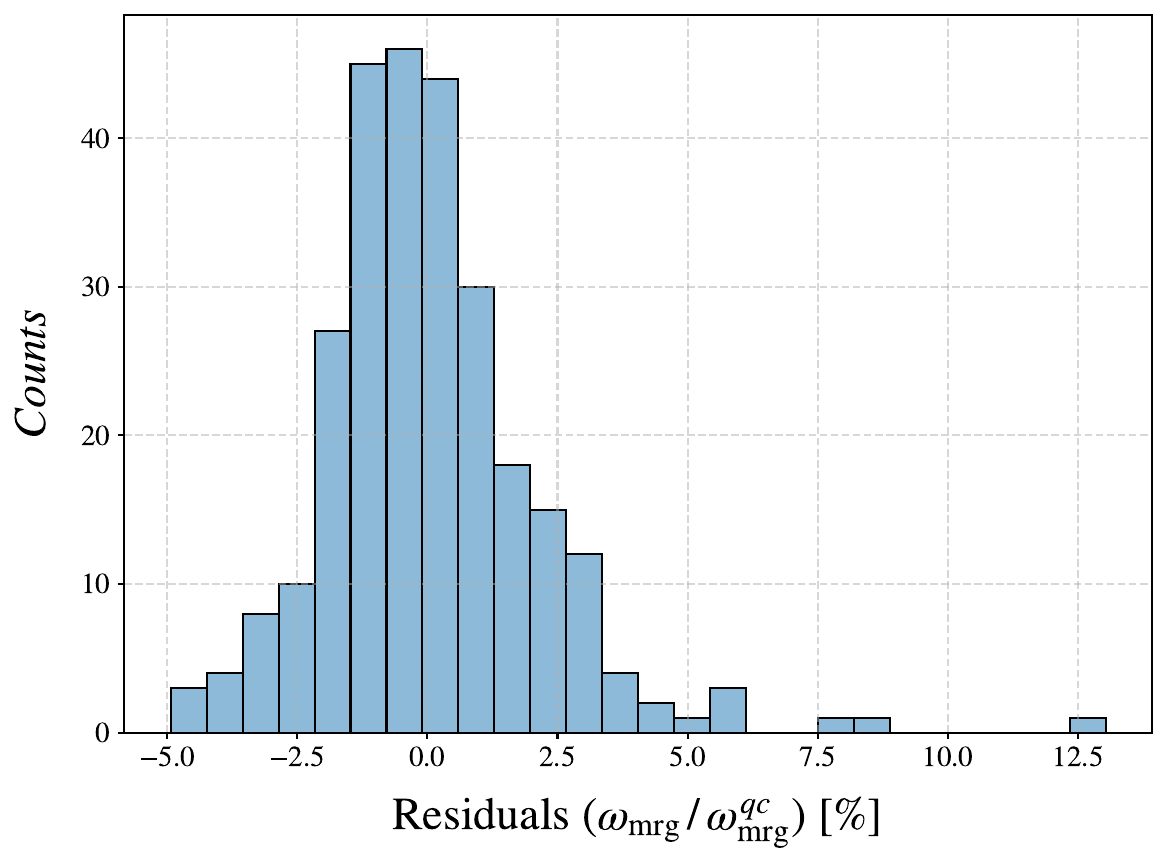}
         \includegraphics[width=0.4\textwidth]{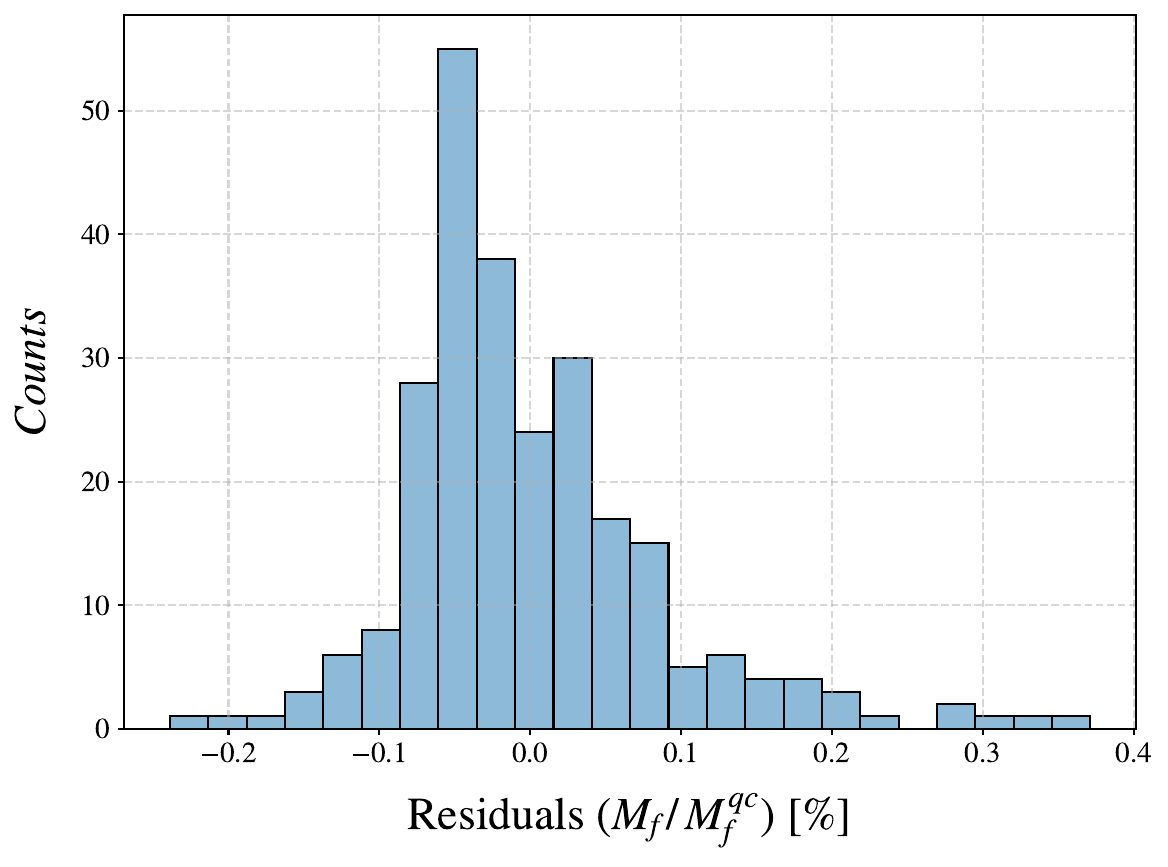}
         \includegraphics[width=0.4\textwidth]{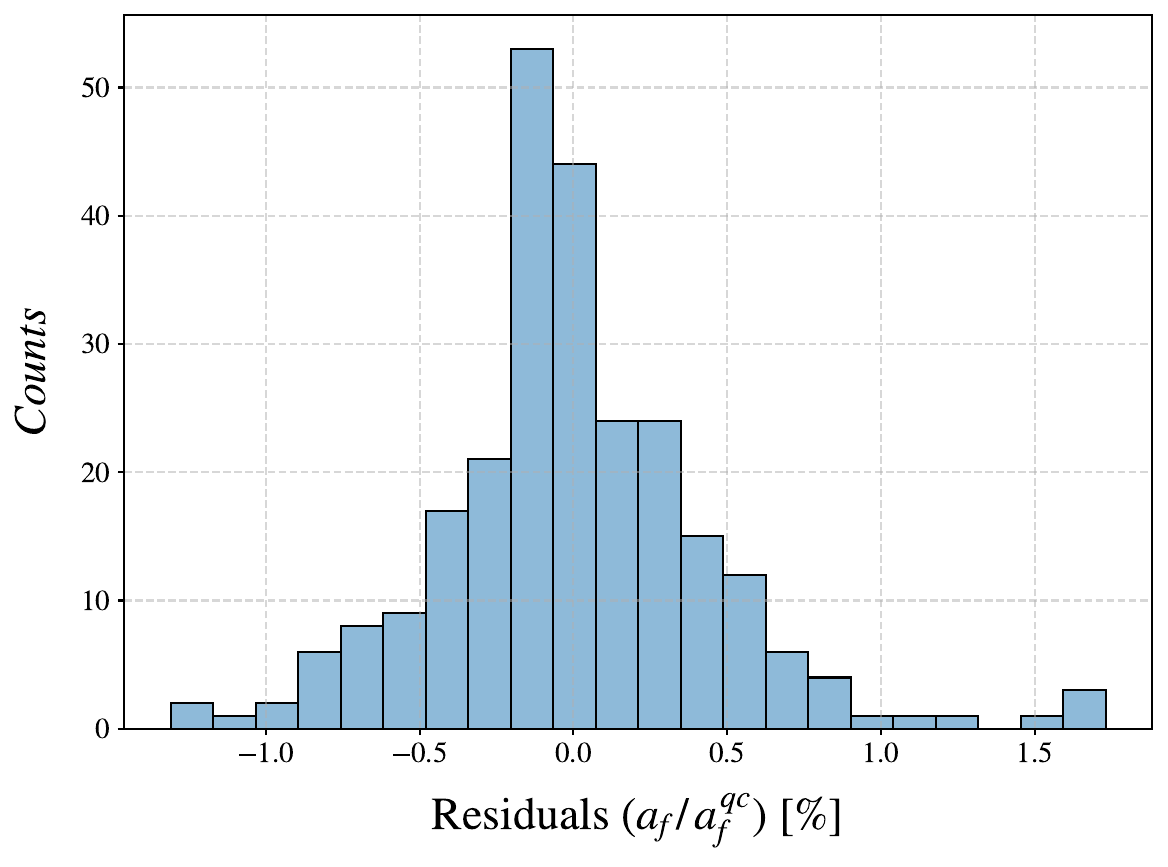}
   \caption{Residuals corresponding to the three variables fits (same combinations as Fig.~\ref{fig:2D_equal_mass}, plus the symmetric mass ratio) for the full nonspinning dataset.}
   \label{fig:3D_unequal_mass}
\end{figure*}

The results of these generalized relationships are shown in Fig.~\ref{fig:2D_equal_mass} for the equal mass case.
Although the additional variable helps in resolving subdominant structures present in our dataset, the residuals improvement compared to the single-variable case presented in the main text is modest.
This feature is not surprising since the precision of the single-variable relationships was already close to the expected numerical accuracy.
Considering that the number of free parameters contained in our new ansatz almost doubled, this shows that $\hat{b}_{\mathrm{mrg}}$ is already capturing close to the whole information content.
The residuals of the unequal mass case, depending on three fitting variables, are instead reported as histograms in Fig.~\ref{fig:3D_unequal_mass}.
In this more general case (except for $a_f$), the second fitting variable has a more significant impact: the same remarkable accuracy obtained for the equal-mass case is now maintained also for the full dataset.
Improvements to these relationships might come from more accurate NR simulations, or more flexible fitting functions.\\

\noindent {\textbf{\textit{Spinning binaries}}.} 
%
\begin{figure*}[thbp]
         
    \includegraphics[width=0.4\textwidth]{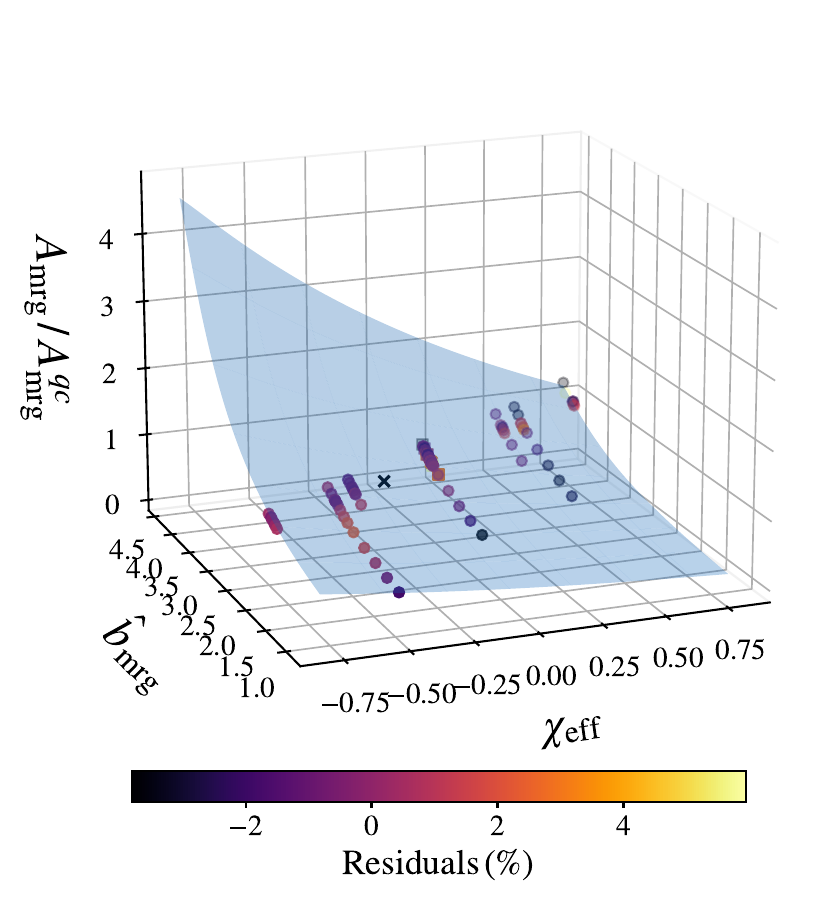}   
    \includegraphics[width=0.4\textwidth]{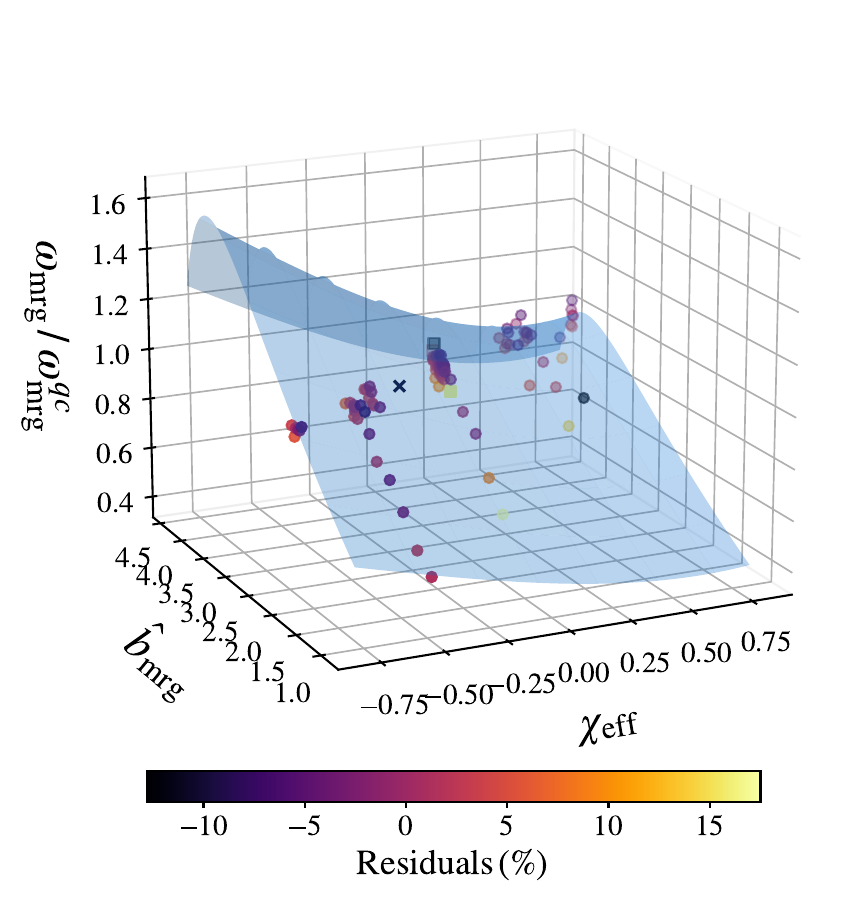}   
    \includegraphics[width=0.4\textwidth]{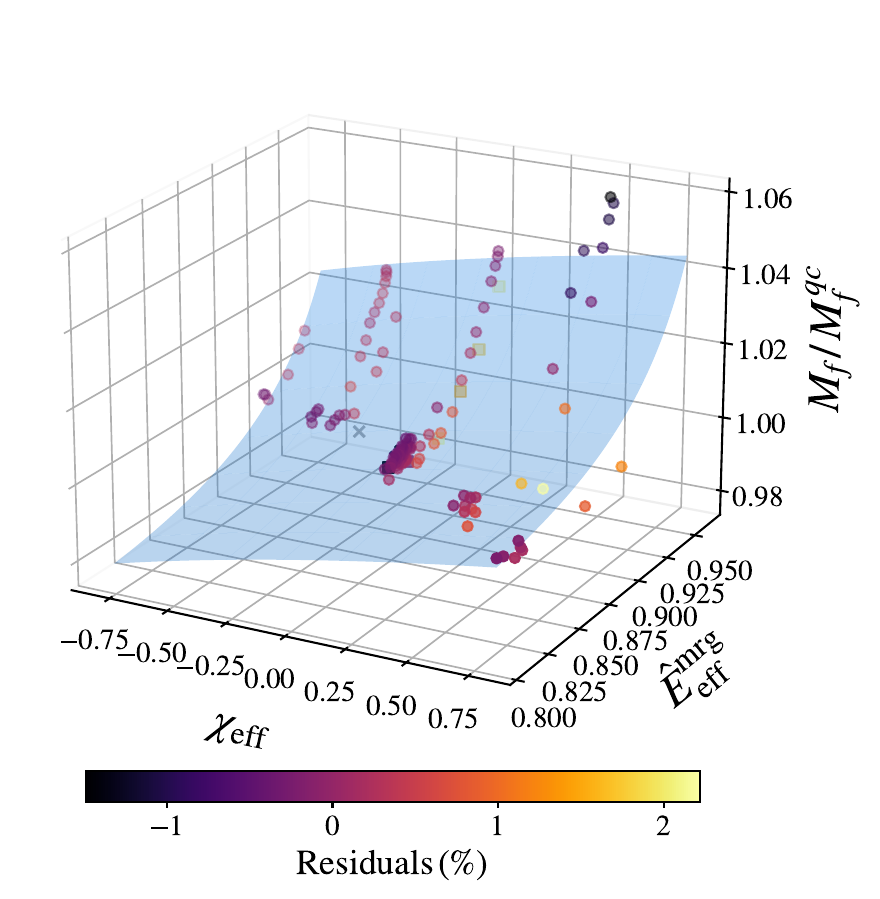}   
    \includegraphics[width=0.4\textwidth]{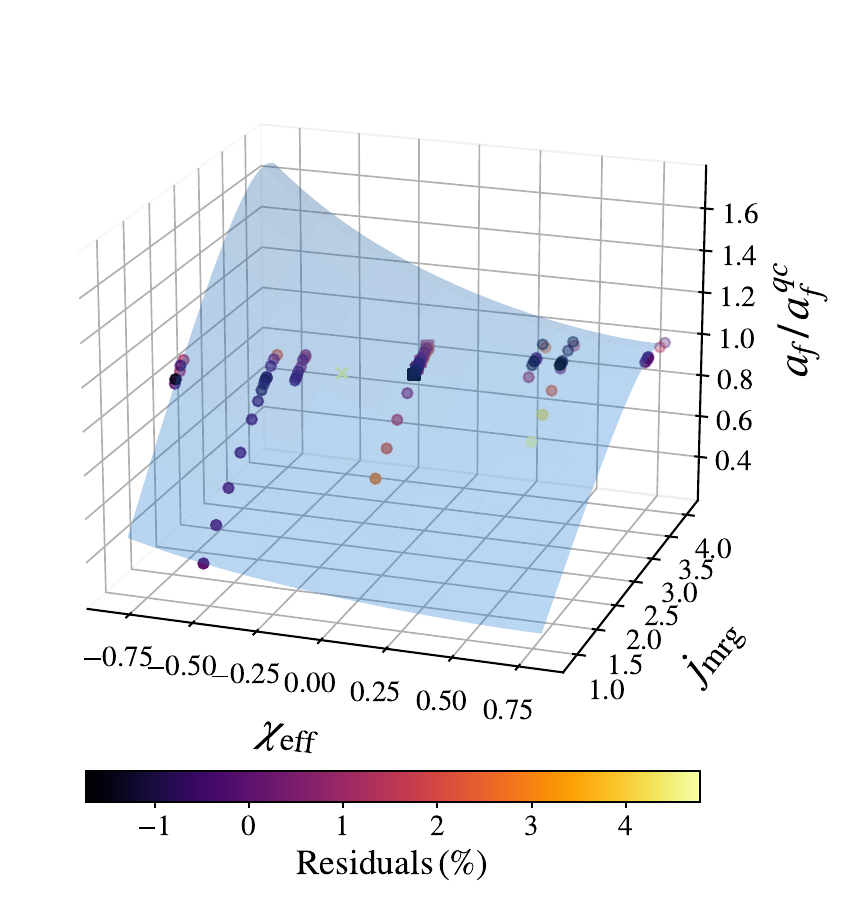}   
   \caption{Two-dimensional relationships of the merger quantities in terms of the dimensionless evolved variables described in the main text, and the effective spin parameter $\chi_{\rm eff}$, for the comparable mass aligned-spins dataset.}
   \label{fig:2D_equal_mass_aligned_spins}
\end{figure*}
\begin{figure*}[thbp]

         \includegraphics[width=0.4\textwidth]{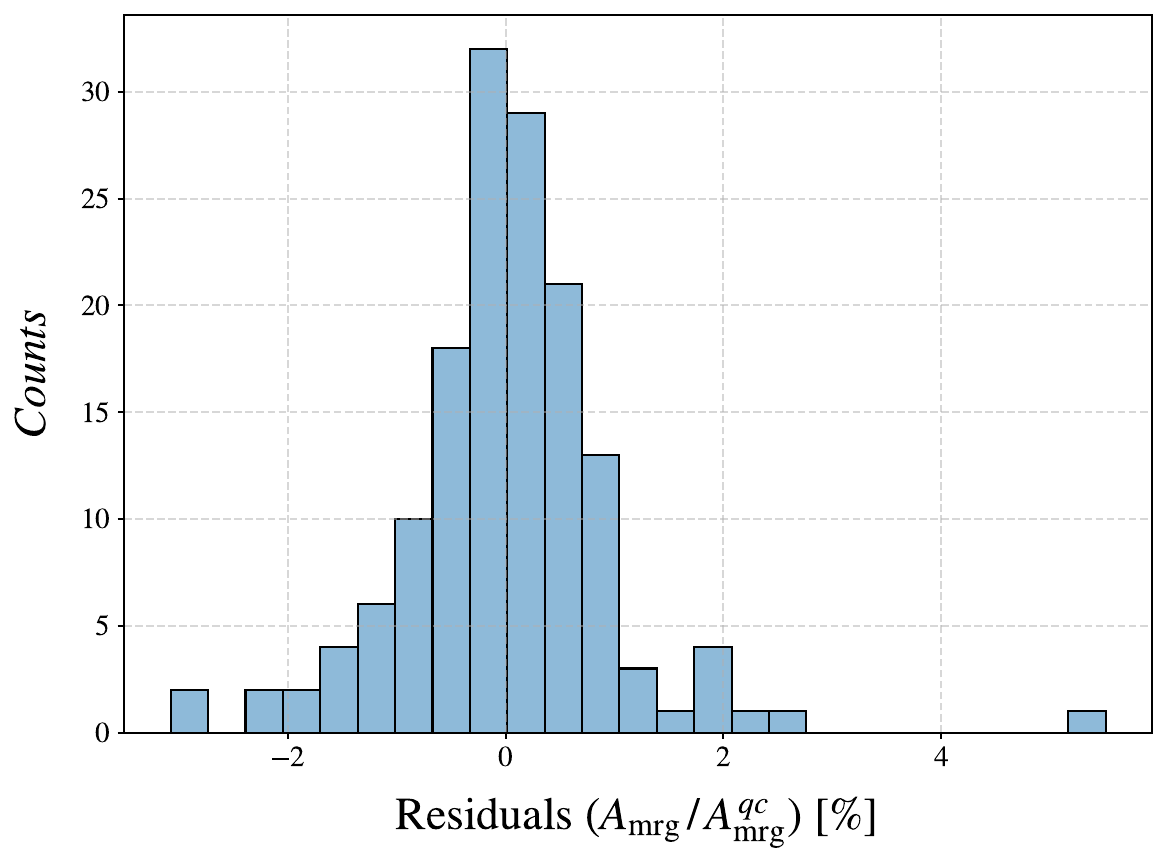}
         \includegraphics[width=0.4\textwidth]{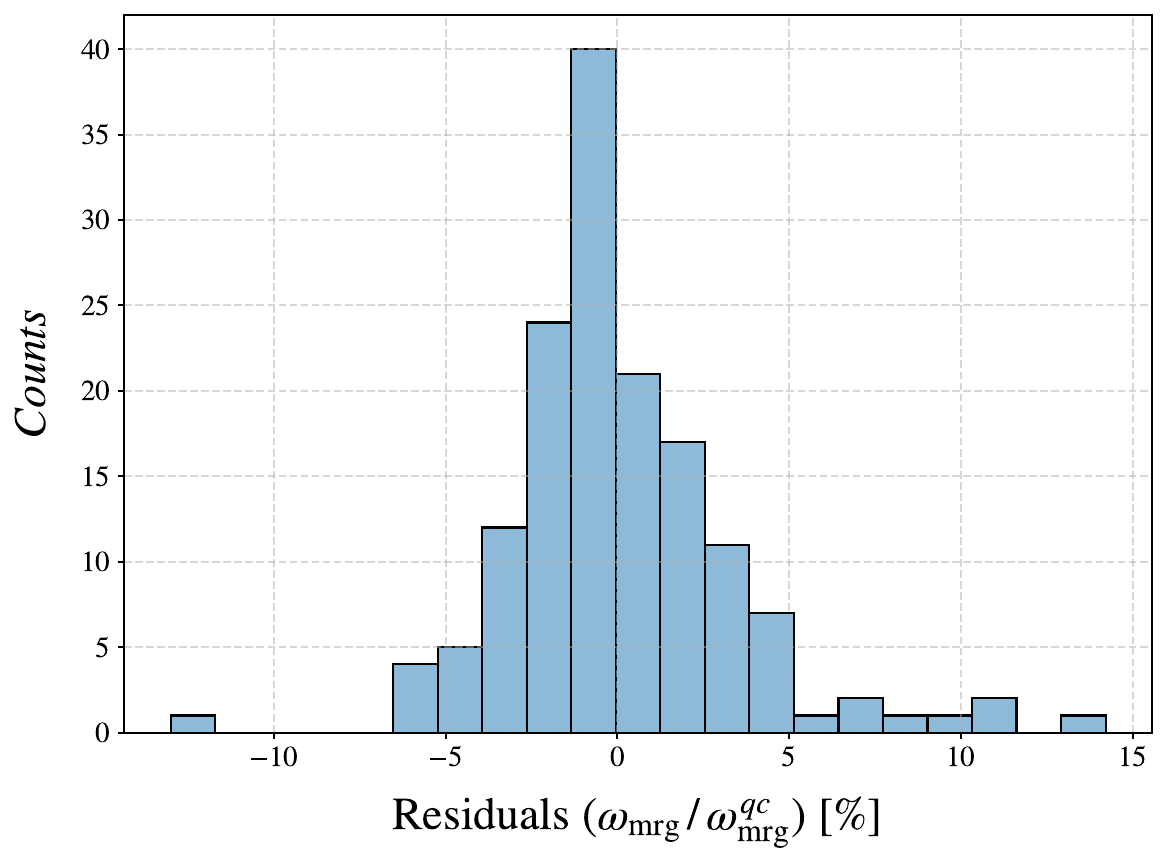}
         \includegraphics[width=0.4\textwidth]{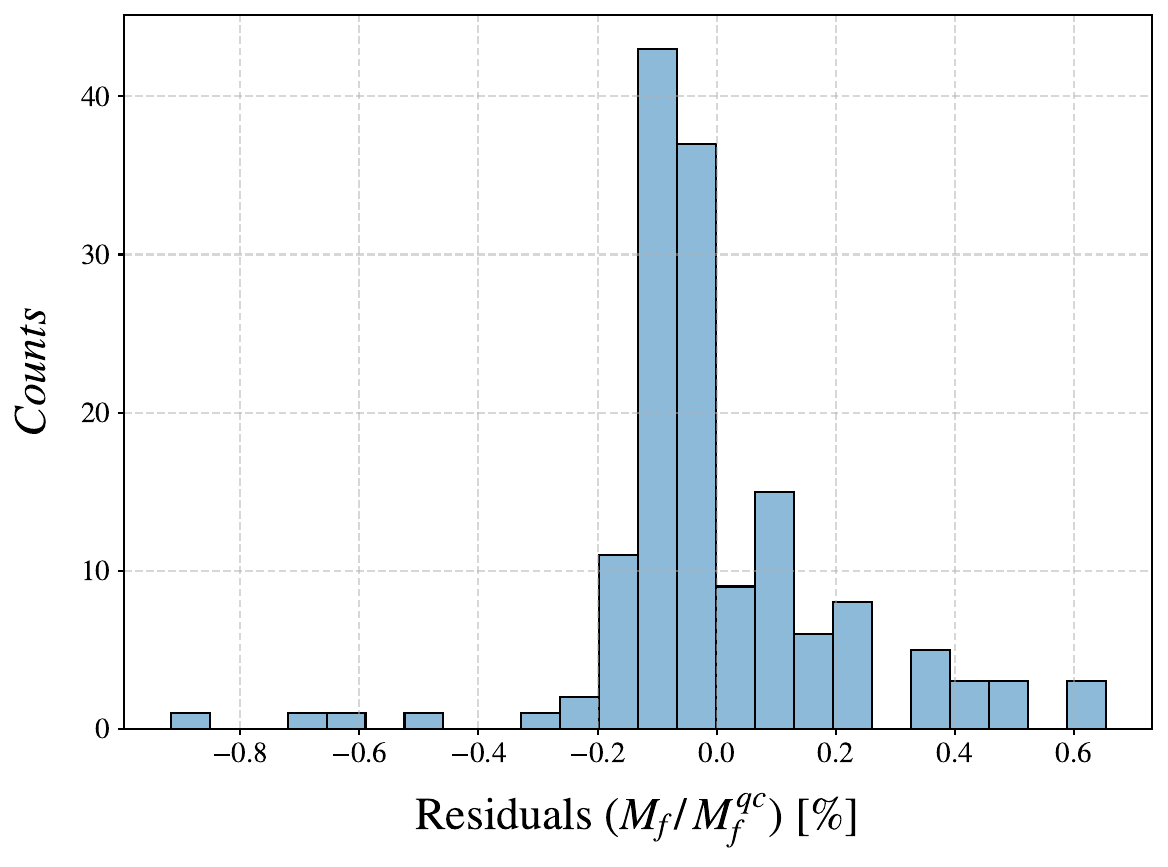}
         \includegraphics[width=0.4\textwidth]{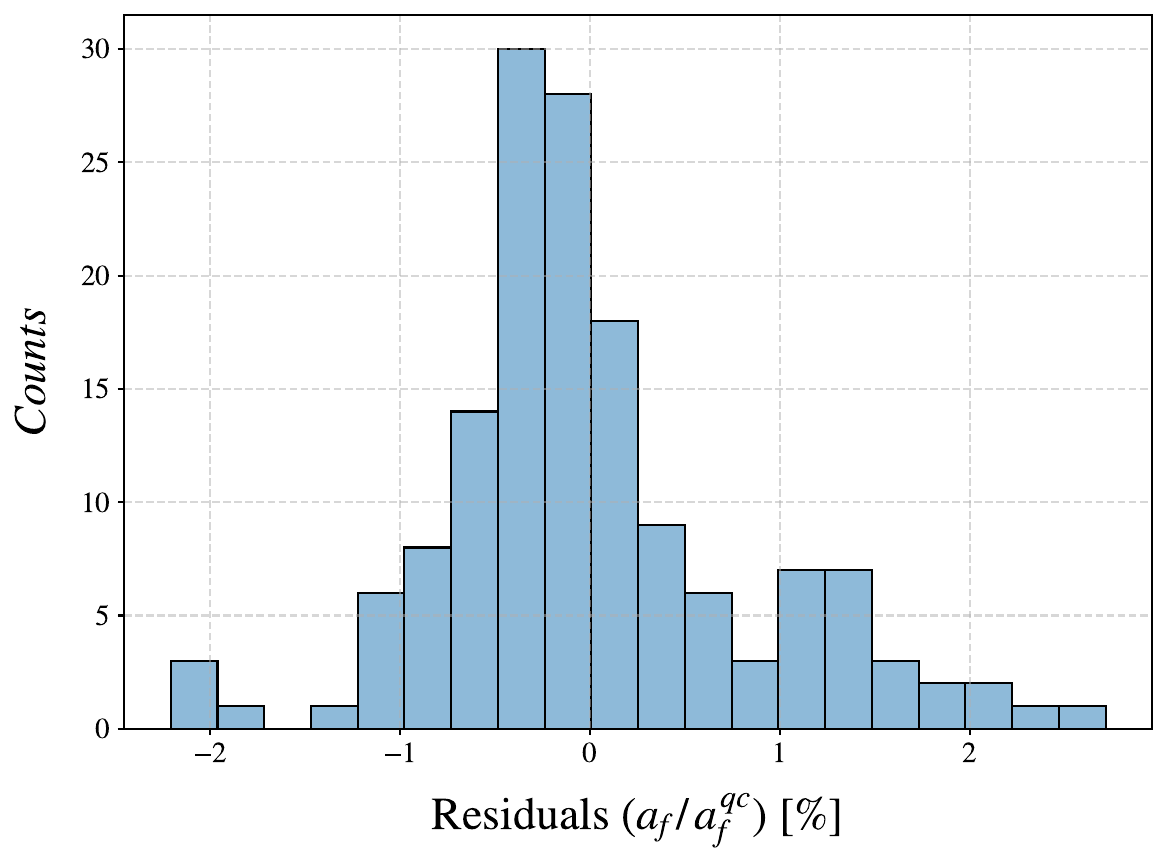}
   \caption{Residuals corresponding to the three variables fits (same combinations as Fig.~\ref{fig:2D_equal_mass}, plus $\chi_{\rm eff}$) for the full equal-mass aligned-spins dataset.}
   \label{fig:3D_equal_mass_aligned_spins}
\end{figure*}
Having modelled the merger structure using variables that encode the binary dynamics, it is straightforward to extend the relationships presented in the main text to the case of  binaries with spins aligned to the orbital angular momentum.
To incorporate the dominant spins effect, we use as fitting variable $\chi_{\rm eff}$, the leading-order spin parameter in a post-Newtonian expasion~\cite{Santamaria:2010yb}.
We repeat the fit described in the main text on the RIT spinning equal-mass dataset with the addition of this variable, but without changing any other element (fit functional form, number of free coefficients per fitting variable, etc.).

We show the results of this fit in Figs.~\ref{fig:2D_equal_mass_aligned_spins},\ref{fig:3D_equal_mass_aligned_spins}.
In principle, the fits should be performed using four fitting variables (two to represent the initial conditions, and two for the binary spins).
In Fig.~\ref{fig:2D_equal_mass_aligned_spins}, we show the results when assuming a single effective parameter to describe the two-dimensional space of the initial conditions, and the single effective parameter $\chi_{\rm eff}$ to incorporate spin effects.
The result shown in the figure, demonstrates how the ``quasi-universal'' relationships presented in the main text hold even in the spinning case, i.e. for the generic case of planar orbits.
The relationships remain in fact single-valued and non-oscillatory, with a smooth functional form.
Residuals on the amplitude remain below $2\%$ for $96\%$ of the cases, with only two outliers beyond $3\%$.
The merger frequency keeps showing more noisy results, but we can still obtain residuals below $5\%$ for $93\%$ of the cases, when still assuming four free parameters per fitting variable.
The final mass keeps being the best modelled parameter, with $99\%$ of cases below the $1\%$ threshold.
Also the remnant spin is very well-constrained, with $93\%$ of datapoints having residuals smaller than $2\%$, and $99\%$ below $3\%$.

Going beyond the quasi-universal case, in Fig.~\ref{fig:3D_equal_mass_aligned_spins} we now report the residuals of the fit employing two variables to represent the space of initial conditions, and $\chi_{\rm eff}$ to incorporate spin effects.
The residuals show a small, but non-negligible improvement compared to the quasi-universal case, eliminating a few outliers.
Notably, although the parameter space has considerably extended compared to the one discussed in the main text, the residuals remain small and comparable to the nonspinning case.
The merger amplitude now shows $99\%$ of cases with residuals below the $2\%$ level, with a single outlier beyond $3\%$.
The frequency doesn't show any large improvement, with $97\%$ of cases below $10\%$.
Instead, the remnant mass can be captured with remarkable accuracy, with $98\%$ of the dataset below $0.5\%$ and all points below $1\%$.
The remnant spin fit also considerably improves in this case, with $97\%$ of cases below $2\%$ and all resuduals below $3\%$.

These relationships could be further extended by considering the subdominant anti-symmetric contributions to aligned-spin binaries, and by incorporating the appropriate time-dependent rotation factors in the case of spins misaligned with the orbital angular momentum (precessing case).
We leave the exploration of these extensions to future work.

\noindent {\textbf{\textit{Simulations dataset}}.}
%
Inspection of the waveform amplitude around merger for highly eccentric simulations contained in the RIT catalog reveals that in cases with typical $e_0 \gtrsim 0.9$, a sharp transition from a smooth morphology to a highly oscillatory one takes place.
This change is due to the $(\ell,m)=(2, \pm 2)$ mode becoming almost purely real, a feature which is also observed in the merger-ringdown of test particles undergoing a radial plunge~\cite{Lousto:1997ge, Martel:2001yf}.
To obtain a smooth quantity amenable to modelling, in these cases it is necessary to consider the combined amplitude of all the modes beyond the dominant one, including $m=0$ modes.
The transition to this ``radial'' behaviour happens sharply for $A_{\mathrm{mrg}}/\nu \simeq 0.8$.
In these cases, multiple $A_{22}$ local maxima are observed during the ringdown phase, and a $t_{\mathrm{mrg}}$ definition taking into account higher modes is required.
Since the purpose of this study is to model the dominant mode, $(\ell,m)=(2, \pm 2)$, we do not include such simulations in our dataset.
For the same reason, we do not include simulations with $q>3$.

The computation of the fluxes not only involves the strain modes, but also their time derivatives. 
Hence, it is important to employ highly accurate waveforms, to avoid outliers stemming from numerical error.
We check which simulations are accurate enough for our purposes by verifying that balance laws are satisfied.
Integrating throughout the evolution, the relations $M_f = E(t_{\mathrm{end}}), J_f = J(t_{\mathrm{end}})$ need to hold, where $t_{\mathrm{end}}$ is the end time of the simulation.
The above relationships might be violated due to numerical error related to finite resolution or to the extrapolation of the waveform to infinity.
We find that on the considered dataset, balance laws are typically 
violated at the $0.1 \%$ ($1 \%$) level for $M_f (J_f)$. These numbers are
fully with the values reported Table~II of~\cite{Reisswig:2009rx} that refer to
the case of quasi-circular binaries.
For some simulations, the above relations are violated to a higher degree. Consequently, we 
apply a ``data quality'' selection cut, excluding from our dataset all simulations for which balance laws are violated by more than $5\%$.
We leave to future work the inclusion of a small amount of simulations displaying a ``radial'' merger behavior, since their modeling requires 
accounting for the challenging set of $m=0$ modes, not always well-resolved in current simulations. 

Finally, we note that error estimation in noncircular binary mergers simulations presents several additional complications compared to the well-studied quasicircular ones.
For example, due to the non-monotonicity of the signal frequency, the metric reconstruction cannot be performed by multiplication in 
the frequency domain with a cutoff~\cite{Reisswig:2009rx}. We refer the reader to Ref.~\cite{Andrade:2023trh} for a more in-depth discussion.
We leave a detailed study of the impact of numerical error on our modeling strategy to the future, once more accurate or multi-resolution numerical data will become available\footnote{RIT simulations, constituting the vast majority of our dataset, are only available at a single resolution, hence currently this error cannot be straightforwardly estimated.}.
However, the smooth structure observed through the proposed method, and the agreement with the behavior observed in the test-mass case~\cite{Albanesi:2023bgi} (similarly to what happens in the quasicircular case~\cite{Damour:2007xr, Nagar:2018zoe}), indicate that a sub-percentage agreement will likely be achieved with higher accuracy data.

The ET dataset has been verified to correspond to the last stages of dynamical captures by showing their agreement with waveforms from dynamical capture configurations ($E^{ADM}_0 > 1$) starting at larger initial separation, for which the initial energy drops below unity at shorter radii due to repeated GW emission from multiple encounters at early times~\cite{Andrade:2023trh}.

The considered simulations and their relevant parameters are given in the tables below. Tab.~\ref{tab:sims} reports the list of full numerical simulations, and Tab.~\ref{tab:sims_RWZ} reports test-mass data.

\clearpage

\begin{longtable*}{llcccccccc}
\caption{List of full numerical simulations included in our dataset, and selected associated parameters.\vspace{0.1cm}}
\label{tab:sims}\\
\toprule
ID & q & $\chi_1$ & $\chi_2$& $e_0$ & $E^{ADM}_0$ & $J^{ADM}_0$ &  $\Eeffmrg$ &  $j_{\mathrm{mrg}}$ &  $\hat{b}_{\mathrm{mrg}}$ \\
\toprule
\endfirsthead

\multicolumn{10}{c}
{{\tablename\ \thetable{} -- continued from previous
page\vspace{0.1cm}}}\\
\toprule
ID & q & $\chi_1$ & $\chi_2$ & $e_0$ & $E^{ADM}_0$ & $J^{ADM}_0$ & $\Eeffmrg$ & $j_{\mathrm{mrg}}$ & $\hat{b}_{\mathrm{mrg}}$ \\
\toprule
\endhead
\hline \multicolumn{10}{r}{{Continued on next page\vspace{0.1cm}}} \\
\hline
\endfoot
\hline\hline
\endlastfoot

    RIT-1090  & 1.0000 &  0.0000 &  0.0000 & 0.0000 & 0.9907 & 1.0028 &    0.8812 &    2.8108 &          3.0935 \\
    RIT-1091  & 1.0000 &  0.0000 &  0.0000 & 0.0020 & 0.9907 & 1.0018 &    0.8808 &    2.8083 &          3.0920 \\
    RIT-1092  & 1.0000 &  0.0000 &  0.0000 & 0.0040 & 0.9907 & 1.0008 &    0.8808 &    2.8079 &          3.0915 \\
    RIT-1093  & 1.0000 &  0.0000 &  0.0000 & 0.0060 & 0.9907 & 0.9998 &    0.8809 &    2.8088 &          3.0922 \\
    RIT-1094  & 1.0000 &  0.0000 &  0.0000 & 0.0080 & 0.9906 & 0.9988 &    0.8817 &    2.8141 &          3.0958 \\
    RIT-1095  & 1.0000 &  0.0000 &  0.0000 & 0.0100 & 0.9906 & 0.9978 &    0.8810 &    2.8099 &          3.0930 \\
    RIT-1096  & 1.0000 &  0.0000 &  0.0000 & 0.0150 & 0.9906 & 0.9953 &    0.8810 &    2.8106 &          3.0939 \\
    RIT-1097  & 1.0000 &  0.0000 &  0.0000 & 0.0200 & 0.9905 & 0.9927 &    0.8807 &    2.8095 &          3.0935 \\
    RIT-1098  & 1.0000 &  0.0000 &  0.0000 & 0.0400 & 0.9903 & 0.9825 &    0.8839 &    2.8241 &          3.1010 \\
    RIT-1099  & 1.0000 &  0.0000 &  0.0000 & 0.0500 & 0.9901 & 0.9774 &    0.8805 &    2.8050 &          3.0891 \\
    RIT-1100  & 1.0000 &  0.0000 &  0.0000 & 0.0600 & 0.9900 & 0.9723 &    0.8801 &    2.8084 &          3.0940 \\
    RIT-1101  & 1.0000 &  0.0000 &  0.0000 & 0.0700 & 0.9899 & 0.9671 &    0.8821 &    2.8226 &          3.1039 \\
    RIT-1102  & 1.0000 &  0.0000 &  0.0000 & 0.0800 & 0.9898 & 0.9619 &    0.8830 &    2.8219 &          3.1008 \\
    RIT-1103  & 1.0000 &  0.0000 &  0.0000 & 0.0900 & 0.9897 & 0.9566 &    0.8817 &    2.8078 &          3.0888 \\
    RIT-1104  & 1.0000 &  0.0000 &  0.0000 & 0.1000 & 0.9896 & 0.9513 &    0.8808 &    2.8070 &          3.0905 \\
    RIT-1105  & 1.0000 &  0.0000 &  0.0000 & 0.1100 & 0.9894 & 0.9460 &    0.8822 &    2.8266 &          3.1081 \\
    RIT-1106  & 1.0000 &  0.0000 &  0.0000 & 0.1200 & 0.9893 & 0.9407 &    0.8831 &    2.8348 &          3.1149 \\
    RIT-1107  & 1.0000 &  0.0000 &  0.0000 & 0.1300 & 0.9892 & 0.9354 &    0.8833 &    2.8278 &          3.1067 \\
    RIT-1108  & 1.0000 &  0.0000 &  0.0000 & 0.1400 & 0.9891 & 0.9300 &    0.8830 &    2.8134 &          3.0915 \\
    RIT-1109  & 1.0000 &  0.0000 &  0.0000 & 0.1450 & 0.9890 & 0.9273 &    0.8818 &    2.8023 &          3.0826 \\
    RIT-1110  & 1.0000 &  0.0000 &  0.0000 & 0.1570 & 0.9889 & 0.9207 &    0.8784 &    2.7928 &          3.0811 \\
    RIT-1111  & 1.0000 &  0.0000 &  0.0000 & 0.1600 & 0.9889 & 0.9191 &    0.8784 &    2.7985 &          3.0874 \\
    RIT-1112  & 1.0000 &  0.0000 &  0.0000 & 0.1700 & 0.9887 & 0.9136 &    0.8790 &    2.8185 &          3.1080 \\
    RIT-1113  & 1.0000 &  0.0000 &  0.0000 & 0.1800 & 0.9886 & 0.9081 &    0.8833 &    2.8486 &          3.1295 \\
    RIT-1114  & 1.0000 &  0.0000 &  0.0000 & 0.1850 & 0.9886 & 0.9053 &    0.8849 &    2.8556 &          3.1329 \\
    RIT-1115  & 1.0000 &  0.0000 &  0.0000 & 0.1950 & 0.9884 & 0.8997 &    0.8863 &    2.8527 &          3.1257 \\
    RIT-1116  & 1.0000 &  0.0000 &  0.0000 & 0.2000 & 0.9884 & 0.8969 &    0.8877 &    2.8519 &          3.1212 \\
    RIT-1117  & 1.0000 &  0.0000 &  0.0000 & 0.2100 & 0.9883 & 0.8913 &    0.8876 &    2.8320 &          3.0998 \\
    RIT-1118  & 1.0000 &  0.0000 &  0.0000 & 0.2350 & 0.9880 & 0.8771 &    0.8752 &    2.7837 &          3.0799 \\
    RIT-1119  & 1.0000 &  0.0000 &  0.0000 & 0.2450 & 0.9879 & 0.8713 &    0.8792 &    2.8352 &          3.1258 \\
    RIT-1120  & 1.0000 &  0.0000 &  0.0000 & 0.2500 & 0.9878 & 0.8685 &    0.8857 &    2.8834 &          3.1612 \\
    RIT-1121  & 1.0000 &  0.0000 &  0.0000 & 0.2650 & 0.9876 & 0.8597 &    0.8832 &    2.8837 &          3.1682 \\
    RIT-1122  & 1.0000 &  0.0000 &  0.0000 & 0.2750 & 0.9875 & 0.8539 &    0.8868 &    2.9052 &          3.1819 \\
    RIT-1123  & 1.0000 &  0.0000 &  0.0000 & 0.2850 & 0.9874 & 0.8479 &    0.8977 &    2.9618 &          3.2138 \\
    RIT-1124  & 1.0000 &  0.0000 &  0.0000 & 0.3000 & 0.9872 & 0.8390 &    0.9049 &    2.9894 &          3.2242 \\
    RIT-1125  & 1.0000 &  0.0000 &  0.0000 & 0.3500 & 0.9866 & 0.8085 &    0.9114 &    2.9560 &          3.1707 \\
    RIT-1201  & 2.0000 &  0.0000 &  0.0000 & 0.0020 & 0.9918 & 0.8910 &    0.8904 &    2.8875 &          3.1629 \\
    RIT-1202  & 2.0000 &  0.0000 &  0.0000 & 0.0040 & 0.9917 & 0.8901 &    0.8882 &    2.8743 &          3.1545 \\
    RIT-1203  & 2.0000 &  0.0000 &  0.0000 & 0.0060 & 0.9917 & 0.8892 &    0.8885 &    2.8760 &          3.1558 \\
    RIT-1204  & 2.0000 &  0.0000 &  0.0000 & 0.0080 & 0.9917 & 0.8883 &    0.8884 &    2.8752 &          3.1552 \\
    RIT-1205  & 2.0000 &  0.0000 &  0.0000 & 0.0100 & 0.9917 & 0.8874 &    0.8885 &    2.8758 &          3.1554 \\
    RIT-1206  & 2.0000 &  0.0000 &  0.0000 & 0.0150 & 0.9916 & 0.8851 &    0.8888 &    2.8770 &          3.1560 \\
    RIT-1207  & 2.0000 &  0.0000 &  0.0000 & 0.0200 & 0.9916 & 0.8829 &    0.8884 &    2.8746 &          3.1544 \\
    RIT-1208  & 2.0000 &  0.0000 &  0.0000 & 0.0300 & 0.9915 & 0.8784 &    0.8903 &    2.8890 &          3.1648 \\
    RIT-1209  & 2.0000 &  0.0000 &  0.0000 & 0.0400 & 0.9914 & 0.8738 &    0.8892 &    2.8830 &          3.1614 \\
    RIT-1210  & 2.0000 &  0.0000 &  0.0000 & 0.0500 & 0.9913 & 0.8693 &    0.8910 &    2.8889 &          3.1629 \\
    RIT-1211  & 2.0000 &  0.0000 &  0.0000 & 0.0600 & 0.9912 & 0.8647 &    0.8879 &    2.8697 &          3.1505 \\
    RIT-1212  & 2.0000 &  0.0000 &  0.0000 & 0.0700 & 0.9910 & 0.8601 &    0.8877 &    2.8763 &          3.1582 \\
    RIT-1213  & 2.0000 &  0.0000 &  0.0000 & 0.0800 & 0.9909 & 0.8554 &    0.8890 &    2.8863 &          3.1656 \\
    RIT-1214  & 2.0000 &  0.0000 &  0.0000 & 0.0900 & 0.9908 & 0.8508 &    0.8926 &    2.8998 &          3.1703 \\
    RIT-1215  & 2.0000 &  0.0000 &  0.0000 & 0.1000 & 0.9907 & 0.8461 &    0.8894 &    2.8741 &          3.1511 \\
    RIT-1216  & 2.0000 &  0.0000 &  0.0000 & 0.1100 & 0.9906 & 0.8414 &    0.8860 &    2.8635 &          3.1489 \\
    RIT-1217  & 2.0000 &  0.0000 &  0.0000 & 0.1200 & 0.9905 & 0.8366 &    0.8910 &    2.9046 &          3.1800 \\
    RIT-1218  & 2.0000 &  0.0000 &  0.0000 & 0.1300 & 0.9904 & 0.8319 &    0.8913 &    2.9046 &          3.1790 \\
    RIT-1219  & 2.0000 &  0.0000 &  0.0000 & 0.1400 & 0.9903 & 0.8271 &    0.8933 &    2.9031 &          3.1718 \\
    RIT-1220  & 2.0000 &  0.0000 &  0.0000 & 0.1500 & 0.9902 & 0.8222 &    0.8912 &    2.8793 &          3.1518 \\
    RIT-1221  & 2.0000 &  0.0000 &  0.0000 & 0.1600 & 0.9901 & 0.8174 &    0.8865 &    2.8626 &          3.1467 \\
    RIT-1222  & 2.0000 &  0.0000 &  0.0000 & 0.1700 & 0.9900 & 0.8125 &    0.8877 &    2.8904 &          3.1738 \\
    RIT-1223  & 2.0000 &  0.0000 &  0.0000 & 0.1800 & 0.9899 & 0.8076 &    0.8897 &    2.9101 &          3.1898 \\
    RIT-1224  & 2.0000 &  0.0000 &  0.0000 & 0.1900 & 0.9898 & 0.8027 &    0.8963 &    2.9433 &          3.2071 \\
    RIT-1226  & 2.0000 &  0.0000 &  0.0000 & 0.2100 & 0.9896 & 0.7927 &    0.8926 &    2.8832 &          3.1522 \\
    RIT-1227  & 2.0000 &  0.0000 &  0.0000 & 0.2200 & 0.9895 & 0.7877 &    0.8903 &    2.8652 &          3.1388 \\
    RIT-1228  & 2.0000 &  0.0000 &  0.0000 & 0.2300 & 0.9894 & 0.7826 &    0.8848 &    2.8608 &          3.1494 \\
    RIT-1229  & 2.0000 &  0.0000 &  0.0000 & 0.2400 & 0.9893 & 0.7775 &    0.8848 &    2.8919 &          3.1837 \\
    RIT-1230  & 2.0000 &  0.0000 &  0.0000 & 0.2500 & 0.9892 & 0.7724 &    0.8931 &    2.9587 &          3.2331 \\
    RIT-1231  & 2.0000 &  0.0000 &  0.0000 & 0.2700 & 0.9890 & 0.7620 &    0.8993 &    2.9990 &          3.2594 \\
    RIT-1232  & 2.0000 &  0.0000 &  0.0000 & 0.3000 & 0.9886 & 0.7462 &    0.9132 &    3.0524 &          3.2775 \\
    RIT-1241  & 1.3333 &  0.0000 &  0.0000 & 0.0000 & 0.9909 & 0.9824 &    0.8824 &    2.8210 &          3.1034 \\
    RIT-1242  & 1.3333 &  0.0000 &  0.0000 & 0.0020 & 0.9909 & 0.9814 &    0.8827 &    2.8224 &          3.1043 \\
    RIT-1243  & 1.3333 &  0.0000 &  0.0000 & 0.0040 & 0.9909 & 0.9805 &    0.8824 &    2.8204 &          3.1029 \\
    RIT-1244  & 1.3333 &  0.0000 &  0.0000 & 0.0060 & 0.9909 & 0.9795 &    0.8824 &    2.8205 &          3.1030 \\
    RIT-1245  & 1.3333 &  0.0000 &  0.0000 & 0.0080 & 0.9908 & 0.9785 &    0.8819 &    2.8178 &          3.1013 \\
    RIT-1246  & 1.3333 &  0.0000 &  0.0000 & 0.0100 & 0.9908 & 0.9775 &    0.8829 &    2.8233 &          3.1048 \\
    RIT-1247  & 1.3333 &  0.0000 &  0.0000 & 0.0150 & 0.9908 & 0.9750 &    0.8825 &    2.8222 &          3.1045 \\
    RIT-1248  & 1.3333 &  0.0000 &  0.0000 & 0.0200 & 0.9907 & 0.9726 &    0.8830 &    2.8255 &          3.1070 \\
    RIT-1249  & 1.3333 &  0.0000 &  0.0000 & 0.0300 & 0.9906 & 0.9676 &    0.8819 &    2.8201 &          3.1039 \\
    RIT-1250  & 1.3333 &  0.0000 &  0.0000 & 0.0400 & 0.9905 & 0.9626 &    0.8831 &    2.8237 &          3.1046 \\
    RIT-1251  & 1.3333 &  0.0000 &  0.0000 & 0.0500 & 0.9904 & 0.9576 &    0.8822 &    2.8177 &          3.1005 \\
    RIT-1252  & 1.3333 &  0.0000 &  0.0000 & 0.0600 & 0.9902 & 0.9525 &    0.8827 &    2.8265 &          3.1086 \\
    RIT-1253  & 1.3333 &  0.0000 &  0.0000 & 0.0700 & 0.9901 & 0.9474 &    0.8822 &    2.8271 &          3.1107 \\
    RIT-1254  & 1.3333 &  0.0000 &  0.0000 & 0.0800 & 0.9900 & 0.9423 &    0.8835 &    2.8296 &          3.1100 \\
    RIT-1255  & 1.3333 &  0.0000 &  0.0000 & 0.0900 & 0.9899 & 0.9372 &    0.8828 &    2.8177 &          3.0989 \\
    RIT-1256  & 1.3333 &  0.0000 &  0.0000 & 0.1000 & 0.9898 & 0.9320 &    0.8835 &    2.8249 &          3.1048 \\
    RIT-1257  & 1.3333 &  0.0000 &  0.0000 & 0.1100 & 0.9897 & 0.9268 &    0.8834 &    2.8361 &          3.1173 \\
    RIT-1258  & 1.3333 &  0.0000 &  0.0000 & 0.1200 & 0.9895 & 0.9216 &    0.8835 &    2.8411 &          3.1228 \\
    RIT-1259  & 1.3333 &  0.0000 &  0.0000 & 0.1300 & 0.9894 & 0.9164 &    0.8875 &    2.8576 &          3.1298 \\
    RIT-1260  & 1.3333 &  0.0000 &  0.0000 & 0.1400 & 0.9893 & 0.9111 &    0.8847 &    2.8275 &          3.1045 \\
    RIT-1261  & 1.3333 &  0.0000 &  0.0000 & 0.1500 & 0.9892 & 0.9058 &    0.8849 &    2.8244 &          3.1004 \\
    RIT-1262  & 1.3333 &  0.0000 &  0.0000 & 0.1600 & 0.9891 & 0.9004 &    0.8799 &    2.8099 &          3.0981 \\
    RIT-1263  & 1.3333 &  0.0000 &  0.0000 & 0.1700 & 0.9890 & 0.8950 &    0.8810 &    2.8340 &          3.1216 \\
    RIT-1264  & 1.3333 &  0.0000 &  0.0000 & 0.1800 & 0.9889 & 0.8896 &    0.8847 &    2.8608 &          3.1411 \\
    RIT-1265  & 1.3333 &  0.0000 &  0.0000 & 0.1900 & 0.9887 & 0.8842 &    0.8886 &    2.8769 &          3.1480 \\
    RIT-1266  & 1.3333 &  0.0000 &  0.0000 & 0.2000 & 0.9886 & 0.8787 &    0.8892 &    2.8648 &          3.1331 \\
    RIT-1267  & 1.3333 &  0.0000 &  0.0000 & 0.2100 & 0.9885 & 0.8732 &    0.8880 &    2.8387 &          3.1077 \\
    RIT-1268  & 1.3333 &  0.0000 &  0.0000 & 0.2200 & 0.9884 & 0.8677 &    0.8832 &    2.8025 &          3.0810 \\
    RIT-1269  & 1.3333 &  0.0000 &  0.0000 & 0.2300 & 0.9883 & 0.8621 &    0.8805 &    2.8054 &          3.0917 \\
    RIT-1270  & 1.3333 &  0.0000 &  0.0000 & 0.2400 & 0.9882 & 0.8565 &    0.8779 &    2.8219 &          3.1167 \\
    RIT-1271  & 1.3333 &  0.0000 &  0.0000 & 0.2500 & 0.9880 & 0.8508 &    0.8867 &    2.8954 &          3.1734 \\
    RIT-1272  & 1.3333 &  0.0000 &  0.0000 & 0.2700 & 0.9878 & 0.8394 &    0.8872 &    2.9120 &          3.1902 \\
    RIT-1273  & 1.3333 &  0.0000 &  0.0000 & 0.3000 & 0.9875 & 0.8220 &    0.9042 &    2.9885 &          3.2265 \\
    RIT-1274  & 1.3333 &  0.0000 &  0.0000 & 0.3500 & 0.9869 & 0.7921 &    0.9132 &    2.9680 &          3.1802 \\
    RIT-1282  & 1.0000 &  0.0000 &  0.0000 & 0.1900 & 0.9943 & 1.2099 &    0.8810 &    2.8193 &          3.1035 \\
    RIT-1283  & 1.0000 &  0.0000 &  0.0000 & 0.2775 & 0.9938 & 1.1427 &    0.8825 &    2.8190 &          3.0991 \\
    RIT-1284  & 1.0000 &  0.0000 &  0.0000 & 0.3276 & 0.9936 & 1.1024 &    0.8797 &    2.8076 &          3.0941 \\
    RIT-1285  & 1.0000 &  0.0000 &  0.0000 & 0.3600 & 0.9934 & 1.0755 &    0.8827 &    2.8292 &          3.1097 \\
    RIT-1286  & 1.0000 &  0.0000 &  0.0000 & 0.3994 & 0.9932 & 1.0419 &    0.8841 &    2.8052 &          3.0797 \\
    RIT-1287  & 1.0000 &  0.0000 &  0.0000 & 0.3916 & 0.9932 & 1.0486 &    0.8826 &    2.8339 &          3.1152 \\
    RIT-1288  & 1.0000 &  0.0000 &  0.0000 & 0.4148 & 0.9931 & 1.0284 &    0.8849 &    2.8346 &          3.1096 \\
    RIT-1289  & 1.0000 &  0.0000 &  0.0000 & 0.4071 & 0.9932 & 1.0351 &    0.8782 &    2.8022 &          3.0922 \\
    RIT-1290  & 1.0000 &  0.0000 &  0.0000 & 0.4300 & 0.9930 & 1.0150 &    0.8810 &    2.8310 &          3.1162 \\
    RIT-1291  & 1.0000 &  0.0000 &  0.0000 & 0.4224 & 0.9931 & 1.0217 &    0.8790 &    2.7721 &          3.0569 \\
    RIT-1292  & 1.0000 &  0.0000 &  0.0000 & 0.4450 & 0.9930 & 1.0015 &    0.8769 &    2.7606 &          3.0496 \\
    RIT-1293  & 1.0000 &  0.0000 &  0.0000 & 0.4375 & 0.9930 & 1.0083 &    0.8872 &    2.8383 &          3.1075 \\
    RIT-1294  & 1.0000 &  0.0000 &  0.0000 & 0.4598 & 0.9929 & 0.9881 &    0.8903 &    2.8707 &          3.1348 \\
    RIT-1295  & 1.0000 &  0.0000 &  0.0000 & 0.4524 & 0.9929 & 0.9948 &    0.8806 &    2.8327 &          3.1192 \\
    RIT-1296  & 1.0000 &  0.0000 &  0.0000 & 0.4744 & 0.9928 & 0.9746 &    0.8745 &    2.7812 &          3.0788 \\
    RIT-1297  & 1.0000 &  0.0000 &  0.0000 & 0.4671 & 0.9928 & 0.9814 &    0.8834 &    2.7850 &          3.0593 \\
    RIT-1298  & 1.0000 &  0.0000 &  0.0000 & 0.4888 & 0.9927 & 0.9612 &    0.8895 &    2.8682 &          3.1341 \\
    RIT-1299  & 1.0000 &  0.0000 &  0.0000 & 0.4816 & 0.9928 & 0.9679 &    0.8863 &    2.8747 &          3.1501 \\
    RIT-1300  & 1.0000 &  0.0000 &  0.0000 & 0.5030 & 0.9927 & 0.9478 &    0.8752 &    2.7620 &          3.0559 \\
    RIT-1301  & 1.0000 &  0.0000 &  0.0000 & 0.4959 & 0.9927 & 0.9545 &    0.8893 &    2.8178 &          3.0797 \\
    RIT-1302  & 1.0000 &  0.0000 &  0.0000 & 0.5170 & 0.9926 & 0.9343 &    0.8868 &    2.8909 &          3.1662 \\
    RIT-1303  & 1.0000 &  0.0000 &  0.0000 & 0.5100 & 0.9926 & 0.9410 &    0.8800 &    2.8455 &          3.1350 \\
    RIT-1304  & 1.0000 &  0.0000 &  0.0000 & 0.5308 & 0.9925 & 0.9209 &    0.8908 &    2.8536 &          3.1148 \\
    RIT-1305  & 1.0000 &  0.0000 &  0.0000 & 0.5239 & 0.9925 & 0.9276 &    0.8898 &    2.8872 &          3.1541 \\
    RIT-1306  & 1.0000 &  0.0000 &  0.0000 & 0.5444 & 0.9924 & 0.9074 &    0.8700 &    2.7304 &          3.0345 \\
    RIT-1307  & 1.0000 &  0.0000 &  0.0000 & 0.5376 & 0.9925 & 0.9142 &    0.8865 &    2.7831 &          3.0490 \\
    RIT-1308  & 1.0000 &  0.0000 &  0.0000 & 0.5578 & 0.9924 & 0.8940 &    0.8952 &    2.9740 &          3.2341 \\
    RIT-1309  & 1.0000 &  0.0000 &  0.0000 & 0.5511 & 0.9924 & 0.9007 &    0.8732 &    2.8194 &          3.1246 \\
    RIT-1310  & 1.0000 &  0.0000 &  0.0000 & 0.5710 & 0.9923 & 0.8805 &    0.9090 &    3.0481 &          3.2762 \\
    RIT-1311  & 1.0000 &  0.0000 &  0.0000 & 0.5644 & 0.9923 & 0.8873 &    0.8972 &    2.9883 &          3.2441 \\
    RIT-1312  & 1.0000 &  0.0000 &  0.0000 & 0.5775 & 0.9923 & 0.8738 &    0.9145 &    3.0661 &          3.2802 \\
    RIT-1313  & 1.0000 &  0.0000 &  0.0000 & 0.5904 & 0.9922 & 0.8604 &    0.9238 &    3.0839 &          3.2740 \\
    RIT-1314  & 1.0000 &  0.0000 &  0.0000 & 0.6156 & 0.9921 & 0.8335 &    0.9313 &    3.0430 &          3.2108 \\
    RIT-1315  & 1.0000 &  0.0000 &  0.0000 & 0.6400 & 0.9919 & 0.8066 &    0.9384 &    2.9915 &          3.1383 \\
    RIT-1316  & 1.0000 &  0.0000 &  0.0000 & 0.6636 & 0.9918 & 0.7797 &    0.9431 &    2.9224 &          3.0542 \\
    RIT-1317  & 1.0000 &  0.0000 &  0.0000 & 0.6975 & 0.9916 & 0.7394 &    0.9493 &    2.8101 &          2.9223 \\
    RIT-1318  & 1.0000 &  0.0000 &  0.0000 & 0.7500 & 0.9914 & 0.6722 &    0.9555 &    2.5943 &          2.6848 \\
    RIT-1319  & 1.0000 &  0.0000 &  0.0000 & 0.7975 & 0.9911 & 0.6050 &    0.9589 &    2.3597 &          2.4355 \\
    RIT-1320  & 1.0000 &  0.0000 &  0.0000 & 0.8400 & 0.9909 & 0.5377 &    0.9608 &    2.1145 &          2.1792 \\
    RIT-1321  & 1.0000 &  0.0000 &  0.0000 & 0.8775 & 0.9907 & 0.4705 &    0.9613 &    1.8592 &          1.9152 \\
    RIT-1330  & 1.1111 &  0.0000 &  0.0000 & 0.1900 & 0.9943 & 1.2077 &    0.8813 &    2.8053 &          3.0874 \\
    RIT-1331  & 1.1111 &  0.0000 &  0.0000 & 0.3600 & 0.9934 & 1.0735 &    0.8812 &    2.8199 &          3.1039 \\
    RIT-1332  & 1.1111 &  0.0000 &  0.0000 & 0.4375 & 0.9930 & 1.0064 &    0.8870 &    2.8459 &          3.1168 \\
    RIT-1333  & 1.1111 &  0.0000 &  0.0000 & 0.4671 & 0.9929 & 0.9796 &    0.8869 &    2.8097 &          3.0773 \\
    RIT-1334  & 1.1111 &  0.0000 &  0.0000 & 0.5100 & 0.9926 & 0.9393 &    0.8754 &    2.8155 &          3.1148 \\
    RIT-1335  & 1.1111 &  0.0000 &  0.0000 & 0.5511 & 0.9924 & 0.8990 &    0.8707 &    2.8002 &          3.1107 \\
    RIT-1336  & 1.1111 &  0.0000 &  0.0000 & 0.5775 & 0.9923 & 0.8722 &    0.9148 &    3.0699 &          3.2837 \\
    RIT-1337  & 1.1111 &  0.0000 &  0.0000 & 0.6156 & 0.9921 & 0.8319 &    0.9322 &    3.0506 &          3.2167 \\
    RIT-1338  & 1.1111 &  0.0000 &  0.0000 & 0.6400 & 0.9920 & 0.8051 &    0.9402 &    3.0044 &          3.1474 \\
    RIT-1339  & 1.1111 &  0.0000 &  0.0000 & 0.6636 & 0.9918 & 0.7783 &    0.9436 &    2.9278 &          3.0588 \\
    RIT-1340  & 1.1111 &  0.0000 &  0.0000 & 0.6975 & 0.9917 & 0.7380 &    0.9497 &    2.8145 &          2.9262 \\
    RIT-1341  & 1.1111 &  0.0000 &  0.0000 & 0.7500 & 0.9914 & 0.6709 &    0.9556 &    2.5977 &          2.6880 \\
    RIT-1342  & 1.1111 &  0.0000 &  0.0000 & 0.7975 & 0.9911 & 0.6038 &    0.9589 &    2.3623 &          2.4382 \\
    RIT-1343  & 1.1111 &  0.0000 &  0.0000 & 0.8400 & 0.9909 & 0.5367 &    0.9608 &    2.1167 &          2.1814 \\
    RIT-1344  & 1.1111 &  0.0000 &  0.0000 & 0.8775 & 0.9907 & 0.4696 &    0.9613 &    1.8610 &          1.9171 \\
    RIT-1353  & 1.2500 &  0.0000 &  0.0000 & 0.1900 & 0.9944 & 1.1961 &    0.8830 &    2.8237 &          3.1040 \\
    RIT-1354  & 1.2500 &  0.0000 &  0.0000 & 0.3600 & 0.9935 & 1.0632 &    0.8801 &    2.8136 &          3.1008 \\
    RIT-1364  & 1.2500 &  0.0000 &  0.0000 & 0.7500 & 0.9915 & 0.6645 &    0.9557 &    2.5986 &          2.6890 \\
    RIT-1365  & 1.2500 &  0.0000 &  0.0000 & 0.7975 & 0.9912 & 0.5980 &    0.9590 &    2.3629 &          2.4389 \\
    RIT-1366  & 1.2500 &  0.0000 &  0.0000 & 0.8400 & 0.9910 & 0.5316 &    0.9608 &    2.1171 &          2.1820 \\
    RIT-1367  & 1.2500 &  0.0000 &  0.0000 & 0.8775 & 0.9908 & 0.4651 &    0.9614 &    1.8612 &          1.9174 \\
    RIT-1376  & 1.4286 &  0.0000 &  0.0000 & 0.1900 & 0.9945 & 1.1733 &    0.8831 &    2.8291 &          3.1117 \\
    RIT-1377  & 1.4286 &  0.0000 &  0.0000 & 0.3600 & 0.9936 & 1.0429 &    0.8783 &    2.7979 &          3.0901 \\
    RIT-1378  & 1.4286 &  0.0000 &  0.0000 & 0.4375 & 0.9932 & 0.9778 &    0.8884 &    2.8689 &          3.1409 \\
    RIT-1379  & 1.4286 &  0.0000 &  0.0000 & 0.4671 & 0.9931 & 0.9517 &    0.8876 &    2.8258 &          3.0957 \\
    RIT-1380  & 1.4286 &  0.0000 &  0.0000 & 0.5100 & 0.9929 & 0.9126 &    0.8802 &    2.8507 &          3.1433 \\
    RIT-1381  & 1.4286 &  0.0000 &  0.0000 & 0.5511 & 0.9926 & 0.8735 &    0.8856 &    2.9036 &          3.1865 \\
    RIT-1382  & 1.4286 &  0.0000 &  0.0000 & 0.5775 & 0.9925 & 0.8474 &    0.9222 &    3.1162 &          3.3147 \\
    RIT-1383  & 1.4286 &  0.0000 &  0.0000 & 0.6156 & 0.9923 & 0.8083 &    0.9333 &    3.0588 &          3.2241 \\
    RIT-1384  & 1.4286 &  0.0000 &  0.0000 & 0.6400 & 0.9922 & 0.7822 &    0.9377 &    2.9921 &          3.1424 \\
    RIT-1385  & 1.4286 &  0.0000 &  0.0000 & 0.6636 & 0.9921 & 0.7561 &    0.9444 &    2.9337 &          3.0643 \\
    RIT-1386  & 1.4286 &  0.0000 &  0.0000 & 0.6975 & 0.9919 & 0.7170 &    0.9502 &    2.8189 &          2.9306 \\
    RIT-1387  & 1.4286 &  0.0000 &  0.0000 & 0.7500 & 0.9916 & 0.6518 &    0.9559 &    2.6004 &          2.6911 \\
    RIT-1388  & 1.4286 &  0.0000 &  0.0000 & 0.7975 & 0.9914 & 0.5867 &    0.9591 &    2.3640 &          2.4403 \\
    RIT-1389  & 1.4286 &  0.0000 &  0.0000 & 0.8400 & 0.9912 & 0.5215 &    0.9609 &    2.1177 &          2.1830 \\
    RIT-1390  & 1.4286 &  0.0000 &  0.0000 & 0.8775 & 0.9910 & 0.4563 &    0.9614 &    1.8616 &          1.9182 \\
    RIT-1399  & 1.6667 &  0.0000 &  0.0000 & 0.1900 & 0.9947 & 1.1353 &    0.8858 &    2.8528 &          3.1333 \\
    RIT-1400  & 1.6667 &  0.0000 &  0.0000 & 0.3600 & 0.9938 & 1.0092 &    0.8841 &    2.8180 &          3.0996 \\
    RIT-1401  & 1.6667 &  0.0000 &  0.0000 & 0.4375 & 0.9934 & 0.9461 &    0.8887 &    2.8852 &          3.1606 \\
    RIT-1402  & 1.6667 &  0.0000 &  0.0000 & 0.4671 & 0.9933 & 0.9209 &    0.8908 &    2.8579 &          3.1252 \\
    RIT-1403  & 1.6667 &  0.0000 &  0.0000 & 0.5100 & 0.9931 & 0.8830 &    0.8840 &    2.8811 &          3.1693 \\
    RIT-1404  & 1.6667 &  0.0000 &  0.0000 & 0.5511 & 0.9929 & 0.8452 &    0.9007 &    3.0131 &          3.2665 \\
    RIT-1405  & 1.6667 &  0.0000 &  0.0000 & 0.5775 & 0.9928 & 0.8200 &    0.9244 &    3.1327 &          3.3283 \\
    RIT-1406  & 1.6667 &  0.0000 &  0.0000 & 0.6156 & 0.9926 & 0.7821 &    0.9345 &    3.0680 &          3.2324 \\
    RIT-1407  & 1.6667 &  0.0000 &  0.0000 & 0.6400 & 0.9924 & 0.7569 &    0.9442 &    3.0304 &          3.1673 \\
    RIT-1408  & 1.6667 &  0.0000 &  0.0000 & 0.6636 & 0.9923 & 0.7316 &    0.9462 &    2.9457 &          3.0736 \\
    RIT-1409  & 1.6667 &  0.0000 &  0.0000 & 0.6975 & 0.9922 & 0.6938 &    0.9536 &    2.8405 &          2.9461 \\
    RIT-1410  & 1.6667 &  0.0000 &  0.0000 & 0.7500 & 0.9919 & 0.6307 &    0.9563 &    2.6036 &          2.6945 \\
    RIT-1411  & 1.6667 &  0.0000 &  0.0000 & 0.7975 & 0.9917 & 0.5677 &    0.9593 &    2.3661 &          2.4428 \\
    RIT-1412  & 1.6667 &  0.0000 &  0.0000 & 0.8400 & 0.9915 & 0.5046 &    0.9610 &    2.1189 &          2.1847 \\
    RIT-1413  & 1.6667 &  0.0000 &  0.0000 & 0.8775 & 0.9913 & 0.4415 &    0.9615 &    1.8633 &          1.9203 \\
    RIT-1422  & 2.0000 &  0.0000 &  0.0000 & 0.1900 & 0.9949 & 1.0755 &    0.8889 &    2.8819 &          3.1612 \\
    RIT-1423  & 2.0000 &  0.0000 &  0.0000 & 0.3600 & 0.9941 & 0.9560 &    0.8912 &    2.8819 &          3.1546 \\
    RIT-1424  & 2.0000 &  0.0000 &  0.0000 & 0.4375 & 0.9938 & 0.8963 &    0.8907 &    2.9129 &          3.1899 \\
    RIT-1425  & 2.0000 &  0.0000 &  0.0000 & 0.4671 & 0.9936 & 0.8724 &    0.8934 &    2.8901 &          3.1574 \\
    RIT-1426  & 2.0000 &  0.0000 &  0.0000 & 0.5100 & 0.9934 & 0.8365 &    0.8899 &    2.9329 &          3.2140 \\
    RIT-1427  & 2.0000 &  0.0000 &  0.0000 & 0.5511 & 0.9932 & 0.8007 &    0.9038 &    3.0511 &          3.3029 \\
    RIT-1428  & 2.0000 &  0.0000 &  0.0000 & 0.5775 & 0.9931 & 0.7768 &    0.9259 &    3.1454 &          3.3408 \\
    RIT-1429  & 2.0000 &  0.0000 &  0.0000 & 0.6156 & 0.9929 & 0.7409 &    0.9401 &    3.1012 &          3.2545 \\
    RIT-1430  & 2.0000 &  0.0000 &  0.0000 & 0.6400 & 0.9928 & 0.7170 &    0.9460 &    3.0417 &          3.1765 \\
    RIT-1431  & 2.0000 &  0.0000 &  0.0000 & 0.6636 & 0.9927 & 0.6931 &    0.9480 &    2.9563 &          3.0821 \\
    RIT-1432  & 2.0000 &  0.0000 &  0.0000 & 0.6975 & 0.9926 & 0.6573 &    0.9530 &    2.8362 &          2.9449 \\
    RIT-1433  & 2.0000 &  0.0000 &  0.0000 & 0.7500 & 0.9923 & 0.5975 &    0.9572 &    2.6086 &          2.6992 \\
    RIT-1434  & 2.0000 &  0.0000 &  0.0000 & 0.7975 & 0.9921 & 0.5378 &    0.9598 &    2.3689 &          2.4458 \\
    RIT-1435  & 2.0000 &  0.0000 &  0.0000 & 0.8400 & 0.9919 & 0.4780 &    0.9612 &    2.1201 &          2.1865 \\
    RIT-1436  & 2.0000 &  0.0000 &  0.0000 & 0.8775 & 0.9917 & 0.4183 &    0.9616 &    1.8627 &          1.9205 \\
    RIT-1445  & 2.5000 &  0.0000 &  0.0000 & 0.1900 & 0.9953 & 0.9886 &    0.8931 &    2.9277 &          3.2058 \\
    RIT-1446  & 2.5000 &  0.0000 &  0.0000 & 0.3600 & 0.9946 & 0.8787 &    0.8905 &    2.8916 &          3.1737 \\
    RIT-1447  & 2.5000 &  0.0000 &  0.0000 & 0.4375 & 0.9943 & 0.8238 &    0.8892 &    2.9052 &          3.1925 \\
    RIT-1448  & 2.5000 &  0.0000 &  0.0000 & 0.4671 & 0.9942 & 0.8018 &    0.9026 &    2.9890 &          3.2451 \\
    RIT-1449  & 2.5000 &  0.0000 &  0.0000 & 0.5100 & 0.9940 & 0.7689 &    0.8937 &    2.9715 &          3.2521 \\
    RIT-1450  & 2.5000 &  0.0000 &  0.0000 & 0.5511 & 0.9938 & 0.7359 &    0.9117 &    3.1204 &          3.3603 \\
    RIT-1451  & 2.5000 &  0.0000 &  0.0000 & 0.5775 & 0.9937 & 0.7140 &    0.9328 &    3.1961 &          3.3791 \\
    RIT-1452  & 2.5000 &  0.0000 &  0.0000 & 0.6156 & 0.9935 & 0.6810 &    0.9421 &    3.1204 &          3.2729 \\
    RIT-1453  & 2.5000 &  0.0000 &  0.0000 & 0.6400 & 0.9934 & 0.6590 &    0.9483 &    3.0620 &          3.1946 \\
    RIT-1454  & 2.5000 &  0.0000 &  0.0000 & 0.6636 & 0.9933 & 0.6371 &    0.9508 &    2.9788 &          3.1012 \\
    RIT-1455  & 2.5000 &  0.0000 &  0.0000 & 0.6975 & 0.9932 & 0.6041 &    0.9550 &    2.8535 &          2.9605 \\
    RIT-1456  & 2.5000 &  0.0000 &  0.0000 & 0.7500 & 0.9929 & 0.5492 &    0.9584 &    2.6206 &          2.7110 \\
    RIT-1457  & 2.5000 &  0.0000 &  0.0000 & 0.7975 & 0.9927 & 0.4943 &    0.9605 &    2.3770 &          2.4547 \\
    RIT-1458  & 2.5000 &  0.0000 &  0.0000 & 0.8400 & 0.9926 & 0.4394 &    0.9614 &    2.1244 &          2.1922 \\
    RIT-1459  & 2.5000 &  0.0000 &  0.0000 & 0.8775 & 0.9924 & 0.3844 &    0.9617 &    1.8658 &          1.9249 \\
    RIT-1468  & 3.0000 &  0.0000 &  0.0000 & 0.1900 & 0.9957 & 0.9075 &    0.8967 &    2.9637 &          3.2406 \\
    RIT-1469  & 3.0000 &  0.0000 &  0.0000 & 0.3600 & 0.9951 & 0.8067 &    0.8985 &    2.9763 &          3.2489 \\
    RIT-1470  & 3.0000 &  0.0000 &  0.0000 & 0.4375 & 0.9947 & 0.7563 &    0.8934 &    2.9121 &          3.1937 \\
    RIT-1471  & 3.0000 &  0.0000 &  0.0000 & 0.4671 & 0.9946 & 0.7361 &    0.9045 &    3.0281 &          3.2872 \\
    RIT-1472  & 3.0000 &  0.0000 &  0.0000 & 0.5100 & 0.9945 & 0.7059 &    0.9033 &    3.0514 &          3.3161 \\
    RIT-1473  & 3.0000 &  0.0000 &  0.0000 & 0.5511 & 0.9943 & 0.6756 &    0.9155 &    3.1599 &          3.3964 \\
    RIT-1474  & 3.0000 &  0.0000 &  0.0000 & 0.5775 & 0.9942 & 0.6555 &    0.9372 &    3.2278 &          3.4033 \\
    RIT-1475  & 3.0000 &  0.0000 &  0.0000 & 0.6156 & 0.9940 & 0.6252 &    0.9464 &    3.1486 &          3.2932 \\
    RIT-1476  & 3.0000 &  0.0000 &  0.0000 & 0.6400 & 0.9940 & 0.6050 &    0.9505 &    3.0767 &          3.2067 \\
    RIT-1477  & 3.0000 &  0.0000 &  0.0000 & 0.6636 & 0.9939 & 0.5849 &    0.9528 &    2.9916 &          3.1119 \\
    RIT-1478  & 3.0000 &  0.0000 &  0.0000 & 0.6975 & 0.9937 & 0.5546 &    0.9564 &    2.8630 &          2.9689 \\
    RIT-1479  & 3.0000 &  0.0000 &  0.0000 & 0.7500 & 0.9935 & 0.5042 &    0.9593 &    2.6266 &          2.7169 \\
    RIT-1480  & 3.0000 &  0.0000 &  0.0000 & 0.7975 & 0.9933 & 0.4538 &    0.9610 &    2.3803 &          2.4587 \\
    RIT-1481  & 3.0000 &  0.0000 &  0.0000 & 0.8400 & 0.9932 & 0.4034 &    0.9617 &    2.1261 &          2.1948 \\
    RIT-1482  & 3.0000 &  0.0000 &  0.0000 & 0.8775 & 0.9930 & 0.3529 &    0.9618 &    1.8665 &          1.9267 \\
    RIT-1740  & 1.0000 & -0.5000 & -0.5000 & 0.1900 & 0.9944 & 0.9750 &    0.9059 &    2.2301 &          2.4031 \\
    RIT-1741  & 1.0000 & -0.5000 & -0.5000 & 0.3600 & 0.9935 & 0.8389 &    0.9075 &    2.2298 &          2.3997 \\
    RIT-1742  & 1.0000 & -0.5000 & -0.5000 & 0.4375 & 0.9930 & 0.7709 &    0.9025 &    2.2387 &          2.4194 \\
    RIT-1743  & 1.0000 & -0.5000 & -0.5000 & 0.4671 & 0.9929 & 0.7436 &    0.9018 &    2.1736 &          2.3502 \\
    RIT-1744  & 1.0000 & -0.5000 & -0.5000 & 0.5100 & 0.9927 & 0.7028 &    0.9328 &    2.4622 &          2.5949 \\
    RIT-1745  & 1.0000 & -0.5000 & -0.5000 & 0.5511 & 0.9924 & 0.6620 &    0.9407 &    2.3881 &          2.5007 \\
    RIT-1746  & 1.0000 & -0.5000 & -0.5000 & 0.5775 & 0.9923 & 0.6347 &    0.9453 &    2.3236 &          2.4241 \\
    RIT-1747  & 1.0000 & -0.5000 & -0.5000 & 0.6156 & 0.9921 & 0.5939 &    0.9484 &    2.1971 &          2.2865 \\
    RIT-1748  & 1.0000 & -0.5000 & -0.5000 & 0.6400 & 0.9920 & 0.5667 &    0.9522 &    2.1208 &          2.2004 \\
    RIT-1749  & 1.0000 & -0.5000 & -0.5000 & 0.6636 & 0.9918 & 0.5395 &    0.9560 &    2.0437 &          2.1141 \\
    RIT-1750  & 1.0000 & -0.5000 & -0.5000 & 0.6975 & 0.9917 & 0.4986 &    0.9589 &    1.9096 &          1.9709 \\
    RIT-1751  & 1.0000 & -0.5000 & -0.5000 & 0.7500 & 0.9914 & 0.4306 &    0.9607 &    1.6644 &          1.7153 \\
    RIT-1752  & 1.0000 & -0.5000 & -0.5000 & 0.7975 & 0.9911 & 0.3625 &    0.9618 &    1.4122 &          1.4542 \\
    RIT-1753  & 1.0000 & -0.5000 & -0.5000 & 0.8400 & 0.9909 & 0.2945 &    0.9620 &    1.1519 &          1.1860 \\
    RIT-1754  & 1.0000 & -0.5000 & -0.5000 & 0.8775 & 0.9907 & 0.2264 &    0.9619 &    0.8874 &          0.9138 \\
    RIT-1763  & 1.0000 & -0.8000 & -0.8000 & 0.1900 & 0.9945 & 0.8338 &    0.9165 &    1.8661 &          1.9931 \\
    RIT-1764  & 1.0000 & -0.8000 & -0.8000 & 0.3600 & 0.9935 & 0.6967 &    0.9179 &    1.8926 &          2.0191 \\
    RIT-1765  & 1.0000 & -0.8000 & -0.8000 & 0.4375 & 0.9931 & 0.6282 &    0.9201 &    1.9713 &          2.0992 \\
    RIT-1766  & 1.0000 & -0.8000 & -0.8000 & 0.4671 & 0.9929 & 0.6008 &    0.9356 &    2.0502 &          2.1557 \\
    RIT-1767  & 1.0000 & -0.8000 & -0.8000 & 0.5100 & 0.9927 & 0.5596 &    0.9442 &    1.9843 &          2.0721 \\
    RIT-1768  & 1.0000 & -0.8000 & -0.8000 & 0.5511 & 0.9925 & 0.5185 &    0.9490 &    1.8751 &          1.9505 \\
    RIT-1786  & 1.0000 &  0.5000 &  0.5000 & 0.4375 & 0.9930 & 1.2483 &    0.8478 &    3.3397 &          3.7866 \\
    RIT-1787  & 1.0000 &  0.5000 &  0.5000 & 0.4671 & 0.9928 & 1.2216 &    0.8393 &    3.2714 &          3.7381 \\
    RIT-1788  & 1.0000 &  0.5000 &  0.5000 & 0.5100 & 0.9926 & 1.1817 &    0.8478 &    3.3036 &          3.7453 \\
    RIT-1789  & 1.0000 &  0.5000 &  0.5000 & 0.5511 & 0.9924 & 1.1418 &    0.8417 &    3.3245 &          3.7901 \\
    RIT-1790  & 1.0000 &  0.5000 &  0.5000 & 0.5775 & 0.9923 & 1.1152 &    0.8438 &    3.2682 &          3.7188 \\
    RIT-1793  & 1.0000 &  0.5000 &  0.5000 & 0.6636 & 0.9918 & 1.0220 &    0.9026 &    3.6512 &          3.9455 \\
    RIT-1794  & 1.0000 &  0.5000 &  0.5000 & 0.6975 & 0.9916 & 0.9821 &    0.9203 &    3.6174 &          3.8516 \\
    RIT-1795  & 1.0000 &  0.5000 &  0.5000 & 0.7500 & 0.9914 & 0.9155 &    0.9415 &    3.4856 &          3.6476 \\
    RIT-1796  & 1.0000 &  0.5000 &  0.5000 & 0.7975 & 0.9911 & 0.8490 &    0.9507 &    3.2846 &          3.4121 \\
    RIT-1797  & 1.0000 &  0.5000 &  0.5000 & 0.8400 & 0.9909 & 0.7824 &    0.9557 &    3.0589 &          3.1650 \\
    RIT-1798  & 1.0000 &  0.5000 &  0.5000 & 0.8775 & 0.9907 & 0.7159 &    0.9595 &    2.8258 &          2.9151 \\
    RIT-1799  & 1.0000 &  0.5000 &  0.5000 & 0.9100 & 0.9905 & 0.6493 &    0.9568 &    2.5513 &          2.6376 \\
    RIT-1807  & 1.0000 &  0.8000 &  0.8000 & 0.4375 & 0.9930 & 1.3921 &    0.8209 &    3.6194 &          4.2072 \\
    RIT-1808  & 1.0000 &  0.8000 &  0.8000 & 0.4671 & 0.9929 & 1.3656 &    0.8067 &    3.5327 &          4.1623 \\
    RIT-1809  & 1.0000 &  0.8000 &  0.8000 & 0.5100 & 0.9926 & 1.3259 &    0.8119 &    3.5650 &          4.1795 \\
    RIT-1810  & 1.0000 &  0.8000 &  0.8000 & 0.5511 & 0.9924 & 1.2862 &    0.8236 &    3.6033 &          4.1779 \\
    RIT-1811  & 1.0000 &  0.8000 &  0.8000 & 0.5775 & 0.9923 & 1.2598 &    0.8266 &    3.6320 &          4.1993 \\
    RIT-1812  & 1.0000 &  0.8000 &  0.8000 & 0.6156 & 0.9921 & 1.2201 &    0.8253 &    3.6168 &          4.1866 \\
    RIT-1813  & 1.0000 &  0.8000 &  0.8000 & 0.6400 & 0.9920 & 1.1936 &    0.8773 &    3.9589 &          4.3721 \\
    RIT-1815  & 1.0000 &  0.8000 &  0.8000 & 0.6975 & 0.9917 & 1.1275 &    0.9078 &    4.1072 &          4.4187 \\
    RIT-1828  & 1.0000 &  0.0000 &  0.8000 & 0.4375 & 0.9930 & 1.2004 &    0.8496 &    3.1979 &          3.6198 \\
    RIT-1829  & 1.0000 &  0.0000 &  0.8000 & 0.4671 & 0.9929 & 1.1737 &    0.8578 &    3.2434 &          3.6443 \\
    RIT-1830  & 1.0000 &  0.0000 &  0.8000 & 0.5100 & 0.9926 & 1.1337 &    0.8586 &    3.2731 &          3.6750 \\
    RIT-1831  & 1.0000 &  0.0000 &  0.8000 & 0.5775 & 0.9923 & 1.0670 &    0.8639 &    3.3222 &          3.7123 \\
    RIT-1832  & 1.0000 &  0.0000 &  0.8000 & 0.6400 & 0.9919 & 1.0003 &    0.9043 &    3.5489 &          3.8296 \\
    RIT-1833  & 1.0000 &  0.0000 &  0.8000 & 0.6975 & 0.9916 & 0.9336 &    0.9296 &    3.4736 &          3.6702 \\
    RIT-1834  & 1.0000 &  0.0000 &  0.8000 & 0.7500 & 0.9914 & 0.8669 &    0.9441 &    3.3061 &          3.4524 \\
    RIT-1835  & 1.0000 &  0.0000 &  0.8000 & 0.7975 & 0.9911 & 0.8002 &    0.9549 &    3.1139 &          3.2241 \\
    RIT-1899  & 1.0000 &  0.0000 & -0.8000 & 0.1900 & 0.9944 & 1.0221 &    0.9026 &    2.3545 &          2.5442 \\
    RIT-1900  & 1.0000 &  0.0000 & -0.8000 & 0.3600 & 0.9935 & 0.8864 &    0.8991 &    2.3277 &          2.5226 \\
    RIT-1901  & 1.0000 &  0.0000 & -0.8000 & 0.4375 & 0.9931 & 0.8185 &    0.9062 &    2.3841 &          2.5685 \\
    RIT-1902  & 1.0000 &  0.0000 & -0.8000 & 0.4671 & 0.9929 & 0.7913 &    0.9106 &    2.4398 &          2.6187 \\
    RIT-1903  & 1.0000 &  0.0000 & -0.8000 & 0.5100 & 0.9927 & 0.7506 &    0.9179 &    2.5412 &          2.7111 \\
    RIT-1904  & 1.0000 &  0.0000 & -0.8000 & 0.5511 & 0.9924 & 0.7098 &    0.9357 &    2.5434 &          2.6741 \\
    RIT-1905  & 1.0000 &  0.0000 & -0.8000 & 0.5775 & 0.9923 & 0.6827 &    0.9410 &    2.4855 &          2.6022 \\
    RIT-1906  & 1.0000 &  0.0000 & -0.8000 & 0.6400 & 0.9920 & 0.6148 &    0.9515 &    2.3075 &          2.3954 \\     
    SXS-1355  & 1.0000 &  0.0000 &  0.0000 & 0.0678 & 0.9921 & 1.0719 &    0.8783 &    2.8044 &          3.0945 \\
    SXS-1356  & 1.0000 &  0.0000 &  0.0000 & 0.1975 & 0.9934 & 1.1467 &    0.8811 &    2.8096 &          3.0925 \\
    SXS-1357  & 1.0000 &  0.0000 &  0.0000 & 0.2211 & 0.9924 & 1.0823 &    0.8821 &    2.8049 &          3.0846 \\
    SXS-1358  & 1.0000 &  0.0000 &  0.0000 & 0.2186 & 0.9923 & 1.0759 &    0.8811 &    2.8094 &          3.0924 \\
    SXS-1359  & 1.0000 &  0.0000 &  0.0000 & 0.2178 & 0.9923 & 1.0721 &    0.8824 &    2.8156 &          3.0955 \\
    SXS-1360  & 1.0000 &  0.0000 &  0.0000 & 0.3636 & 0.9924 & 1.0682 &    0.8788 &    2.7941 &          3.0818 \\
    SXS-1361  & 1.0000 &  0.0000 &  0.0000 & 0.3326 & 0.9923 & 1.0668 &    0.8802 &    2.8009 &          3.0855 \\
    SXS-1362  & 1.0000 &  0.0000 &  0.0000 & 0.0000 & 0.9925 & 1.0619 &    0.8828 &    2.8098 &          3.0881 \\
    SXS-1363  & 1.0000 &  0.0000 &  0.0000 & 0.0000 & 0.9925 & 1.0608 &    0.8823 &    2.7970 &          3.0754 \\
    ET-37       & 1.0000 &  0.0000 &  0.0000 & -      & 0.9943 & 0.7432 &    0.9621 &    2.8421 &          2.9259 \\
    ET-42       & 1.0000 &  0.0000 &  0.0000 & -      & 0.9943 & 0.8263 &    0.9456 &    3.0592 &          3.1909 \\
    ET-42\_q150 & 1.5000 &  0.0000 &  0.0000 & -      & 0.9950 & 0.8263 &    0.9382 &    3.1313 &          3.2875 \\
    ET-42\_q200 & 2.0000 &  0.0000 &  0.0000 & -      & 0.9964 & 0.8263 &    0.9056 &    3.0975 &          3.3479 \\
    ET-42\_q215 & 2.1500 &  0.0000 &  0.0000 & -      & 0.9969 & 0.8263 &    0.9140 &    3.0779 &          3.3041 \\
    ET-44       & 1.0000 &  0.0000 &  0.0000 & -      & 0.9943 & 0.8579 &    0.9301 &    3.0833 &          3.2566 \\
    ET-46       & 1.0000 &  0.0000 &  0.0000 & -      & 0.9943 & 0.8883 &    0.9123 &    3.0702 &          3.2907 \\
    ET-48       & 1.0000 &  0.0000 &  0.0000 & -      & 0.9943 & 0.9177 &    0.8728 &    2.7504 &          3.0493 \\
    ET-50       & 1.0000 &  0.0000 &  0.0000 & -      & 0.9943 & 0.9460 &    0.8874 &    2.9056 &          3.1806 \\

\end{longtable*}

\begin{table*}[t]
\caption{List of test-mass data included in our dataset, and selected associated parameters. All evolutions start at apastron.\vspace{0.1cm}}
\label{tab:sims_RWZ}
\begin{tabular}{lcccc}

\toprule
ID & $e_0$ & $\Eeffmrg$ &  $j_{\mathrm{mrg}}$ &  $\hat{b}_{\mathrm{mrg}}$ \\
\toprule

RWZ-1  & 0.0000 & 0.9423 & 3.4577 & 3.6695 \\
RWZ-2  & 0.0500 & 0.9425 & 3.4602 & 3.6714 \\
RWZ-3  & 0.1000 & 0.9430 & 3.4671 & 3.6768 \\
RWZ-4  & 0.1500 & 0.9433 & 3.4720 & 3.6805 \\
RWZ-5  & 0.2000 & 0.9444 & 3.4859 & 3.6911 \\
RWZ-6  & 0.2500 & 0.9451 & 3.4936 & 3.6967 \\
RWZ-7  & 0.3000 & 0.9463 & 3.5085 & 3.7077 \\
RWZ-8  & 0.3500 & 0.9477 & 3.5247 & 3.7193 \\
RWZ-9  & 0.4000 & 0.9504 & 3.5548 & 3.7403 \\
RWZ-10 & 0.4500 & 0.9513 & 3.5649 & 3.7473 \\
RWZ-11 & 0.5000 & 0.9546 & 3.5989 & 3.7702 \\
RWZ-12 & 0.5500 & 0.9563 & 3.6165 & 3.7817 \\
RWZ-13 & 0.6000 & 0.9593 & 3.6453 & 3.8001 \\
RWZ-14 & 0.6500 & 0.9626 & 3.6774 & 3.8203 \\
RWZ-15 & 0.7000 & 0.9666 & 3.7149 & 3.8433 \\
RWZ-16 & 0.7500 & 0.9711 & 3.7562 & 3.8681 \\
RWZ-17 & 0.8000 & 0.9752 & 3.7921 & 3.8885 \\
RWZ-18 & 0.8500 & 0.9785 & 3.8212 & 3.9050 \\
RWZ-19 & 0.9000 & 0.9841 & 3.8696 & 3.9321 \\
RWZ-20 & 0.9500 & 0.9880 & 3.9017 & 3.9492 \\
    
\hline\hline

\end{tabular}
\end{table*}

\clearpage

\bibliography{local,references}

\begin{thebibliography}{145}%
\makeatletter
\providecommand \@ifxundefined [1]{%
 \@ifx{#1\undefined}
}%
\providecommand \@ifnum [1]{%
 \ifnum #1\expandafter \@firstoftwo
 \else \expandafter \@secondoftwo
 \fi
}%
\providecommand \@ifx [1]{%
 \ifx #1\expandafter \@firstoftwo
 \else \expandafter \@secondoftwo
 \fi
}%
\providecommand \natexlab [1]{#1}%
\providecommand \enquote  [1]{``#1''}%
\providecommand \bibnamefont  [1]{#1}%
\providecommand \bibfnamefont [1]{#1}%
\providecommand \citenamefont [1]{#1}%
\providecommand \href@noop [0]{\@secondoftwo}%
\providecommand \href [0]{\begingroup \@sanitize@url \@href}%
\providecommand \@href[1]{\@@startlink{#1}\@@href}%
\providecommand \@@href[1]{\endgroup#1\@@endlink}%
\providecommand \@sanitize@url [0]{\catcode `\\12\catcode `\$12\catcode
  `\&12\catcode `\#12\catcode `\^12\catcode `\_12\catcode `\%12\relax}%
\providecommand \@@startlink[1]{}%
\providecommand \@@endlink[0]{}%
\providecommand \url  [0]{\begingroup\@sanitize@url \@url }%
\providecommand \@url [1]{\endgroup\@href {#1}{\urlprefix }}%
\providecommand \urlprefix  [0]{URL }%
\providecommand \Eprint [0]{\href }%
\providecommand \doibase [0]{https://doi.org/}%
\providecommand \selectlanguage [0]{\@gobble}%
\providecommand \bibinfo  [0]{\@secondoftwo}%
\providecommand \bibfield  [0]{\@secondoftwo}%
\providecommand \translation [1]{[#1]}%
\providecommand \BibitemOpen [0]{}%
\providecommand \bibitemStop [0]{}%
\providecommand \bibitemNoStop [0]{.\EOS\space}%
\providecommand \EOS [0]{\spacefactor3000\relax}%
\providecommand \BibitemShut  [1]{\csname bibitem#1\endcsname}%
\let\auto@bib@innerbib\@empty
\bibitem [{\citenamefont {Mandel}\ and\ \citenamefont
  {Farmer}(2022)}]{Mandel:2018hfr}%
  \BibitemOpen
  \bibfield  {author} {\bibinfo {author} {\bibfnamefont {I.}~\bibnamefont
  {Mandel}}\ and\ \bibinfo {author} {\bibfnamefont {A.}~\bibnamefont
  {Farmer}},\ }\href {https://doi.org/10.1016/j.physrep.2022.01.003} {\bibfield
   {journal} {\bibinfo  {journal} {Phys. Rept.}\ }\textbf {\bibinfo {volume}
  {955}},\ \bibinfo {pages} {1} (\bibinfo {year} {2022})},\ \Eprint
  {https://arxiv.org/abs/1806.05820} {arXiv:1806.05820 [astro-ph.HE]}
  \BibitemShut {NoStop}%
\bibitem [{\citenamefont {Mapelli}(2021)}]{Mapelli:2021taw}%
  \BibitemOpen
  \bibfield  {author} {\bibinfo {author} {\bibfnamefont {M.}~\bibnamefont
  {Mapelli}},\ }\href@noop {} {\  (\bibinfo {year} {2021})},\ \Eprint
  {https://arxiv.org/abs/2106.00699} {arXiv:2106.00699 [astro-ph.HE]}
  \BibitemShut {NoStop}%
\bibitem [{\citenamefont {Abbott}\ \emph
  {et~al.}(2021{\natexlab{a}})\citenamefont {Abbott} \emph
  {et~al.}}]{LIGOScientific:2021sio}%
  \BibitemOpen
  \bibfield  {author} {\bibinfo {author} {\bibfnamefont {R.}~\bibnamefont
  {Abbott}} \emph {et~al.} (\bibinfo {collaboration} {LIGO Scientific, VIRGO,
  KAGRA}),\ }\href@noop {} {\  (\bibinfo {year} {2021}{\natexlab{a}})},\
  \Eprint {https://arxiv.org/abs/2112.06861} {arXiv:2112.06861 [gr-qc]}
  \BibitemShut {NoStop}%
\bibitem [{\citenamefont {Narayan}\ \emph {et~al.}(2023)\citenamefont
  {Narayan}, \citenamefont {Johnson-McDaniel},\ and\ \citenamefont
  {Gupta}}]{Narayan:2023vhm}%
  \BibitemOpen
  \bibfield  {author} {\bibinfo {author} {\bibfnamefont {P.}~\bibnamefont
  {Narayan}}, \bibinfo {author} {\bibfnamefont {N.~K.}\ \bibnamefont
  {Johnson-McDaniel}},\ and\ \bibinfo {author} {\bibfnamefont {A.}~\bibnamefont
  {Gupta}},\ }\href@noop {} {\  (\bibinfo {year} {2023})},\ \Eprint
  {https://arxiv.org/abs/2306.04068} {arXiv:2306.04068 [gr-qc]} \BibitemShut
  {NoStop}%
\bibitem [{\citenamefont {Aasi}\ \emph {et~al.}(2015)\citenamefont {Aasi} \emph
  {et~al.}}]{TheLIGOScientific:2014jea}%
  \BibitemOpen
  \bibfield  {author} {\bibinfo {author} {\bibfnamefont {J.}~\bibnamefont
  {Aasi}} \emph {et~al.} (\bibinfo {collaboration} {LIGO Scientific}),\ }\href
  {https://doi.org/10.1088/0264-9381/32/7/074001} {\bibfield  {journal}
  {\bibinfo  {journal} {Class. Quant. Grav.}\ }\textbf {\bibinfo {volume}
  {32}},\ \bibinfo {pages} {074001} (\bibinfo {year} {2015})},\ \Eprint
  {https://arxiv.org/abs/1411.4547} {arXiv:1411.4547 [gr-qc]} \BibitemShut
  {NoStop}%
\bibitem [{\citenamefont {Acernese}\ \emph {et~al.}(2015)\citenamefont
  {Acernese} \emph {et~al.}}]{TheVirgo:2014hva}%
  \BibitemOpen
  \bibfield  {author} {\bibinfo {author} {\bibfnamefont {F.}~\bibnamefont
  {Acernese}} \emph {et~al.} (\bibinfo {collaboration} {VIRGO}),\ }\href
  {https://doi.org/10.1088/0264-9381/32/2/024001} {\bibfield  {journal}
  {\bibinfo  {journal} {Class. Quant. Grav.}\ }\textbf {\bibinfo {volume}
  {32}},\ \bibinfo {pages} {024001} (\bibinfo {year} {2015})},\ \Eprint
  {https://arxiv.org/abs/1408.3978} {arXiv:1408.3978 [gr-qc]} \BibitemShut
  {NoStop}%
\bibitem [{\citenamefont {Amaro-Seoane}\ \emph {et~al.}(2017)\citenamefont
  {Amaro-Seoane} \emph {et~al.}}]{LISA:2017pwj}%
  \BibitemOpen
  \bibfield  {author} {\bibinfo {author} {\bibfnamefont {P.}~\bibnamefont
  {Amaro-Seoane}} \emph {et~al.} (\bibinfo {collaboration} {LISA}),\
  }\href@noop {} {\  (\bibinfo {year} {2017})},\ \Eprint
  {https://arxiv.org/abs/1702.00786} {arXiv:1702.00786 [astro-ph.IM]}
  \BibitemShut {NoStop}%
\bibitem [{\citenamefont {Antoniadis}\ \emph {et~al.}(2023)\citenamefont
  {Antoniadis} \emph {et~al.}}]{Antoniadis:2023rey}%
  \BibitemOpen
  \bibfield  {author} {\bibinfo {author} {\bibfnamefont {J.}~\bibnamefont
  {Antoniadis}} \emph {et~al.},\ }\href@noop {} {\  (\bibinfo {year} {2023})},\
  \Eprint {https://arxiv.org/abs/2306.16214} {arXiv:2306.16214 [astro-ph.HE]}
  \BibitemShut {NoStop}%
\bibitem [{\citenamefont {Agazie}\ \emph {et~al.}(2023)\citenamefont {Agazie}
  \emph {et~al.}}]{NANOGrav:2023gor}%
  \BibitemOpen
  \bibfield  {author} {\bibinfo {author} {\bibfnamefont {G.}~\bibnamefont
  {Agazie}} \emph {et~al.} (\bibinfo {collaboration} {NANOGrav}),\ }\href
  {https://doi.org/10.3847/2041-8213/acdac6} {\bibfield  {journal} {\bibinfo
  {journal} {Astrophys. J. Lett.}\ }\textbf {\bibinfo {volume} {951}},\
  \bibinfo {pages} {L8} (\bibinfo {year} {2023})},\ \Eprint
  {https://arxiv.org/abs/2306.16213} {arXiv:2306.16213 [astro-ph.HE]}
  \BibitemShut {NoStop}%
\bibitem [{\citenamefont {Reardon}\ \emph {et~al.}(2023)\citenamefont {Reardon}
  \emph {et~al.}}]{Reardon:2023gzh}%
  \BibitemOpen
  \bibfield  {author} {\bibinfo {author} {\bibfnamefont {D.~J.}\ \bibnamefont
  {Reardon}} \emph {et~al.},\ }\href {https://doi.org/10.3847/2041-8213/acdd02}
  {\bibfield  {journal} {\bibinfo  {journal} {Astrophys. J. Lett.}\ }\textbf
  {\bibinfo {volume} {951}},\ \bibinfo {pages} {L6} (\bibinfo {year} {2023})},\
  \Eprint {https://arxiv.org/abs/2306.16215} {arXiv:2306.16215 [astro-ph.HE]}
  \BibitemShut {NoStop}%
\bibitem [{\citenamefont {Xu}\ \emph {et~al.}(2023)\citenamefont {Xu} \emph
  {et~al.}}]{Xu:2023wog}%
  \BibitemOpen
  \bibfield  {author} {\bibinfo {author} {\bibfnamefont {H.}~\bibnamefont {Xu}}
  \emph {et~al.},\ }\href {https://doi.org/10.1088/1674-4527/acdfa5} {\bibfield
   {journal} {\bibinfo  {journal} {Res. Astron. Astrophys.}\ }\textbf {\bibinfo
  {volume} {23}},\ \bibinfo {pages} {075024} (\bibinfo {year} {2023})},\
  \Eprint {https://arxiv.org/abs/2306.16216} {arXiv:2306.16216 [astro-ph.HE]}
  \BibitemShut {NoStop}%
\bibitem [{\citenamefont {Abbott}\ \emph {et~al.}(2019)\citenamefont {Abbott}
  \emph {et~al.}}]{LIGOScientific:2019dag}%
  \BibitemOpen
  \bibfield  {author} {\bibinfo {author} {\bibfnamefont {B.~P.}\ \bibnamefont
  {Abbott}} \emph {et~al.} (\bibinfo {collaboration} {LIGO Scientific,
  Virgo}),\ }\href {https://doi.org/10.3847/1538-4357/ab3c2d} {\bibfield
  {journal} {\bibinfo  {journal} {Astrophys. J.}\ }\textbf {\bibinfo {volume}
  {883}},\ \bibinfo {pages} {149} (\bibinfo {year} {2019})},\ \Eprint
  {https://arxiv.org/abs/1907.09384} {arXiv:1907.09384 [astro-ph.HE]}
  \BibitemShut {NoStop}%
\bibitem [{\citenamefont {Romero-Shaw}\ \emph {et~al.}(2019)\citenamefont
  {Romero-Shaw}, \citenamefont {Lasky},\ and\ \citenamefont
  {Thrane}}]{Romero-Shaw:2019itr}%
  \BibitemOpen
  \bibfield  {author} {\bibinfo {author} {\bibfnamefont {I.~M.}\ \bibnamefont
  {Romero-Shaw}}, \bibinfo {author} {\bibfnamefont {P.~D.}\ \bibnamefont
  {Lasky}},\ and\ \bibinfo {author} {\bibfnamefont {E.}~\bibnamefont
  {Thrane}},\ }\href {https://doi.org/10.1093/mnras/stz2996} {\bibfield
  {journal} {\bibinfo  {journal} {Mon. Not. Roy. Astron. Soc.}\ }\textbf
  {\bibinfo {volume} {490}},\ \bibinfo {pages} {5210} (\bibinfo {year}
  {2019})},\ \Eprint {https://arxiv.org/abs/1909.05466} {arXiv:1909.05466
  [astro-ph.HE]} \BibitemShut {NoStop}%
\bibitem [{\citenamefont {Nitz}\ \emph {et~al.}(2019)\citenamefont {Nitz},
  \citenamefont {Lenon},\ and\ \citenamefont {Brown}}]{Nitz:2019spj}%
  \BibitemOpen
  \bibfield  {author} {\bibinfo {author} {\bibfnamefont {A.~H.}\ \bibnamefont
  {Nitz}}, \bibinfo {author} {\bibfnamefont {A.}~\bibnamefont {Lenon}},\ and\
  \bibinfo {author} {\bibfnamefont {D.~A.}\ \bibnamefont {Brown}},\ }\href
  {https://doi.org/10.3847/1538-4357/ab6611} {\bibfield  {journal} {\bibinfo
  {journal} {Astrophys. J.}\ }\textbf {\bibinfo {volume} {890}},\ \bibinfo
  {pages} {1} (\bibinfo {year} {2019})},\ \Eprint
  {https://arxiv.org/abs/1912.05464} {arXiv:1912.05464 [astro-ph.HE]}
  \BibitemShut {NoStop}%
\bibitem [{\citenamefont {Gayathri}\ \emph {et~al.}(2022)\citenamefont
  {Gayathri}, \citenamefont {Healy}, \citenamefont {Lange}, \citenamefont
  {O'Brien}, \citenamefont {Szczepanczyk}, \citenamefont {Bartos},
  \citenamefont {Campanelli}, \citenamefont {Klimenko}, \citenamefont
  {Lousto},\ and\ \citenamefont {O'Shaughnessy}}]{Gayathri:2020coq}%
  \BibitemOpen
  \bibfield  {author} {\bibinfo {author} {\bibfnamefont {V.}~\bibnamefont
  {Gayathri}}, \bibinfo {author} {\bibfnamefont {J.}~\bibnamefont {Healy}},
  \bibinfo {author} {\bibfnamefont {J.}~\bibnamefont {Lange}}, \bibinfo
  {author} {\bibfnamefont {B.}~\bibnamefont {O'Brien}}, \bibinfo {author}
  {\bibfnamefont {M.}~\bibnamefont {Szczepanczyk}}, \bibinfo {author}
  {\bibfnamefont {I.}~\bibnamefont {Bartos}}, \bibinfo {author} {\bibfnamefont
  {M.}~\bibnamefont {Campanelli}}, \bibinfo {author} {\bibfnamefont
  {S.}~\bibnamefont {Klimenko}}, \bibinfo {author} {\bibfnamefont {C.~O.}\
  \bibnamefont {Lousto}},\ and\ \bibinfo {author} {\bibfnamefont
  {R.}~\bibnamefont {O'Shaughnessy}},\ }\href
  {https://doi.org/10.1038/s41550-021-01568-w} {\bibfield  {journal} {\bibinfo
  {journal} {Nature Astron.}\ }\textbf {\bibinfo {volume} {6}},\ \bibinfo
  {pages} {344} (\bibinfo {year} {2022})},\ \Eprint
  {https://arxiv.org/abs/2009.05461} {arXiv:2009.05461 [astro-ph.HE]}
  \BibitemShut {NoStop}%
\bibitem [{\citenamefont {Ramos-Buades}\ \emph {et~al.}(2020)\citenamefont
  {Ramos-Buades}, \citenamefont {Tiwari}, \citenamefont {Haney},\ and\
  \citenamefont {Husa}}]{Ramos-Buades:2020eju}%
  \BibitemOpen
  \bibfield  {author} {\bibinfo {author} {\bibfnamefont {A.}~\bibnamefont
  {Ramos-Buades}}, \bibinfo {author} {\bibfnamefont {S.}~\bibnamefont
  {Tiwari}}, \bibinfo {author} {\bibfnamefont {M.}~\bibnamefont {Haney}},\ and\
  \bibinfo {author} {\bibfnamefont {S.}~\bibnamefont {Husa}},\ }\href
  {https://doi.org/10.1103/PhysRevD.102.043005} {\bibfield  {journal} {\bibinfo
   {journal} {Phys. Rev. D}\ }\textbf {\bibinfo {volume} {102}},\ \bibinfo
  {pages} {043005} (\bibinfo {year} {2020})},\ \Eprint
  {https://arxiv.org/abs/2005.14016} {arXiv:2005.14016 [gr-qc]} \BibitemShut
  {NoStop}%
\bibitem [{\citenamefont {Veske}\ \emph {et~al.}(2021)\citenamefont {Veske},
  \citenamefont {Sullivan}, \citenamefont {M\'arka}, \citenamefont {Bartos},
  \citenamefont {Corley}, \citenamefont {Samsing}, \citenamefont {Buscicchio},\
  and\ \citenamefont {M\'arka}}]{Veske:2020idq}%
  \BibitemOpen
  \bibfield  {author} {\bibinfo {author} {\bibfnamefont {D.}~\bibnamefont
  {Veske}}, \bibinfo {author} {\bibfnamefont {A.~G.}\ \bibnamefont {Sullivan}},
  \bibinfo {author} {\bibfnamefont {Z.}~\bibnamefont {M\'arka}}, \bibinfo
  {author} {\bibfnamefont {I.}~\bibnamefont {Bartos}}, \bibinfo {author}
  {\bibfnamefont {K.~R.}\ \bibnamefont {Corley}}, \bibinfo {author}
  {\bibfnamefont {J.}~\bibnamefont {Samsing}}, \bibinfo {author} {\bibfnamefont
  {R.}~\bibnamefont {Buscicchio}},\ and\ \bibinfo {author} {\bibfnamefont
  {S.}~\bibnamefont {M\'arka}},\ }\href
  {https://doi.org/10.3847/2041-8213/abd721} {\bibfield  {journal} {\bibinfo
  {journal} {Astrophys. J. Lett.}\ }\textbf {\bibinfo {volume} {907}},\
  \bibinfo {pages} {L48} (\bibinfo {year} {2021})},\ \Eprint
  {https://arxiv.org/abs/2011.06591} {arXiv:2011.06591 [astro-ph.HE]}
  \BibitemShut {NoStop}%
\bibitem [{\citenamefont {Romero-Shaw}\ \emph {et~al.}(2020)\citenamefont
  {Romero-Shaw}, \citenamefont {Lasky}, \citenamefont {Thrane},\ and\
  \citenamefont {Bustillo}}]{Romero-Shaw:2020thy}%
  \BibitemOpen
  \bibfield  {author} {\bibinfo {author} {\bibfnamefont {I.~M.}\ \bibnamefont
  {Romero-Shaw}}, \bibinfo {author} {\bibfnamefont {P.~D.}\ \bibnamefont
  {Lasky}}, \bibinfo {author} {\bibfnamefont {E.}~\bibnamefont {Thrane}},\ and\
  \bibinfo {author} {\bibfnamefont {J.~C.}\ \bibnamefont {Bustillo}},\ }\href
  {https://doi.org/10.3847/2041-8213/abbe26} {\bibfield  {journal} {\bibinfo
  {journal} {Astrophys. J. Lett.}\ }\textbf {\bibinfo {volume} {903}},\
  \bibinfo {pages} {L5} (\bibinfo {year} {2020})},\ \Eprint
  {https://arxiv.org/abs/2009.04771} {arXiv:2009.04771 [astro-ph.HE]}
  \BibitemShut {NoStop}%
\bibitem [{\citenamefont {Veske}\ \emph {et~al.}(2020)\citenamefont {Veske},
  \citenamefont {M\'arka}, \citenamefont {Sullivan}, \citenamefont {Bartos},
  \citenamefont {Corley}, \citenamefont {Samsing},\ and\ \citenamefont
  {M\'arka}}]{Veske:2020zch}%
  \BibitemOpen
  \bibfield  {author} {\bibinfo {author} {\bibfnamefont {D.}~\bibnamefont
  {Veske}}, \bibinfo {author} {\bibfnamefont {Z.}~\bibnamefont {M\'arka}},
  \bibinfo {author} {\bibfnamefont {A.~G.}\ \bibnamefont {Sullivan}}, \bibinfo
  {author} {\bibfnamefont {I.}~\bibnamefont {Bartos}}, \bibinfo {author}
  {\bibfnamefont {K.~R.}\ \bibnamefont {Corley}}, \bibinfo {author}
  {\bibfnamefont {J.}~\bibnamefont {Samsing}},\ and\ \bibinfo {author}
  {\bibfnamefont {S.}~\bibnamefont {M\'arka}},\ }\href
  {https://doi.org/10.1093/mnrasl/slaa123} {\bibfield  {journal} {\bibinfo
  {journal} {Mon. Not. Roy. Astron. Soc.}\ }\textbf {\bibinfo {volume} {498}},\
  \bibinfo {pages} {L46} (\bibinfo {year} {2020})},\ \Eprint
  {https://arxiv.org/abs/2002.12346} {arXiv:2002.12346 [astro-ph.HE]}
  \BibitemShut {NoStop}%
\bibitem [{\citenamefont {Romero-Shaw}\ \emph {et~al.}(2021)\citenamefont
  {Romero-Shaw}, \citenamefont {Lasky},\ and\ \citenamefont
  {Thrane}}]{Romero-Shaw:2021ual}%
  \BibitemOpen
  \bibfield  {author} {\bibinfo {author} {\bibfnamefont {I.~M.}\ \bibnamefont
  {Romero-Shaw}}, \bibinfo {author} {\bibfnamefont {P.~D.}\ \bibnamefont
  {Lasky}},\ and\ \bibinfo {author} {\bibfnamefont {E.}~\bibnamefont
  {Thrane}},\ }\href {https://doi.org/10.3847/2041-8213/ac3138} {\bibfield
  {journal} {\bibinfo  {journal} {Astrophys. J. Lett.}\ }\textbf {\bibinfo
  {volume} {921}},\ \bibinfo {pages} {L31} (\bibinfo {year} {2021})},\ \Eprint
  {https://arxiv.org/abs/2108.01284} {arXiv:2108.01284 [astro-ph.HE]}
  \BibitemShut {NoStop}%
\bibitem [{\citenamefont {O'Shea}\ and\ \citenamefont
  {Kumar}(2021)}]{OShea:2021ugg}%
  \BibitemOpen
  \bibfield  {author} {\bibinfo {author} {\bibfnamefont {E.}~\bibnamefont
  {O'Shea}}\ and\ \bibinfo {author} {\bibfnamefont {P.}~\bibnamefont {Kumar}},\
  }\href@noop {} {\  (\bibinfo {year} {2021})},\ \Eprint
  {https://arxiv.org/abs/2107.07981} {arXiv:2107.07981 [astro-ph.HE]}
  \BibitemShut {NoStop}%
\bibitem [{\citenamefont {Abbott}\ \emph {et~al.}(2022)\citenamefont {Abbott}
  \emph {et~al.}}]{LIGOScientific:2021tfm}%
  \BibitemOpen
  \bibfield  {author} {\bibinfo {author} {\bibfnamefont {R.}~\bibnamefont
  {Abbott}} \emph {et~al.} (\bibinfo {collaboration} {LIGO Scientific, VIRGO,
  KAGRA}),\ }\href {https://doi.org/10.1051/0004-6361/202141452} {\bibfield
  {journal} {\bibinfo  {journal} {Astron. Astrophys.}\ }\textbf {\bibinfo
  {volume} {659}},\ \bibinfo {pages} {A84} (\bibinfo {year} {2022})},\ \Eprint
  {https://arxiv.org/abs/2105.15120} {arXiv:2105.15120 [astro-ph.HE]}
  \BibitemShut {NoStop}%
\bibitem [{\citenamefont {Gayathri}\ \emph {et~al.}(2021)\citenamefont
  {Gayathri}, \citenamefont {Yang}, \citenamefont {Tagawa}, \citenamefont
  {Haiman},\ and\ \citenamefont {Bartos}}]{Gayathri:2021xwb}%
  \BibitemOpen
  \bibfield  {author} {\bibinfo {author} {\bibfnamefont {V.}~\bibnamefont
  {Gayathri}}, \bibinfo {author} {\bibfnamefont {Y.}~\bibnamefont {Yang}},
  \bibinfo {author} {\bibfnamefont {H.}~\bibnamefont {Tagawa}}, \bibinfo
  {author} {\bibfnamefont {Z.}~\bibnamefont {Haiman}},\ and\ \bibinfo {author}
  {\bibfnamefont {I.}~\bibnamefont {Bartos}},\ }\href
  {https://doi.org/10.3847/2041-8213/ac2cc1} {\bibfield  {journal} {\bibinfo
  {journal} {Astrophys. J. Lett.}\ }\textbf {\bibinfo {volume} {920}},\
  \bibinfo {pages} {L42} (\bibinfo {year} {2021})},\ \Eprint
  {https://arxiv.org/abs/2104.10253} {arXiv:2104.10253 [gr-qc]} \BibitemShut
  {NoStop}%
\bibitem [{\citenamefont {Iglesias}\ \emph {et~al.}(2022)\citenamefont
  {Iglesias} \emph {et~al.}}]{Iglesias:2022xfc}%
  \BibitemOpen
  \bibfield  {author} {\bibinfo {author} {\bibfnamefont {H.~L.}\ \bibnamefont
  {Iglesias}} \emph {et~al.},\ }\href@noop {} {\  (\bibinfo {year} {2022})},\
  \Eprint {https://arxiv.org/abs/2208.01766} {arXiv:2208.01766 [gr-qc]}
  \BibitemShut {NoStop}%
\bibitem [{\citenamefont {Romero-Shaw}\ \emph {et~al.}(2022)\citenamefont
  {Romero-Shaw}, \citenamefont {Lasky},\ and\ \citenamefont
  {Thrane}}]{Romero-Shaw:2022xko}%
  \BibitemOpen
  \bibfield  {author} {\bibinfo {author} {\bibfnamefont {I.~M.}\ \bibnamefont
  {Romero-Shaw}}, \bibinfo {author} {\bibfnamefont {P.~D.}\ \bibnamefont
  {Lasky}},\ and\ \bibinfo {author} {\bibfnamefont {E.}~\bibnamefont
  {Thrane}},\ }\href {https://doi.org/10.3847/1538-4357/ac9798} {\bibfield
  {journal} {\bibinfo  {journal} {Astrophys. J.}\ }\textbf {\bibinfo {volume}
  {940}},\ \bibinfo {pages} {171} (\bibinfo {year} {2022})},\ \Eprint
  {https://arxiv.org/abs/2206.14695} {arXiv:2206.14695 [astro-ph.HE]}
  \BibitemShut {NoStop}%
\bibitem [{\citenamefont {Ebersold}\ \emph {et~al.}(2022)\citenamefont
  {Ebersold}, \citenamefont {Tiwari}, \citenamefont {Smith}, \citenamefont
  {Bae}, \citenamefont {Kang}, \citenamefont {Williams}, \citenamefont
  {Gopakumar}, \citenamefont {Heng},\ and\ \citenamefont
  {Haney}}]{Ebersold:2022zvz}%
  \BibitemOpen
  \bibfield  {author} {\bibinfo {author} {\bibfnamefont {M.}~\bibnamefont
  {Ebersold}}, \bibinfo {author} {\bibfnamefont {S.}~\bibnamefont {Tiwari}},
  \bibinfo {author} {\bibfnamefont {L.}~\bibnamefont {Smith}}, \bibinfo
  {author} {\bibfnamefont {Y.-B.}\ \bibnamefont {Bae}}, \bibinfo {author}
  {\bibfnamefont {G.}~\bibnamefont {Kang}}, \bibinfo {author} {\bibfnamefont
  {D.}~\bibnamefont {Williams}}, \bibinfo {author} {\bibfnamefont
  {A.}~\bibnamefont {Gopakumar}}, \bibinfo {author} {\bibfnamefont {I.~S.}\
  \bibnamefont {Heng}},\ and\ \bibinfo {author} {\bibfnamefont
  {M.}~\bibnamefont {Haney}},\ }\href
  {https://doi.org/10.1103/PhysRevD.106.104014} {\bibfield  {journal} {\bibinfo
   {journal} {Phys. Rev. D}\ }\textbf {\bibinfo {volume} {106}},\ \bibinfo
  {pages} {104014} (\bibinfo {year} {2022})},\ \Eprint
  {https://arxiv.org/abs/2208.07762} {arXiv:2208.07762 [gr-qc]} \BibitemShut
  {NoStop}%
\bibitem [{\citenamefont {Dandapat}\ \emph {et~al.}(2023)\citenamefont
  {Dandapat}, \citenamefont {Ebersold}, \citenamefont {Susobhanan},
  \citenamefont {Rana}, \citenamefont {Gopakumar}, \citenamefont {Tiwari},
  \citenamefont {Haney}, \citenamefont {Lee},\ and\ \citenamefont
  {Kolhe}}]{Dandapat:2023zzn}%
  \BibitemOpen
  \bibfield  {author} {\bibinfo {author} {\bibfnamefont {S.}~\bibnamefont
  {Dandapat}}, \bibinfo {author} {\bibfnamefont {M.}~\bibnamefont {Ebersold}},
  \bibinfo {author} {\bibfnamefont {A.}~\bibnamefont {Susobhanan}}, \bibinfo
  {author} {\bibfnamefont {P.}~\bibnamefont {Rana}}, \bibinfo {author}
  {\bibfnamefont {A.}~\bibnamefont {Gopakumar}}, \bibinfo {author}
  {\bibfnamefont {S.}~\bibnamefont {Tiwari}}, \bibinfo {author} {\bibfnamefont
  {M.}~\bibnamefont {Haney}}, \bibinfo {author} {\bibfnamefont {H.~M.}\
  \bibnamefont {Lee}},\ and\ \bibinfo {author} {\bibfnamefont {N.}~\bibnamefont
  {Kolhe}},\ }\href {https://doi.org/10.1103/PhysRevD.108.024013} {\bibfield
  {journal} {\bibinfo  {journal} {Phys. Rev. D}\ }\textbf {\bibinfo {volume}
  {108}},\ \bibinfo {pages} {024013} (\bibinfo {year} {2023})},\ \Eprint
  {https://arxiv.org/abs/2305.19318} {arXiv:2305.19318 [gr-qc]} \BibitemShut
  {NoStop}%
\bibitem [{\citenamefont {Garg}\ \emph {et~al.}(2023)\citenamefont {Garg},
  \citenamefont {Tiwari}, \citenamefont {Derdzinski}, \citenamefont {Baker},
  \citenamefont {Marsat},\ and\ \citenamefont {Mayer}}]{Garg:2023lfg}%
  \BibitemOpen
  \bibfield  {author} {\bibinfo {author} {\bibfnamefont {M.}~\bibnamefont
  {Garg}}, \bibinfo {author} {\bibfnamefont {S.}~\bibnamefont {Tiwari}},
  \bibinfo {author} {\bibfnamefont {A.}~\bibnamefont {Derdzinski}}, \bibinfo
  {author} {\bibfnamefont {J.}~\bibnamefont {Baker}}, \bibinfo {author}
  {\bibfnamefont {S.}~\bibnamefont {Marsat}},\ and\ \bibinfo {author}
  {\bibfnamefont {L.}~\bibnamefont {Mayer}},\ }\href@noop {} {\  (\bibinfo
  {year} {2023})},\ \Eprint {https://arxiv.org/abs/2307.13367}
  {arXiv:2307.13367 [astro-ph.GA]} \BibitemShut {NoStop}%
\bibitem [{\citenamefont {Ramos-Buades}\ \emph {et~al.}(2023)\citenamefont
  {Ramos-Buades}, \citenamefont {Buonanno},\ and\ \citenamefont
  {Gair}}]{Ramos-Buades:2023yhy}%
  \BibitemOpen
  \bibfield  {author} {\bibinfo {author} {\bibfnamefont {A.}~\bibnamefont
  {Ramos-Buades}}, \bibinfo {author} {\bibfnamefont {A.}~\bibnamefont
  {Buonanno}},\ and\ \bibinfo {author} {\bibfnamefont {J.}~\bibnamefont
  {Gair}},\ }\href@noop {} {\  (\bibinfo {year} {2023})},\ \Eprint
  {https://arxiv.org/abs/2309.15528} {arXiv:2309.15528 [gr-qc]} \BibitemShut
  {NoStop}%
\bibitem [{\citenamefont {Gayathri}\ \emph {et~al.}(2020)\citenamefont
  {Gayathri}, \citenamefont {Bartos}, \citenamefont {Haiman}, \citenamefont
  {Klimenko}, \citenamefont {Kocsis}, \citenamefont {Marka},\ and\
  \citenamefont {Yang}}]{Gayathri:2019kop}%
  \BibitemOpen
  \bibfield  {author} {\bibinfo {author} {\bibfnamefont {V.}~\bibnamefont
  {Gayathri}}, \bibinfo {author} {\bibfnamefont {I.}~\bibnamefont {Bartos}},
  \bibinfo {author} {\bibfnamefont {Z.}~\bibnamefont {Haiman}}, \bibinfo
  {author} {\bibfnamefont {S.}~\bibnamefont {Klimenko}}, \bibinfo {author}
  {\bibfnamefont {B.}~\bibnamefont {Kocsis}}, \bibinfo {author} {\bibfnamefont
  {S.}~\bibnamefont {Marka}},\ and\ \bibinfo {author} {\bibfnamefont
  {Y.}~\bibnamefont {Yang}},\ }\href {https://doi.org/10.3847/2041-8213/ab745d}
  {\bibfield  {journal} {\bibinfo  {journal} {Astrophys. J. Lett.}\ }\textbf
  {\bibinfo {volume} {890}},\ \bibinfo {pages} {L20} (\bibinfo {year}
  {2020})},\ \Eprint {https://arxiv.org/abs/1911.11142} {arXiv:1911.11142
  [astro-ph.HE]} \BibitemShut {NoStop}%
\bibitem [{\citenamefont {Abbott}\ \emph {et~al.}(2020)\citenamefont {Abbott}
  \emph {et~al.}}]{LIGOScientific:2020ufj}%
  \BibitemOpen
  \bibfield  {author} {\bibinfo {author} {\bibfnamefont {R.}~\bibnamefont
  {Abbott}} \emph {et~al.} (\bibinfo {collaboration} {LIGO Scientific,
  Virgo}),\ }\href {https://doi.org/10.3847/2041-8213/aba493} {\bibfield
  {journal} {\bibinfo  {journal} {Astrophys. J. Lett.}\ }\textbf {\bibinfo
  {volume} {900}},\ \bibinfo {pages} {L13} (\bibinfo {year} {2020})},\ \Eprint
  {https://arxiv.org/abs/2009.01190} {arXiv:2009.01190 [astro-ph.HE]}
  \BibitemShut {NoStop}%
\bibitem [{\citenamefont {Tagawa}\ \emph
  {et~al.}(2021{\natexlab{a}})\citenamefont {Tagawa}, \citenamefont {Kocsis},
  \citenamefont {Haiman}, \citenamefont {Bartos}, \citenamefont {Omukai},\ and\
  \citenamefont {Samsing}}]{Tagawa:2020qll}%
  \BibitemOpen
  \bibfield  {author} {\bibinfo {author} {\bibfnamefont {H.}~\bibnamefont
  {Tagawa}}, \bibinfo {author} {\bibfnamefont {B.}~\bibnamefont {Kocsis}},
  \bibinfo {author} {\bibfnamefont {Z.}~\bibnamefont {Haiman}}, \bibinfo
  {author} {\bibfnamefont {I.}~\bibnamefont {Bartos}}, \bibinfo {author}
  {\bibfnamefont {K.}~\bibnamefont {Omukai}},\ and\ \bibinfo {author}
  {\bibfnamefont {J.}~\bibnamefont {Samsing}},\ }\href
  {https://doi.org/10.3847/1538-4357/abd555} {\bibfield  {journal} {\bibinfo
  {journal} {Astrophys. J.}\ }\textbf {\bibinfo {volume} {908}},\ \bibinfo
  {pages} {194} (\bibinfo {year} {2021}{\natexlab{a}})},\ \Eprint
  {https://arxiv.org/abs/2012.00011} {arXiv:2012.00011 [astro-ph.HE]}
  \BibitemShut {NoStop}%
\bibitem [{\citenamefont {Tagawa}\ \emph
  {et~al.}(2021{\natexlab{b}})\citenamefont {Tagawa}, \citenamefont {Haiman},
  \citenamefont {Bartos}, \citenamefont {Kocsis},\ and\ \citenamefont
  {Omukai}}]{Tagawa:2021ofj}%
  \BibitemOpen
  \bibfield  {author} {\bibinfo {author} {\bibfnamefont {H.}~\bibnamefont
  {Tagawa}}, \bibinfo {author} {\bibfnamefont {Z.}~\bibnamefont {Haiman}},
  \bibinfo {author} {\bibfnamefont {I.}~\bibnamefont {Bartos}}, \bibinfo
  {author} {\bibfnamefont {B.}~\bibnamefont {Kocsis}},\ and\ \bibinfo {author}
  {\bibfnamefont {K.}~\bibnamefont {Omukai}},\ }\href
  {https://doi.org/10.1093/mnras/stab2315} {\bibfield  {journal} {\bibinfo
  {journal} {Mon. Not. Roy. Astron. Soc.}\ }\textbf {\bibinfo {volume} {507}},\
  \bibinfo {pages} {3362} (\bibinfo {year} {2021}{\natexlab{b}})},\ \Eprint
  {https://arxiv.org/abs/2104.09510} {arXiv:2104.09510 [astro-ph.HE]}
  \BibitemShut {NoStop}%
\bibitem [{\citenamefont {O'Brien}\ \emph {et~al.}(2021)\citenamefont
  {O'Brien}, \citenamefont {Szczepanczyk}, \citenamefont {Gayathri},
  \citenamefont {Bartos}, \citenamefont {Vedovato}, \citenamefont {Prodi},
  \citenamefont {Mitselmakher},\ and\ \citenamefont
  {Klimenko}}]{OBrien:2021sua}%
  \BibitemOpen
  \bibfield  {author} {\bibinfo {author} {\bibfnamefont {B.}~\bibnamefont
  {O'Brien}}, \bibinfo {author} {\bibfnamefont {M.}~\bibnamefont
  {Szczepanczyk}}, \bibinfo {author} {\bibfnamefont {V.}~\bibnamefont
  {Gayathri}}, \bibinfo {author} {\bibfnamefont {I.}~\bibnamefont {Bartos}},
  \bibinfo {author} {\bibfnamefont {G.}~\bibnamefont {Vedovato}}, \bibinfo
  {author} {\bibfnamefont {G.}~\bibnamefont {Prodi}}, \bibinfo {author}
  {\bibfnamefont {G.}~\bibnamefont {Mitselmakher}},\ and\ \bibinfo {author}
  {\bibfnamefont {S.}~\bibnamefont {Klimenko}},\ }\href
  {https://doi.org/10.1103/PhysRevD.104.082003} {\bibfield  {journal} {\bibinfo
   {journal} {Phys. Rev. D}\ }\textbf {\bibinfo {volume} {104}},\ \bibinfo
  {pages} {082003} (\bibinfo {year} {2021})},\ \Eprint
  {https://arxiv.org/abs/2106.00605} {arXiv:2106.00605 [gr-qc]} \BibitemShut
  {NoStop}%
\bibitem [{\citenamefont {Barrera}\ and\ \citenamefont
  {Bartos}(2022)}]{Barrera:2022yfj}%
  \BibitemOpen
  \bibfield  {author} {\bibinfo {author} {\bibfnamefont {O.}~\bibnamefont
  {Barrera}}\ and\ \bibinfo {author} {\bibfnamefont {I.}~\bibnamefont
  {Bartos}},\ }\href {https://doi.org/10.3847/2041-8213/ac5f47} {\bibfield
  {journal} {\bibinfo  {journal} {Astrophys. J. Lett.}\ }\textbf {\bibinfo
  {volume} {929}},\ \bibinfo {pages} {L1} (\bibinfo {year} {2022})},\ \Eprint
  {https://arxiv.org/abs/2201.09943} {arXiv:2201.09943 [astro-ph.HE]}
  \BibitemShut {NoStop}%
\bibitem [{\citenamefont {Gayathri}\ \emph {et~al.}(2023)\citenamefont
  {Gayathri}, \citenamefont {Wysocki}, \citenamefont {Yang}, \citenamefont
  {Delfavero}, \citenamefont {Shaughnessy}, \citenamefont {Haiman},
  \citenamefont {Tagawa},\ and\ \citenamefont {Bartos}}]{Gayathri:2023bha}%
  \BibitemOpen
  \bibfield  {author} {\bibinfo {author} {\bibfnamefont {V.}~\bibnamefont
  {Gayathri}}, \bibinfo {author} {\bibfnamefont {D.}~\bibnamefont {Wysocki}},
  \bibinfo {author} {\bibfnamefont {Y.}~\bibnamefont {Yang}}, \bibinfo {author}
  {\bibfnamefont {V.}~\bibnamefont {Delfavero}}, \bibinfo {author}
  {\bibfnamefont {R.~O.}\ \bibnamefont {Shaughnessy}}, \bibinfo {author}
  {\bibfnamefont {Z.}~\bibnamefont {Haiman}}, \bibinfo {author} {\bibfnamefont
  {H.}~\bibnamefont {Tagawa}},\ and\ \bibinfo {author} {\bibfnamefont
  {I.}~\bibnamefont {Bartos}},\ }\href
  {https://doi.org/10.3847/2041-8213/acbfb8} {\bibfield  {journal} {\bibinfo
  {journal} {Astrophys. J. Lett.}\ }\textbf {\bibinfo {volume} {945}},\
  \bibinfo {pages} {L29} (\bibinfo {year} {2023})},\ \Eprint
  {https://arxiv.org/abs/2301.04187} {arXiv:2301.04187 [gr-qc]} \BibitemShut
  {NoStop}%
\bibitem [{\citenamefont {Samsing}(2018)}]{Samsing:2017xmd}%
  \BibitemOpen
  \bibfield  {author} {\bibinfo {author} {\bibfnamefont {J.}~\bibnamefont
  {Samsing}},\ }\href {https://doi.org/10.1103/PhysRevD.97.103014} {\bibfield
  {journal} {\bibinfo  {journal} {Phys. Rev.}\ }\textbf {\bibinfo {volume}
  {D97}},\ \bibinfo {pages} {103014} (\bibinfo {year} {2018})},\ \Eprint
  {https://arxiv.org/abs/1711.07452} {arXiv:1711.07452 [astro-ph.HE]}
  \BibitemShut {NoStop}%
\bibitem [{\citenamefont {Zevin}\ \emph {et~al.}(2019)\citenamefont {Zevin},
  \citenamefont {Samsing}, \citenamefont {Rodriguez}, \citenamefont {Haster},\
  and\ \citenamefont {Ramirez-Ruiz}}]{Zevin:2018kzq}%
  \BibitemOpen
  \bibfield  {author} {\bibinfo {author} {\bibfnamefont {M.}~\bibnamefont
  {Zevin}}, \bibinfo {author} {\bibfnamefont {J.}~\bibnamefont {Samsing}},
  \bibinfo {author} {\bibfnamefont {C.}~\bibnamefont {Rodriguez}}, \bibinfo
  {author} {\bibfnamefont {C.-J.}\ \bibnamefont {Haster}},\ and\ \bibinfo
  {author} {\bibfnamefont {E.}~\bibnamefont {Ramirez-Ruiz}},\ }\href
  {https://doi.org/10.3847/1538-4357/aaf6ec} {\bibfield  {journal} {\bibinfo
  {journal} {Astrophys. J.}\ }\textbf {\bibinfo {volume} {871}},\ \bibinfo
  {pages} {91} (\bibinfo {year} {2019})},\ \Eprint
  {https://arxiv.org/abs/1810.00901} {arXiv:1810.00901 [astro-ph.HE]}
  \BibitemShut {NoStop}%
\bibitem [{\citenamefont {Tagawa}\ \emph
  {et~al.}(2021{\natexlab{c}})\citenamefont {Tagawa}, \citenamefont {Kocsis},
  \citenamefont {Haiman}, \citenamefont {Bartos}, \citenamefont {Omukai},\ and\
  \citenamefont {Samsing}}]{Tagawa:2020jnc}%
  \BibitemOpen
  \bibfield  {author} {\bibinfo {author} {\bibfnamefont {H.}~\bibnamefont
  {Tagawa}}, \bibinfo {author} {\bibfnamefont {B.}~\bibnamefont {Kocsis}},
  \bibinfo {author} {\bibfnamefont {Z.}~\bibnamefont {Haiman}}, \bibinfo
  {author} {\bibfnamefont {I.}~\bibnamefont {Bartos}}, \bibinfo {author}
  {\bibfnamefont {K.}~\bibnamefont {Omukai}},\ and\ \bibinfo {author}
  {\bibfnamefont {J.}~\bibnamefont {Samsing}},\ }\href
  {https://doi.org/10.3847/2041-8213/abd4d3} {\bibfield  {journal} {\bibinfo
  {journal} {Astrophys. J. Lett.}\ }\textbf {\bibinfo {volume} {907}},\
  \bibinfo {pages} {L20} (\bibinfo {year} {2021}{\natexlab{c}})},\ \Eprint
  {https://arxiv.org/abs/2010.10526} {arXiv:2010.10526 [astro-ph.HE]}
  \BibitemShut {NoStop}%
\bibitem [{\citenamefont {Samsing}\ \emph {et~al.}(2022)\citenamefont
  {Samsing}, \citenamefont {Bartos}, \citenamefont {D'Orazio}, \citenamefont
  {Haiman}, \citenamefont {Kocsis}, \citenamefont {Leigh}, \citenamefont {Liu},
  \citenamefont {Pessah},\ and\ \citenamefont {Tagawa}}]{Samsing:2020tda}%
  \BibitemOpen
  \bibfield  {author} {\bibinfo {author} {\bibfnamefont {J.}~\bibnamefont
  {Samsing}}, \bibinfo {author} {\bibfnamefont {I.}~\bibnamefont {Bartos}},
  \bibinfo {author} {\bibfnamefont {D.~J.}\ \bibnamefont {D'Orazio}}, \bibinfo
  {author} {\bibfnamefont {Z.}~\bibnamefont {Haiman}}, \bibinfo {author}
  {\bibfnamefont {B.}~\bibnamefont {Kocsis}}, \bibinfo {author} {\bibfnamefont
  {N.~W.~C.}\ \bibnamefont {Leigh}}, \bibinfo {author} {\bibfnamefont
  {B.}~\bibnamefont {Liu}}, \bibinfo {author} {\bibfnamefont {M.~E.}\
  \bibnamefont {Pessah}},\ and\ \bibinfo {author} {\bibfnamefont
  {H.}~\bibnamefont {Tagawa}},\ }\href
  {https://doi.org/10.1038/s41586-021-04333-1} {\bibfield  {journal} {\bibinfo
  {journal} {Nature}\ }\textbf {\bibinfo {volume} {603}},\ \bibinfo {pages}
  {237} (\bibinfo {year} {2022})},\ \Eprint {https://arxiv.org/abs/2010.09765}
  {arXiv:2010.09765 [astro-ph.HE]} \BibitemShut {NoStop}%
\bibitem [{\citenamefont {Zevin}\ \emph {et~al.}(2021)\citenamefont {Zevin},
  \citenamefont {Romero-Shaw}, \citenamefont {Kremer}, \citenamefont {Thrane},\
  and\ \citenamefont {Lasky}}]{Zevin:2021rtf}%
  \BibitemOpen
  \bibfield  {author} {\bibinfo {author} {\bibfnamefont {M.}~\bibnamefont
  {Zevin}}, \bibinfo {author} {\bibfnamefont {I.~M.}\ \bibnamefont
  {Romero-Shaw}}, \bibinfo {author} {\bibfnamefont {K.}~\bibnamefont {Kremer}},
  \bibinfo {author} {\bibfnamefont {E.}~\bibnamefont {Thrane}},\ and\ \bibinfo
  {author} {\bibfnamefont {P.~D.}\ \bibnamefont {Lasky}},\ }\href
  {https://doi.org/10.3847/2041-8213/ac32dc} {\bibfield  {journal} {\bibinfo
  {journal} {Astrophys. J. Lett.}\ }\textbf {\bibinfo {volume} {921}},\
  \bibinfo {pages} {L43} (\bibinfo {year} {2021})},\ \Eprint
  {https://arxiv.org/abs/2106.09042} {arXiv:2106.09042 [astro-ph.HE]}
  \BibitemShut {NoStop}%
\bibitem [{\citenamefont {{Fragione}}\ \emph {et~al.}(2021)\citenamefont
  {{Fragione}}, \citenamefont {{Kocsis}}, \citenamefont {{Rasio}},\ and\
  \citenamefont {{Silk}}}]{2021arXiv210704639F}%
  \BibitemOpen
  \bibfield  {author} {\bibinfo {author} {\bibfnamefont {G.}~\bibnamefont
  {{Fragione}}}, \bibinfo {author} {\bibfnamefont {B.}~\bibnamefont
  {{Kocsis}}}, \bibinfo {author} {\bibfnamefont {F.~A.}\ \bibnamefont
  {{Rasio}}},\ and\ \bibinfo {author} {\bibfnamefont {J.}~\bibnamefont
  {{Silk}}},\ }\href@noop {} {\bibfield  {journal} {\bibinfo  {journal} {arXiv
  e-prints}\ ,\ \bibinfo {eid} {arXiv:2107.04639}} (\bibinfo {year} {2021})},\
  \Eprint {https://arxiv.org/abs/2107.04639} {arXiv:2107.04639 [astro-ph.GA]}
  \BibitemShut {NoStop}%
\bibitem [{\citenamefont {Chattopadhyay}\ \emph {et~al.}(2023)\citenamefont
  {Chattopadhyay}, \citenamefont {Stegmann}, \citenamefont {Antonini},
  \citenamefont {Barber},\ and\ \citenamefont
  {Romero-Shaw}}]{Chattopadhyay:2023pil}%
  \BibitemOpen
  \bibfield  {author} {\bibinfo {author} {\bibfnamefont {D.}~\bibnamefont
  {Chattopadhyay}}, \bibinfo {author} {\bibfnamefont {J.}~\bibnamefont
  {Stegmann}}, \bibinfo {author} {\bibfnamefont {F.}~\bibnamefont {Antonini}},
  \bibinfo {author} {\bibfnamefont {J.}~\bibnamefont {Barber}},\ and\ \bibinfo
  {author} {\bibfnamefont {I.~M.}\ \bibnamefont {Romero-Shaw}},\ }\href@noop {}
  {\  (\bibinfo {year} {2023})},\ \Eprint {https://arxiv.org/abs/2308.10884}
  {arXiv:2308.10884 [astro-ph.HE]} \BibitemShut {NoStop}%
\bibitem [{\citenamefont {Gamba}\ \emph {et~al.}(2022)\citenamefont {Gamba},
  \citenamefont {Breschi}, \citenamefont {Carullo}, \citenamefont {Rettegno},
  \citenamefont {Albanesi}, \citenamefont {Bernuzzi},\ and\ \citenamefont
  {Nagar}}]{Gamba:2021gap}%
  \BibitemOpen
  \bibfield  {author} {\bibinfo {author} {\bibfnamefont {R.}~\bibnamefont
  {Gamba}}, \bibinfo {author} {\bibfnamefont {M.}~\bibnamefont {Breschi}},
  \bibinfo {author} {\bibfnamefont {G.}~\bibnamefont {Carullo}}, \bibinfo
  {author} {\bibfnamefont {P.}~\bibnamefont {Rettegno}}, \bibinfo {author}
  {\bibfnamefont {S.}~\bibnamefont {Albanesi}}, \bibinfo {author}
  {\bibfnamefont {S.}~\bibnamefont {Bernuzzi}},\ and\ \bibinfo {author}
  {\bibfnamefont {A.}~\bibnamefont {Nagar}},\ }\href
  {https://doi.org/10.1038/s41550-022-01813-w} {\bibfield  {journal} {\bibinfo
  {journal} {Nat. Astron.}\ } (\bibinfo {year} {2022})},\ \Eprint
  {https://arxiv.org/abs/2106.05575} {arXiv:2106.05575 [gr-qc]} \BibitemShut
  {NoStop}%
\bibitem [{\citenamefont {Calder\'on~Bustillo}\ \emph
  {et~al.}(2021)\citenamefont {Calder\'on~Bustillo}, \citenamefont
  {Sanchis-Gual}, \citenamefont {Torres-Forn\'e},\ and\ \citenamefont
  {Font}}]{CalderonBustillo:2020xms}%
  \BibitemOpen
  \bibfield  {author} {\bibinfo {author} {\bibfnamefont {J.}~\bibnamefont
  {Calder\'on~Bustillo}}, \bibinfo {author} {\bibfnamefont {N.}~\bibnamefont
  {Sanchis-Gual}}, \bibinfo {author} {\bibfnamefont {A.}~\bibnamefont
  {Torres-Forn\'e}},\ and\ \bibinfo {author} {\bibfnamefont {J.~A.}\
  \bibnamefont {Font}},\ }\href
  {https://doi.org/10.1103/PhysRevLett.126.201101} {\bibfield  {journal}
  {\bibinfo  {journal} {Phys. Rev. Lett.}\ }\textbf {\bibinfo {volume} {126}},\
  \bibinfo {pages} {201101} (\bibinfo {year} {2021})},\ \Eprint
  {https://arxiv.org/abs/2009.01066} {arXiv:2009.01066 [gr-qc]} \BibitemShut
  {NoStop}%
\bibitem [{\citenamefont {Gerosa}\ and\ \citenamefont
  {Fishbach}(2021)}]{Gerosa:2021mno}%
  \BibitemOpen
  \bibfield  {author} {\bibinfo {author} {\bibfnamefont {D.}~\bibnamefont
  {Gerosa}}\ and\ \bibinfo {author} {\bibfnamefont {M.}~\bibnamefont
  {Fishbach}},\ }\href {https://doi.org/10.1038/s41550-021-01398-w} {\bibfield
  {journal} {\bibinfo  {journal} {Nature Astron.}\ }\textbf {\bibinfo {volume}
  {5}},\ \bibinfo {pages} {749} (\bibinfo {year} {2021})},\ \Eprint
  {https://arxiv.org/abs/2105.03439} {arXiv:2105.03439 [astro-ph.HE]}
  \BibitemShut {NoStop}%
\bibitem [{\citenamefont {Arun}\ \emph {et~al.}(2009)\citenamefont {Arun},
  \citenamefont {Blanchet}, \citenamefont {Iyer},\ and\ \citenamefont
  {Sinha}}]{Arun:2009mc}%
  \BibitemOpen
  \bibfield  {author} {\bibinfo {author} {\bibfnamefont {K.~G.}\ \bibnamefont
  {Arun}}, \bibinfo {author} {\bibfnamefont {L.}~\bibnamefont {Blanchet}},
  \bibinfo {author} {\bibfnamefont {B.~R.}\ \bibnamefont {Iyer}},\ and\
  \bibinfo {author} {\bibfnamefont {S.}~\bibnamefont {Sinha}},\ }\href
  {https://doi.org/10.1103/PhysRevD.80.124018} {\bibfield  {journal} {\bibinfo
  {journal} {Phys. Rev. D}\ }\textbf {\bibinfo {volume} {80}},\ \bibinfo
  {pages} {124018} (\bibinfo {year} {2009})},\ \Eprint
  {https://arxiv.org/abs/0908.3854} {arXiv:0908.3854 [gr-qc]} \BibitemShut
  {NoStop}%
\bibitem [{\citenamefont {Huerta}\ \emph {et~al.}(2014)\citenamefont {Huerta},
  \citenamefont {Kumar}, \citenamefont {McWilliams}, \citenamefont
  {O'Shaughnessy},\ and\ \citenamefont {Yunes}}]{Huerta:2014eca}%
  \BibitemOpen
  \bibfield  {author} {\bibinfo {author} {\bibfnamefont {E.~A.}\ \bibnamefont
  {Huerta}}, \bibinfo {author} {\bibfnamefont {P.}~\bibnamefont {Kumar}},
  \bibinfo {author} {\bibfnamefont {S.~T.}\ \bibnamefont {McWilliams}},
  \bibinfo {author} {\bibfnamefont {R.}~\bibnamefont {O'Shaughnessy}},\ and\
  \bibinfo {author} {\bibfnamefont {N.}~\bibnamefont {Yunes}},\ }\href
  {https://doi.org/10.1103/PhysRevD.90.084016} {\bibfield  {journal} {\bibinfo
  {journal} {Phys. Rev. D}\ }\textbf {\bibinfo {volume} {90}},\ \bibinfo
  {pages} {084016} (\bibinfo {year} {2014})},\ \Eprint
  {https://arxiv.org/abs/1408.3406} {arXiv:1408.3406 [gr-qc]} \BibitemShut
  {NoStop}%
\bibitem [{\citenamefont {Moore}\ \emph {et~al.}(2016)\citenamefont {Moore},
  \citenamefont {Favata}, \citenamefont {Arun},\ and\ \citenamefont
  {Mishra}}]{Moore:2016qxz}%
  \BibitemOpen
  \bibfield  {author} {\bibinfo {author} {\bibfnamefont {B.}~\bibnamefont
  {Moore}}, \bibinfo {author} {\bibfnamefont {M.}~\bibnamefont {Favata}},
  \bibinfo {author} {\bibfnamefont {K.~G.}\ \bibnamefont {Arun}},\ and\
  \bibinfo {author} {\bibfnamefont {C.~K.}\ \bibnamefont {Mishra}},\ }\href
  {https://doi.org/10.1103/PhysRevD.93.124061} {\bibfield  {journal} {\bibinfo
  {journal} {Phys. Rev. D}\ }\textbf {\bibinfo {volume} {93}},\ \bibinfo
  {pages} {124061} (\bibinfo {year} {2016})},\ \Eprint
  {https://arxiv.org/abs/1605.00304} {arXiv:1605.00304 [gr-qc]} \BibitemShut
  {NoStop}%
\bibitem [{\citenamefont {Chiaramello}\ and\ \citenamefont
  {Nagar}(2020)}]{Chiaramello:2020ehz}%
  \BibitemOpen
  \bibfield  {author} {\bibinfo {author} {\bibfnamefont {D.}~\bibnamefont
  {Chiaramello}}\ and\ \bibinfo {author} {\bibfnamefont {A.}~\bibnamefont
  {Nagar}},\ }\href {https://doi.org/10.1103/PhysRevD.101.101501} {\bibfield
  {journal} {\bibinfo  {journal} {Phys. Rev. D}\ }\textbf {\bibinfo {volume}
  {101}},\ \bibinfo {pages} {101501} (\bibinfo {year} {2020})},\ \Eprint
  {https://arxiv.org/abs/2001.11736} {arXiv:2001.11736 [gr-qc]} \BibitemShut
  {NoStop}%
\bibitem [{\citenamefont {Nagar}\ \emph
  {et~al.}(2021{\natexlab{a}})\citenamefont {Nagar}, \citenamefont {Rettegno},
  \citenamefont {Gamba},\ and\ \citenamefont {Bernuzzi}}]{Nagar:2020xsk}%
  \BibitemOpen
  \bibfield  {author} {\bibinfo {author} {\bibfnamefont {A.}~\bibnamefont
  {Nagar}}, \bibinfo {author} {\bibfnamefont {P.}~\bibnamefont {Rettegno}},
  \bibinfo {author} {\bibfnamefont {R.}~\bibnamefont {Gamba}},\ and\ \bibinfo
  {author} {\bibfnamefont {S.}~\bibnamefont {Bernuzzi}},\ }\href
  {https://doi.org/10.1103/PhysRevD.103.064013} {\bibfield  {journal} {\bibinfo
   {journal} {Phys. Rev. D}\ }\textbf {\bibinfo {volume} {103}},\ \bibinfo
  {pages} {064013} (\bibinfo {year} {2021}{\natexlab{a}})},\ \Eprint
  {https://arxiv.org/abs/2009.12857} {arXiv:2009.12857 [gr-qc]} \BibitemShut
  {NoStop}%
\bibitem [{\citenamefont {Placidi}\ \emph {et~al.}(2022)\citenamefont
  {Placidi}, \citenamefont {Albanesi}, \citenamefont {Nagar}, \citenamefont
  {Orselli}, \citenamefont {Bernuzzi},\ and\ \citenamefont
  {Grignani}}]{Placidi:2021rkh}%
  \BibitemOpen
  \bibfield  {author} {\bibinfo {author} {\bibfnamefont {A.}~\bibnamefont
  {Placidi}}, \bibinfo {author} {\bibfnamefont {S.}~\bibnamefont {Albanesi}},
  \bibinfo {author} {\bibfnamefont {A.}~\bibnamefont {Nagar}}, \bibinfo
  {author} {\bibfnamefont {M.}~\bibnamefont {Orselli}}, \bibinfo {author}
  {\bibfnamefont {S.}~\bibnamefont {Bernuzzi}},\ and\ \bibinfo {author}
  {\bibfnamefont {G.}~\bibnamefont {Grignani}},\ }\href
  {https://doi.org/10.1103/PhysRevD.105.104030} {\bibfield  {journal} {\bibinfo
   {journal} {Phys. Rev. D}\ }\textbf {\bibinfo {volume} {105}},\ \bibinfo
  {pages} {104030} (\bibinfo {year} {2022})},\ \Eprint
  {https://arxiv.org/abs/2112.05448} {arXiv:2112.05448 [gr-qc]} \BibitemShut
  {NoStop}%
\bibitem [{\citenamefont {Paul}\ and\ \citenamefont
  {Mishra}(2023)}]{Paul:2022xfy}%
  \BibitemOpen
  \bibfield  {author} {\bibinfo {author} {\bibfnamefont {K.}~\bibnamefont
  {Paul}}\ and\ \bibinfo {author} {\bibfnamefont {C.~K.}\ \bibnamefont
  {Mishra}},\ }\href {https://doi.org/10.1103/PhysRevD.108.024023} {\bibfield
  {journal} {\bibinfo  {journal} {Phys. Rev. D}\ }\textbf {\bibinfo {volume}
  {108}},\ \bibinfo {pages} {024023} (\bibinfo {year} {2023})},\ \Eprint
  {https://arxiv.org/abs/2211.04155} {arXiv:2211.04155 [gr-qc]} \BibitemShut
  {NoStop}%
\bibitem [{\citenamefont {Albanesi}\ \emph {et~al.}(2022)\citenamefont
  {Albanesi}, \citenamefont {Placidi}, \citenamefont {Nagar}, \citenamefont
  {Orselli},\ and\ \citenamefont {Bernuzzi}}]{Albanesi:2022xge}%
  \BibitemOpen
  \bibfield  {author} {\bibinfo {author} {\bibfnamefont {S.}~\bibnamefont
  {Albanesi}}, \bibinfo {author} {\bibfnamefont {A.}~\bibnamefont {Placidi}},
  \bibinfo {author} {\bibfnamefont {A.}~\bibnamefont {Nagar}}, \bibinfo
  {author} {\bibfnamefont {M.}~\bibnamefont {Orselli}},\ and\ \bibinfo {author}
  {\bibfnamefont {S.}~\bibnamefont {Bernuzzi}},\ }\href
  {https://doi.org/10.1103/PhysRevD.105.L121503} {\bibfield  {journal}
  {\bibinfo  {journal} {Phys. Rev. D}\ }\textbf {\bibinfo {volume} {105}},\
  \bibinfo {pages} {L121503} (\bibinfo {year} {2022})},\ \Eprint
  {https://arxiv.org/abs/2203.16286} {arXiv:2203.16286 [gr-qc]} \BibitemShut
  {NoStop}%
\bibitem [{\citenamefont {Huerta}\ \emph {et~al.}(2018)\citenamefont {Huerta}
  \emph {et~al.}}]{Huerta:2017kez}%
  \BibitemOpen
  \bibfield  {author} {\bibinfo {author} {\bibfnamefont {E.~A.}\ \bibnamefont
  {Huerta}} \emph {et~al.},\ }\href
  {https://doi.org/10.1103/PhysRevD.97.024031} {\bibfield  {journal} {\bibinfo
  {journal} {Phys. Rev.}\ }\textbf {\bibinfo {volume} {D97}},\ \bibinfo {pages}
  {024031} (\bibinfo {year} {2018})},\ \Eprint
  {https://arxiv.org/abs/1711.06276} {arXiv:1711.06276 [gr-qc]} \BibitemShut
  {NoStop}%
\bibitem [{\citenamefont {Cao}\ and\ \citenamefont {Han}(2017)}]{Cao:2017ndf}%
  \BibitemOpen
  \bibfield  {author} {\bibinfo {author} {\bibfnamefont {Z.}~\bibnamefont
  {Cao}}\ and\ \bibinfo {author} {\bibfnamefont {W.-B.}\ \bibnamefont {Han}},\
  }\href {https://doi.org/10.1103/PhysRevD.96.044028} {\bibfield  {journal}
  {\bibinfo  {journal} {Phys. Rev.}\ }\textbf {\bibinfo {volume} {D96}},\
  \bibinfo {pages} {044028} (\bibinfo {year} {2017})},\ \Eprint
  {https://arxiv.org/abs/1708.00166} {arXiv:1708.00166 [gr-qc]} \BibitemShut
  {NoStop}%
\bibitem [{\citenamefont {Hinder}\ \emph {et~al.}(2017)\citenamefont {Hinder},
  \citenamefont {Kidder},\ and\ \citenamefont {Pfeiffer}}]{Hinder:2017sxy}%
  \BibitemOpen
  \bibfield  {author} {\bibinfo {author} {\bibfnamefont {I.}~\bibnamefont
  {Hinder}}, \bibinfo {author} {\bibfnamefont {L.~E.}\ \bibnamefont {Kidder}},\
  and\ \bibinfo {author} {\bibfnamefont {H.~P.}\ \bibnamefont {Pfeiffer}},\
  }\href@noop {} {\  (\bibinfo {year} {2017})},\ \Eprint
  {https://arxiv.org/abs/1709.02007} {arXiv:1709.02007 [gr-qc]} \BibitemShut
  {NoStop}%
\bibitem [{\citenamefont {Nagar}\ \emph
  {et~al.}(2021{\natexlab{b}})\citenamefont {Nagar}, \citenamefont {Bonino},\
  and\ \citenamefont {Rettegno}}]{Nagar:2021gss}%
  \BibitemOpen
  \bibfield  {author} {\bibinfo {author} {\bibfnamefont {A.}~\bibnamefont
  {Nagar}}, \bibinfo {author} {\bibfnamefont {A.}~\bibnamefont {Bonino}},\ and\
  \bibinfo {author} {\bibfnamefont {P.}~\bibnamefont {Rettegno}},\ }\href
  {https://doi.org/10.1103/PhysRevD.103.104021} {\bibfield  {journal} {\bibinfo
   {journal} {Phys. Rev. D}\ }\textbf {\bibinfo {volume} {103}},\ \bibinfo
  {pages} {104021} (\bibinfo {year} {2021}{\natexlab{b}})},\ \Eprint
  {https://arxiv.org/abs/2101.08624} {arXiv:2101.08624 [gr-qc]} \BibitemShut
  {NoStop}%
\bibitem [{\citenamefont {Ramos-Buades}\ \emph {et~al.}(2022)\citenamefont
  {Ramos-Buades}, \citenamefont {Buonanno}, \citenamefont {Khalil},\ and\
  \citenamefont {Ossokine}}]{Ramos-Buades:2021adz}%
  \BibitemOpen
  \bibfield  {author} {\bibinfo {author} {\bibfnamefont {A.}~\bibnamefont
  {Ramos-Buades}}, \bibinfo {author} {\bibfnamefont {A.}~\bibnamefont
  {Buonanno}}, \bibinfo {author} {\bibfnamefont {M.}~\bibnamefont {Khalil}},\
  and\ \bibinfo {author} {\bibfnamefont {S.}~\bibnamefont {Ossokine}},\ }\href
  {https://doi.org/10.1103/PhysRevD.105.044035} {\bibfield  {journal} {\bibinfo
   {journal} {Phys. Rev. D}\ }\textbf {\bibinfo {volume} {105}},\ \bibinfo
  {pages} {044035} (\bibinfo {year} {2022})},\ \Eprint
  {https://arxiv.org/abs/2112.06952} {arXiv:2112.06952 [gr-qc]} \BibitemShut
  {NoStop}%
\bibitem [{\citenamefont {Liu}\ \emph {et~al.}(2021)\citenamefont {Liu},
  \citenamefont {Cao},\ and\ \citenamefont {Zhu}}]{Liu:2021pkr}%
  \BibitemOpen
  \bibfield  {author} {\bibinfo {author} {\bibfnamefont {X.}~\bibnamefont
  {Liu}}, \bibinfo {author} {\bibfnamefont {Z.}~\bibnamefont {Cao}},\ and\
  \bibinfo {author} {\bibfnamefont {Z.-H.}\ \bibnamefont {Zhu}},\ }\href@noop
  {} {\  (\bibinfo {year} {2021})},\ \Eprint {https://arxiv.org/abs/2102.08614}
  {arXiv:2102.08614 [gr-qc]} \BibitemShut {NoStop}%
\bibitem [{\citenamefont {Chattaraj}\ \emph {et~al.}(2022)\citenamefont
  {Chattaraj}, \citenamefont {RoyChowdhury}, \citenamefont {Divyajyoti},
  \citenamefont {Mishra},\ and\ \citenamefont {Gupta}}]{Chattaraj:2022tay}%
  \BibitemOpen
  \bibfield  {author} {\bibinfo {author} {\bibfnamefont {A.}~\bibnamefont
  {Chattaraj}}, \bibinfo {author} {\bibfnamefont {T.}~\bibnamefont
  {RoyChowdhury}}, \bibinfo {author} {\bibnamefont {Divyajyoti}}, \bibinfo
  {author} {\bibfnamefont {C.~K.}\ \bibnamefont {Mishra}},\ and\ \bibinfo
  {author} {\bibfnamefont {A.}~\bibnamefont {Gupta}},\ }\href
  {https://doi.org/10.1103/PhysRevD.106.124008} {\bibfield  {journal} {\bibinfo
   {journal} {Phys. Rev. D}\ }\textbf {\bibinfo {volume} {106}},\ \bibinfo
  {pages} {124008} (\bibinfo {year} {2022})},\ \Eprint
  {https://arxiv.org/abs/2204.02377} {arXiv:2204.02377 [gr-qc]} \BibitemShut
  {NoStop}%
\bibitem [{\citenamefont {Liu}\ \emph {et~al.}(2023)\citenamefont {Liu},
  \citenamefont {Cao},\ and\ \citenamefont {Shao}}]{Liu:2023dgl}%
  \BibitemOpen
  \bibfield  {author} {\bibinfo {author} {\bibfnamefont {X.}~\bibnamefont
  {Liu}}, \bibinfo {author} {\bibfnamefont {Z.}~\bibnamefont {Cao}},\ and\
  \bibinfo {author} {\bibfnamefont {L.}~\bibnamefont {Shao}},\ }\href
  {https://doi.org/10.1142/S0218271823500153} {\bibfield  {journal} {\bibinfo
  {journal} {Int. J. Mod. Phys. D}\ }\textbf {\bibinfo {volume} {32}},\
  \bibinfo {pages} {2350015} (\bibinfo {year} {2023})},\ \Eprint
  {https://arxiv.org/abs/2306.15277} {arXiv:2306.15277 [gr-qc]} \BibitemShut
  {NoStop}%
\bibitem [{\citenamefont {Islam}\ \emph {et~al.}(2021)\citenamefont {Islam},
  \citenamefont {Varma}, \citenamefont {Lodman}, \citenamefont {Field},
  \citenamefont {Khanna}, \citenamefont {Scheel}, \citenamefont {Pfeiffer},
  \citenamefont {Gerosa},\ and\ \citenamefont {Kidder}}]{Islam:2021mha}%
  \BibitemOpen
  \bibfield  {author} {\bibinfo {author} {\bibfnamefont {T.}~\bibnamefont
  {Islam}}, \bibinfo {author} {\bibfnamefont {V.}~\bibnamefont {Varma}},
  \bibinfo {author} {\bibfnamefont {J.}~\bibnamefont {Lodman}}, \bibinfo
  {author} {\bibfnamefont {S.~E.}\ \bibnamefont {Field}}, \bibinfo {author}
  {\bibfnamefont {G.}~\bibnamefont {Khanna}}, \bibinfo {author} {\bibfnamefont
  {M.~A.}\ \bibnamefont {Scheel}}, \bibinfo {author} {\bibfnamefont {H.~P.}\
  \bibnamefont {Pfeiffer}}, \bibinfo {author} {\bibfnamefont {D.}~\bibnamefont
  {Gerosa}},\ and\ \bibinfo {author} {\bibfnamefont {L.~E.}\ \bibnamefont
  {Kidder}},\ }\href {https://doi.org/10.1103/PhysRevD.103.064022} {\bibfield
  {journal} {\bibinfo  {journal} {Phys. Rev. D}\ }\textbf {\bibinfo {volume}
  {103}},\ \bibinfo {pages} {064022} (\bibinfo {year} {2021})},\ \Eprint
  {https://arxiv.org/abs/2101.11798} {arXiv:2101.11798 [gr-qc]} \BibitemShut
  {NoStop}%
\bibitem [{\citenamefont {Damour}\ \emph {et~al.}(2014)\citenamefont {Damour},
  \citenamefont {Guercilena}, \citenamefont {Hinder}, \citenamefont {Hopper},
  \citenamefont {Nagar},\ and\ \citenamefont {Rezzolla}}]{Damour:2014afa}%
  \BibitemOpen
  \bibfield  {author} {\bibinfo {author} {\bibfnamefont {T.}~\bibnamefont
  {Damour}}, \bibinfo {author} {\bibfnamefont {F.}~\bibnamefont {Guercilena}},
  \bibinfo {author} {\bibfnamefont {I.}~\bibnamefont {Hinder}}, \bibinfo
  {author} {\bibfnamefont {S.}~\bibnamefont {Hopper}}, \bibinfo {author}
  {\bibfnamefont {A.}~\bibnamefont {Nagar}},\ and\ \bibinfo {author}
  {\bibfnamefont {L.}~\bibnamefont {Rezzolla}},\ }\href
  {https://doi.org/10.1103/PhysRevD.89.081503} {\bibfield  {journal} {\bibinfo
  {journal} {Phys. Rev. D}\ }\textbf {\bibinfo {volume} {89}},\ \bibinfo
  {pages} {081503} (\bibinfo {year} {2014})},\ \Eprint
  {https://arxiv.org/abs/1402.7307} {arXiv:1402.7307 [gr-qc]} \BibitemShut
  {NoStop}%
\bibitem [{\citenamefont {Hopper}\ \emph {et~al.}(2023)\citenamefont {Hopper},
  \citenamefont {Nagar},\ and\ \citenamefont {Rettegno}}]{Hopper:2022rwo}%
  \BibitemOpen
  \bibfield  {author} {\bibinfo {author} {\bibfnamefont {S.}~\bibnamefont
  {Hopper}}, \bibinfo {author} {\bibfnamefont {A.}~\bibnamefont {Nagar}},\ and\
  \bibinfo {author} {\bibfnamefont {P.}~\bibnamefont {Rettegno}},\ }\href
  {https://doi.org/10.1103/PhysRevD.107.124034} {\bibfield  {journal} {\bibinfo
   {journal} {Phys. Rev. D}\ }\textbf {\bibinfo {volume} {107}},\ \bibinfo
  {pages} {124034} (\bibinfo {year} {2023})},\ \Eprint
  {https://arxiv.org/abs/2204.10299} {arXiv:2204.10299 [gr-qc]} \BibitemShut
  {NoStop}%
\bibitem [{\citenamefont {Damour}(2016)}]{Damour:2016gwp}%
  \BibitemOpen
  \bibfield  {author} {\bibinfo {author} {\bibfnamefont {T.}~\bibnamefont
  {Damour}},\ }\href {https://doi.org/10.1103/PhysRevD.94.104015} {\bibfield
  {journal} {\bibinfo  {journal} {Phys. Rev.}\ }\textbf {\bibinfo {volume}
  {D94}},\ \bibinfo {pages} {104015} (\bibinfo {year} {2016})},\ \Eprint
  {https://arxiv.org/abs/1609.00354} {arXiv:1609.00354 [gr-qc]} \BibitemShut
  {NoStop}%
\bibitem [{\citenamefont {Damour}(2018)}]{Damour:2017zjx}%
  \BibitemOpen
  \bibfield  {author} {\bibinfo {author} {\bibfnamefont {T.}~\bibnamefont
  {Damour}},\ }\href {https://doi.org/10.1103/PhysRevD.97.044038} {\bibfield
  {journal} {\bibinfo  {journal} {Phys. Rev.}\ }\textbf {\bibinfo {volume}
  {D97}},\ \bibinfo {pages} {044038} (\bibinfo {year} {2018})},\ \Eprint
  {https://arxiv.org/abs/1710.10599} {arXiv:1710.10599 [gr-qc]} \BibitemShut
  {NoStop}%
\bibitem [{\citenamefont {Damour}(2020)}]{Damour:2019lcq}%
  \BibitemOpen
  \bibfield  {author} {\bibinfo {author} {\bibfnamefont {T.}~\bibnamefont
  {Damour}},\ }\href {https://doi.org/10.1103/PhysRevD.102.024060} {\bibfield
  {journal} {\bibinfo  {journal} {Phys. Rev. D}\ }\textbf {\bibinfo {volume}
  {102}},\ \bibinfo {pages} {024060} (\bibinfo {year} {2020})},\ \Eprint
  {https://arxiv.org/abs/1912.02139} {arXiv:1912.02139 [gr-qc]} \BibitemShut
  {NoStop}%
\bibitem [{\citenamefont {Bern}\ \emph {et~al.}(2019)\citenamefont {Bern},
  \citenamefont {Cheung}, \citenamefont {Roiban}, \citenamefont {Shen},
  \citenamefont {Solon},\ and\ \citenamefont {Zeng}}]{Bern:2019nnu}%
  \BibitemOpen
  \bibfield  {author} {\bibinfo {author} {\bibfnamefont {Z.}~\bibnamefont
  {Bern}}, \bibinfo {author} {\bibfnamefont {C.}~\bibnamefont {Cheung}},
  \bibinfo {author} {\bibfnamefont {R.}~\bibnamefont {Roiban}}, \bibinfo
  {author} {\bibfnamefont {C.-H.}\ \bibnamefont {Shen}}, \bibinfo {author}
  {\bibfnamefont {M.~P.}\ \bibnamefont {Solon}},\ and\ \bibinfo {author}
  {\bibfnamefont {M.}~\bibnamefont {Zeng}},\ }\href
  {https://doi.org/10.1103/PhysRevLett.122.201603} {\bibfield  {journal}
  {\bibinfo  {journal} {Phys. Rev. Lett.}\ }\textbf {\bibinfo {volume} {122}},\
  \bibinfo {pages} {201603} (\bibinfo {year} {2019})},\ \Eprint
  {https://arxiv.org/abs/1901.04424} {arXiv:1901.04424 [hep-th]} \BibitemShut
  {NoStop}%
\bibitem [{\citenamefont {Bern}\ \emph {et~al.}(2021)\citenamefont {Bern},
  \citenamefont {Parra-Martinez}, \citenamefont {Roiban}, \citenamefont {Ruf},
  \citenamefont {Shen}, \citenamefont {Solon},\ and\ \citenamefont
  {Zeng}}]{Bern:2021dqo}%
  \BibitemOpen
  \bibfield  {author} {\bibinfo {author} {\bibfnamefont {Z.}~\bibnamefont
  {Bern}}, \bibinfo {author} {\bibfnamefont {J.}~\bibnamefont
  {Parra-Martinez}}, \bibinfo {author} {\bibfnamefont {R.}~\bibnamefont
  {Roiban}}, \bibinfo {author} {\bibfnamefont {M.~S.}\ \bibnamefont {Ruf}},
  \bibinfo {author} {\bibfnamefont {C.-H.}\ \bibnamefont {Shen}}, \bibinfo
  {author} {\bibfnamefont {M.~P.}\ \bibnamefont {Solon}},\ and\ \bibinfo
  {author} {\bibfnamefont {M.}~\bibnamefont {Zeng}},\ }\href
  {https://doi.org/10.1103/PhysRevLett.126.171601} {\bibfield  {journal}
  {\bibinfo  {journal} {Phys. Rev. Lett.}\ }\textbf {\bibinfo {volume} {126}},\
  \bibinfo {pages} {171601} (\bibinfo {year} {2021})},\ \Eprint
  {https://arxiv.org/abs/2101.07254} {arXiv:2101.07254 [hep-th]} \BibitemShut
  {NoStop}%
\bibitem [{\citenamefont {Dlapa}\ \emph {et~al.}(2023)\citenamefont {Dlapa},
  \citenamefont {K\"alin}, \citenamefont {Liu}, \citenamefont {Neef},\ and\
  \citenamefont {Porto}}]{Dlapa:2022lmu}%
  \BibitemOpen
  \bibfield  {author} {\bibinfo {author} {\bibfnamefont {C.}~\bibnamefont
  {Dlapa}}, \bibinfo {author} {\bibfnamefont {G.}~\bibnamefont {K\"alin}},
  \bibinfo {author} {\bibfnamefont {Z.}~\bibnamefont {Liu}}, \bibinfo {author}
  {\bibfnamefont {J.}~\bibnamefont {Neef}},\ and\ \bibinfo {author}
  {\bibfnamefont {R.~A.}\ \bibnamefont {Porto}},\ }\href
  {https://doi.org/10.1103/PhysRevLett.130.101401} {\bibfield  {journal}
  {\bibinfo  {journal} {Phys. Rev. Lett.}\ }\textbf {\bibinfo {volume} {130}},\
  \bibinfo {pages} {101401} (\bibinfo {year} {2023})},\ \Eprint
  {https://arxiv.org/abs/2210.05541} {arXiv:2210.05541 [hep-th]} \BibitemShut
  {NoStop}%
\bibitem [{\citenamefont {Dlapa}\ \emph
  {et~al.}(2022{\natexlab{a}})\citenamefont {Dlapa}, \citenamefont {K\"alin},
  \citenamefont {Liu},\ and\ \citenamefont {Porto}}]{Dlapa:2021npj}%
  \BibitemOpen
  \bibfield  {author} {\bibinfo {author} {\bibfnamefont {C.}~\bibnamefont
  {Dlapa}}, \bibinfo {author} {\bibfnamefont {G.}~\bibnamefont {K\"alin}},
  \bibinfo {author} {\bibfnamefont {Z.}~\bibnamefont {Liu}},\ and\ \bibinfo
  {author} {\bibfnamefont {R.~A.}\ \bibnamefont {Porto}},\ }\href
  {https://doi.org/10.1016/j.physletb.2022.137203} {\bibfield  {journal}
  {\bibinfo  {journal} {Phys. Lett. B}\ }\textbf {\bibinfo {volume} {831}},\
  \bibinfo {pages} {137203} (\bibinfo {year} {2022}{\natexlab{a}})},\ \Eprint
  {https://arxiv.org/abs/2106.08276} {arXiv:2106.08276 [hep-th]} \BibitemShut
  {NoStop}%
\bibitem [{\citenamefont {Dlapa}\ \emph
  {et~al.}(2022{\natexlab{b}})\citenamefont {Dlapa}, \citenamefont {K\"alin},
  \citenamefont {Liu},\ and\ \citenamefont {Porto}}]{Dlapa:2021vgp}%
  \BibitemOpen
  \bibfield  {author} {\bibinfo {author} {\bibfnamefont {C.}~\bibnamefont
  {Dlapa}}, \bibinfo {author} {\bibfnamefont {G.}~\bibnamefont {K\"alin}},
  \bibinfo {author} {\bibfnamefont {Z.}~\bibnamefont {Liu}},\ and\ \bibinfo
  {author} {\bibfnamefont {R.~A.}\ \bibnamefont {Porto}},\ }\href
  {https://doi.org/10.1103/PhysRevLett.128.161104} {\bibfield  {journal}
  {\bibinfo  {journal} {Phys. Rev. Lett.}\ }\textbf {\bibinfo {volume} {128}},\
  \bibinfo {pages} {161104} (\bibinfo {year} {2022}{\natexlab{b}})},\ \Eprint
  {https://arxiv.org/abs/2112.11296} {arXiv:2112.11296 [hep-th]} \BibitemShut
  {NoStop}%
\bibitem [{\citenamefont {Bern}\ \emph {et~al.}(2022)\citenamefont {Bern},
  \citenamefont {Parra-Martinez}, \citenamefont {Roiban}, \citenamefont {Ruf},
  \citenamefont {Shen}, \citenamefont {Solon},\ and\ \citenamefont
  {Zeng}}]{Bern:2021yeh}%
  \BibitemOpen
  \bibfield  {author} {\bibinfo {author} {\bibfnamefont {Z.}~\bibnamefont
  {Bern}}, \bibinfo {author} {\bibfnamefont {J.}~\bibnamefont
  {Parra-Martinez}}, \bibinfo {author} {\bibfnamefont {R.}~\bibnamefont
  {Roiban}}, \bibinfo {author} {\bibfnamefont {M.~S.}\ \bibnamefont {Ruf}},
  \bibinfo {author} {\bibfnamefont {C.-H.}\ \bibnamefont {Shen}}, \bibinfo
  {author} {\bibfnamefont {M.~P.}\ \bibnamefont {Solon}},\ and\ \bibinfo
  {author} {\bibfnamefont {M.}~\bibnamefont {Zeng}},\ }\href
  {https://doi.org/10.1103/PhysRevLett.128.161103} {\bibfield  {journal}
  {\bibinfo  {journal} {Phys. Rev. Lett.}\ }\textbf {\bibinfo {volume} {128}},\
  \bibinfo {pages} {161103} (\bibinfo {year} {2022})},\ \Eprint
  {https://arxiv.org/abs/2112.10750} {arXiv:2112.10750 [hep-th]} \BibitemShut
  {NoStop}%
\bibitem [{\citenamefont {Damour}\ and\ \citenamefont
  {Rettegno}(2023)}]{Damour:2022ybd}%
  \BibitemOpen
  \bibfield  {author} {\bibinfo {author} {\bibfnamefont {T.}~\bibnamefont
  {Damour}}\ and\ \bibinfo {author} {\bibfnamefont {P.}~\bibnamefont
  {Rettegno}},\ }\href {https://doi.org/10.1103/PhysRevD.107.064051} {\bibfield
   {journal} {\bibinfo  {journal} {Phys. Rev. D}\ }\textbf {\bibinfo {volume}
  {107}},\ \bibinfo {pages} {064051} (\bibinfo {year} {2023})},\ \Eprint
  {https://arxiv.org/abs/2211.01399} {arXiv:2211.01399 [gr-qc]} \BibitemShut
  {NoStop}%
\bibitem [{\citenamefont {Khalil}\ \emph {et~al.}(2022)\citenamefont {Khalil},
  \citenamefont {Buonanno}, \citenamefont {Steinhoff},\ and\ \citenamefont
  {Vines}}]{Khalil:2022ylj}%
  \BibitemOpen
  \bibfield  {author} {\bibinfo {author} {\bibfnamefont {M.}~\bibnamefont
  {Khalil}}, \bibinfo {author} {\bibfnamefont {A.}~\bibnamefont {Buonanno}},
  \bibinfo {author} {\bibfnamefont {J.}~\bibnamefont {Steinhoff}},\ and\
  \bibinfo {author} {\bibfnamefont {J.}~\bibnamefont {Vines}},\ }\href
  {https://doi.org/10.1103/PhysRevD.106.024042} {\bibfield  {journal} {\bibinfo
   {journal} {Phys. Rev. D}\ }\textbf {\bibinfo {volume} {106}},\ \bibinfo
  {pages} {024042} (\bibinfo {year} {2022})},\ \Eprint
  {https://arxiv.org/abs/2204.05047} {arXiv:2204.05047 [gr-qc]} \BibitemShut
  {NoStop}%
\bibitem [{\citenamefont {Rettegno}\ \emph {et~al.}(2023)\citenamefont
  {Rettegno}, \citenamefont {Pratten}, \citenamefont {Thomas}, \citenamefont
  {Schmidt},\ and\ \citenamefont {Damour}}]{Rettegno:2023ghr}%
  \BibitemOpen
  \bibfield  {author} {\bibinfo {author} {\bibfnamefont {P.}~\bibnamefont
  {Rettegno}}, \bibinfo {author} {\bibfnamefont {G.}~\bibnamefont {Pratten}},
  \bibinfo {author} {\bibfnamefont {L.}~\bibnamefont {Thomas}}, \bibinfo
  {author} {\bibfnamefont {P.}~\bibnamefont {Schmidt}},\ and\ \bibinfo {author}
  {\bibfnamefont {T.}~\bibnamefont {Damour}},\ }\href@noop {} {\  (\bibinfo
  {year} {2023})},\ \Eprint {https://arxiv.org/abs/2307.06999}
  {arXiv:2307.06999 [gr-qc]} \BibitemShut {NoStop}%
\bibitem [{\citenamefont {Abbott}\ \emph
  {et~al.}(2021{\natexlab{b}})\citenamefont {Abbott} \emph
  {et~al.}}]{LIGOScientific:2021djp}%
  \BibitemOpen
  \bibfield  {author} {\bibinfo {author} {\bibfnamefont {R.}~\bibnamefont
  {Abbott}} \emph {et~al.} (\bibinfo {collaboration} {LIGO Scientific, VIRGO,
  KAGRA}),\ }\href@noop {} {\  (\bibinfo {year} {2021}{\natexlab{b}})},\
  \Eprint {https://arxiv.org/abs/2111.03606} {arXiv:2111.03606 [gr-qc]}
  \BibitemShut {NoStop}%
\bibitem [{\citenamefont {East}\ \emph {et~al.}(2013)\citenamefont {East},
  \citenamefont {McWilliams}, \citenamefont {Levin},\ and\ \citenamefont
  {Pretorius}}]{East:2012xq}%
  \BibitemOpen
  \bibfield  {author} {\bibinfo {author} {\bibfnamefont {W.~E.}\ \bibnamefont
  {East}}, \bibinfo {author} {\bibfnamefont {S.~T.}\ \bibnamefont
  {McWilliams}}, \bibinfo {author} {\bibfnamefont {J.}~\bibnamefont {Levin}},\
  and\ \bibinfo {author} {\bibfnamefont {F.}~\bibnamefont {Pretorius}},\ }\href
  {https://doi.org/10.1103/PhysRevD.87.043004} {\bibfield  {journal} {\bibinfo
  {journal} {Phys. Rev.}\ }\textbf {\bibinfo {volume} {D87}},\ \bibinfo {pages}
  {043004} (\bibinfo {year} {2013})},\ \Eprint
  {https://arxiv.org/abs/1212.0837} {arXiv:1212.0837 [gr-qc]} \BibitemShut
  {NoStop}%
\bibitem [{\citenamefont {Mora}\ and\ \citenamefont
  {Will}(2002)}]{Mora:2002gf}%
  \BibitemOpen
  \bibfield  {author} {\bibinfo {author} {\bibfnamefont {T.}~\bibnamefont
  {Mora}}\ and\ \bibinfo {author} {\bibfnamefont {C.~M.}\ \bibnamefont
  {Will}},\ }\href {https://doi.org/10.1103/PhysRevD.66.101501} {\bibfield
  {journal} {\bibinfo  {journal} {Phys. Rev. D}\ }\textbf {\bibinfo {volume}
  {66}},\ \bibinfo {pages} {101501} (\bibinfo {year} {2002})},\ \Eprint
  {https://arxiv.org/abs/gr-qc/0208089} {arXiv:gr-qc/0208089} \BibitemShut
  {NoStop}%
\bibitem [{\citenamefont {Bonino}\ \emph {et~al.}(2023)\citenamefont {Bonino},
  \citenamefont {Gamba}, \citenamefont {Schmidt}, \citenamefont {Nagar},
  \citenamefont {Pratten}, \citenamefont {Breschi}, \citenamefont {Rettegno},\
  and\ \citenamefont {Bernuzzi}}]{Bonino:2022hkj}%
  \BibitemOpen
  \bibfield  {author} {\bibinfo {author} {\bibfnamefont {A.}~\bibnamefont
  {Bonino}}, \bibinfo {author} {\bibfnamefont {R.}~\bibnamefont {Gamba}},
  \bibinfo {author} {\bibfnamefont {P.}~\bibnamefont {Schmidt}}, \bibinfo
  {author} {\bibfnamefont {A.}~\bibnamefont {Nagar}}, \bibinfo {author}
  {\bibfnamefont {G.}~\bibnamefont {Pratten}}, \bibinfo {author} {\bibfnamefont
  {M.}~\bibnamefont {Breschi}}, \bibinfo {author} {\bibfnamefont
  {P.}~\bibnamefont {Rettegno}},\ and\ \bibinfo {author} {\bibfnamefont
  {S.}~\bibnamefont {Bernuzzi}},\ }\href
  {https://doi.org/10.1103/PhysRevD.107.064024} {\bibfield  {journal} {\bibinfo
   {journal} {Phys. Rev. D}\ }\textbf {\bibinfo {volume} {107}},\ \bibinfo
  {pages} {064024} (\bibinfo {year} {2023})},\ \Eprint
  {https://arxiv.org/abs/2207.10474} {arXiv:2207.10474 [gr-qc]} \BibitemShut
  {NoStop}%
\bibitem [{\citenamefont {Shaikh}\ \emph {et~al.}(2023)\citenamefont {Shaikh},
  \citenamefont {Varma}, \citenamefont {Pfeiffer}, \citenamefont
  {Ramos-Buades},\ and\ \citenamefont {van~de Meent}}]{Shaikh:2023ypz}%
  \BibitemOpen
  \bibfield  {author} {\bibinfo {author} {\bibfnamefont {M.~A.}\ \bibnamefont
  {Shaikh}}, \bibinfo {author} {\bibfnamefont {V.}~\bibnamefont {Varma}},
  \bibinfo {author} {\bibfnamefont {H.~P.}\ \bibnamefont {Pfeiffer}}, \bibinfo
  {author} {\bibfnamefont {A.}~\bibnamefont {Ramos-Buades}},\ and\ \bibinfo
  {author} {\bibfnamefont {M.}~\bibnamefont {van~de Meent}},\ }\href@noop {} {\
   (\bibinfo {year} {2023})},\ \Eprint {https://arxiv.org/abs/2302.11257}
  {arXiv:2302.11257 [gr-qc]} \BibitemShut {NoStop}%
\bibitem [{\citenamefont {K\"alin}\ and\ \citenamefont
  {Porto}(2020{\natexlab{a}})}]{Kalin:2019inp}%
  \BibitemOpen
  \bibfield  {author} {\bibinfo {author} {\bibfnamefont {G.}~\bibnamefont
  {K\"alin}}\ and\ \bibinfo {author} {\bibfnamefont {R.~A.}\ \bibnamefont
  {Porto}},\ }\href {https://doi.org/10.1007/JHEP02(2020)120} {\bibfield
  {journal} {\bibinfo  {journal} {JHEP}\ }\textbf {\bibinfo {volume} {02}},\
  \bibinfo {pages} {120}},\ \Eprint {https://arxiv.org/abs/1911.09130}
  {arXiv:1911.09130 [hep-th]} \BibitemShut {NoStop}%
\bibitem [{\citenamefont {Santamaria}\ \emph {et~al.}(2010)\citenamefont
  {Santamaria}, \citenamefont {Ohme}, \citenamefont {Ajith}, \citenamefont
  {Br{\"u}gmann}, \citenamefont {Dorband} \emph {et~al.}}]{Santamaria:2010yb}%
  \BibitemOpen
  \bibfield  {author} {\bibinfo {author} {\bibfnamefont {L.}~\bibnamefont
  {Santamaria}}, \bibinfo {author} {\bibfnamefont {F.}~\bibnamefont {Ohme}},
  \bibinfo {author} {\bibfnamefont {P.}~\bibnamefont {Ajith}}, \bibinfo
  {author} {\bibfnamefont {B.}~\bibnamefont {Br{\"u}gmann}}, \bibinfo {author}
  {\bibfnamefont {N.}~\bibnamefont {Dorband}}, \emph {et~al.},\ }\href
  {https://doi.org/10.1103/PhysRevD.82.064016} {\bibfield  {journal} {\bibinfo
  {journal} {Phys.Rev.}\ }\textbf {\bibinfo {volume} {D82}},\ \bibinfo {pages}
  {064016} (\bibinfo {year} {2010})},\ \Eprint
  {https://arxiv.org/abs/1005.3306} {arXiv:1005.3306 [gr-qc]} \BibitemShut
  {NoStop}%
\bibitem [{\citenamefont {Lousto}\ and\ \citenamefont
  {Price}(1998)}]{Lousto:1997ge}%
  \BibitemOpen
  \bibfield  {author} {\bibinfo {author} {\bibfnamefont {C.~O.}\ \bibnamefont
  {Lousto}}\ and\ \bibinfo {author} {\bibfnamefont {R.~H.}\ \bibnamefont
  {Price}},\ }\href {https://doi.org/10.1103/PhysRevD.57.1073} {\bibfield
  {journal} {\bibinfo  {journal} {Phys. Rev. D}\ }\textbf {\bibinfo {volume}
  {57}},\ \bibinfo {pages} {1073} (\bibinfo {year} {1998})},\ \Eprint
  {https://arxiv.org/abs/gr-qc/9708022} {arXiv:gr-qc/9708022} \BibitemShut
  {NoStop}%
\bibitem [{\citenamefont {Martel}\ and\ \citenamefont
  {Poisson}(2002)}]{Martel:2001yf}%
  \BibitemOpen
  \bibfield  {author} {\bibinfo {author} {\bibfnamefont {K.}~\bibnamefont
  {Martel}}\ and\ \bibinfo {author} {\bibfnamefont {E.}~\bibnamefont
  {Poisson}},\ }\href {https://doi.org/10.1103/PhysRevD.66.084001} {\bibfield
  {journal} {\bibinfo  {journal} {Phys. Rev.}\ }\textbf {\bibinfo {volume}
  {D66}},\ \bibinfo {pages} {084001} (\bibinfo {year} {2002})},\ \Eprint
  {https://arxiv.org/abs/gr-qc/0107104} {arXiv:gr-qc/0107104} \BibitemShut
  {NoStop}%
\bibitem [{\citenamefont {Reisswig}\ \emph {et~al.}(2010)\citenamefont
  {Reisswig}, \citenamefont {Bishop}, \citenamefont {Pollney},\ and\
  \citenamefont {Szilagyi}}]{Reisswig:2009rx}%
  \BibitemOpen
  \bibfield  {author} {\bibinfo {author} {\bibfnamefont {C.}~\bibnamefont
  {Reisswig}}, \bibinfo {author} {\bibfnamefont {N.~T.}\ \bibnamefont
  {Bishop}}, \bibinfo {author} {\bibfnamefont {D.}~\bibnamefont {Pollney}},\
  and\ \bibinfo {author} {\bibfnamefont {B.}~\bibnamefont {Szilagyi}},\ }\href
  {https://doi.org/10.1088/0264-9381/27/7/075014} {\bibfield  {journal}
  {\bibinfo  {journal} {Class. Quant. Grav.}\ }\textbf {\bibinfo {volume}
  {27}},\ \bibinfo {pages} {075014} (\bibinfo {year} {2010})},\ \Eprint
  {https://arxiv.org/abs/0912.1285} {arXiv:0912.1285 [gr-qc]} \BibitemShut
  {NoStop}%
\bibitem [{\citenamefont {Damour}\ and\ \citenamefont
  {Nagar}(2007)}]{Damour:2007xr}%
  \BibitemOpen
  \bibfield  {author} {\bibinfo {author} {\bibfnamefont {T.}~\bibnamefont
  {Damour}}\ and\ \bibinfo {author} {\bibfnamefont {A.}~\bibnamefont {Nagar}},\
  }\href {https://doi.org/10.1103/PhysRevD.76.064028} {\bibfield  {journal}
  {\bibinfo  {journal} {Phys. Rev.}\ }\textbf {\bibinfo {volume} {D76}},\
  \bibinfo {pages} {064028} (\bibinfo {year} {2007})},\ \Eprint
  {https://arxiv.org/abs/0705.2519} {arXiv:0705.2519 [gr-qc]} \BibitemShut
  {NoStop}%
\bibitem [{\citenamefont {Forteza}\ \emph {et~al.}(2023)\citenamefont
  {Forteza}, \citenamefont {Bhagwat}, \citenamefont {Kumar},\ and\
  \citenamefont {Pani}}]{Forteza:2022tgq}%
  \BibitemOpen
  \bibfield  {author} {\bibinfo {author} {\bibfnamefont {X.~J.}\ \bibnamefont
  {Forteza}}, \bibinfo {author} {\bibfnamefont {S.}~\bibnamefont {Bhagwat}},
  \bibinfo {author} {\bibfnamefont {S.}~\bibnamefont {Kumar}},\ and\ \bibinfo
  {author} {\bibfnamefont {P.}~\bibnamefont {Pani}},\ }\href
  {https://doi.org/10.1103/PhysRevLett.130.021001} {\bibfield  {journal}
  {\bibinfo  {journal} {Phys. Rev. Lett.}\ }\textbf {\bibinfo {volume} {130}},\
  \bibinfo {pages} {021001} (\bibinfo {year} {2023})},\ \Eprint
  {https://arxiv.org/abs/2205.14910} {arXiv:2205.14910 [gr-qc]} \BibitemShut
  {NoStop}%
\bibitem [{\citenamefont {Albanesi}\ \emph {et~al.}(2023)\citenamefont
  {Albanesi}, \citenamefont {Bernuzzi}, \citenamefont {Damour}, \citenamefont
  {Nagar},\ and\ \citenamefont {Placidi}}]{Albanesi:2023bgi}%
  \BibitemOpen
  \bibfield  {author} {\bibinfo {author} {\bibfnamefont {S.}~\bibnamefont
  {Albanesi}}, \bibinfo {author} {\bibfnamefont {S.}~\bibnamefont {Bernuzzi}},
  \bibinfo {author} {\bibfnamefont {T.}~\bibnamefont {Damour}}, \bibinfo
  {author} {\bibfnamefont {A.}~\bibnamefont {Nagar}},\ and\ \bibinfo {author}
  {\bibfnamefont {A.}~\bibnamefont {Placidi}},\ }\href@noop {} {\  (\bibinfo
  {year} {2023})},\ \Eprint {https://arxiv.org/abs/2305.19336}
  {arXiv:2305.19336 [gr-qc]} \BibitemShut {NoStop}%
\bibitem [{\citenamefont {Andrade}\ \emph {et~al.}(2023)\citenamefont {Andrade}
  \emph {et~al.}}]{Andrade:2023trh}%
  \BibitemOpen
  \bibfield  {author} {\bibinfo {author} {\bibfnamefont {T.}~\bibnamefont
  {Andrade}} \emph {et~al.},\ }\href@noop {} {\  (\bibinfo {year} {2023})},\
  \Eprint {https://arxiv.org/abs/2307.08697} {arXiv:2307.08697 [gr-qc]}
  \BibitemShut {NoStop}%
\bibitem [{\citenamefont {Buonanno}\ and\ \citenamefont
  {Damour}(1999)}]{Buonanno:1998gg}%
  \BibitemOpen
  \bibfield  {author} {\bibinfo {author} {\bibfnamefont {A.}~\bibnamefont
  {Buonanno}}\ and\ \bibinfo {author} {\bibfnamefont {T.}~\bibnamefont
  {Damour}},\ }\href {https://doi.org/10.1103/PhysRevD.59.084006} {\bibfield
  {journal} {\bibinfo  {journal} {Phys. Rev.}\ }\textbf {\bibinfo {volume}
  {D59}},\ \bibinfo {pages} {084006} (\bibinfo {year} {1999})},\ \Eprint
  {https://arxiv.org/abs/gr-qc/9811091} {arXiv:gr-qc/9811091} \BibitemShut
  {NoStop}%
\bibitem [{\citenamefont {K\"alin}\ and\ \citenamefont
  {Porto}(2020{\natexlab{b}})}]{Kalin:2019rwq}%
  \BibitemOpen
  \bibfield  {author} {\bibinfo {author} {\bibfnamefont {G.}~\bibnamefont
  {K\"alin}}\ and\ \bibinfo {author} {\bibfnamefont {R.~A.}\ \bibnamefont
  {Porto}},\ }\href {https://doi.org/10.1007/JHEP01(2020)072} {\bibfield
  {journal} {\bibinfo  {journal} {JHEP}\ }\textbf {\bibinfo {volume} {01}},\
  \bibinfo {pages} {072}},\ \Eprint {https://arxiv.org/abs/1910.03008}
  {arXiv:1910.03008 [hep-th]} \BibitemShut {NoStop}%
\bibitem [{\citenamefont {Healy}\ \emph {et~al.}(2014)\citenamefont {Healy},
  \citenamefont {Lousto},\ and\ \citenamefont {Zlochower}}]{Healy:2014yta}%
  \BibitemOpen
  \bibfield  {author} {\bibinfo {author} {\bibfnamefont {J.}~\bibnamefont
  {Healy}}, \bibinfo {author} {\bibfnamefont {C.~O.}\ \bibnamefont {Lousto}},\
  and\ \bibinfo {author} {\bibfnamefont {Y.}~\bibnamefont {Zlochower}},\ }\href
  {https://doi.org/10.1103/PhysRevD.90.104004} {\bibfield  {journal} {\bibinfo
  {journal} {Phys. Rev.}\ }\textbf {\bibinfo {volume} {D90}},\ \bibinfo {pages}
  {104004} (\bibinfo {year} {2014})},\ \Eprint
  {https://arxiv.org/abs/1406.7295} {arXiv:1406.7295 [gr-qc]} \BibitemShut
  {NoStop}%
\bibitem [{\citenamefont {Healy}\ and\ \citenamefont
  {Lousto}(2017)}]{Healy:2016lce}%
  \BibitemOpen
  \bibfield  {author} {\bibinfo {author} {\bibfnamefont {J.}~\bibnamefont
  {Healy}}\ and\ \bibinfo {author} {\bibfnamefont {C.~O.}\ \bibnamefont
  {Lousto}},\ }\href {https://doi.org/10.1103/PhysRevD.95.024037} {\bibfield
  {journal} {\bibinfo  {journal} {Phys. Rev.}\ }\textbf {\bibinfo {volume}
  {D95}},\ \bibinfo {pages} {024037} (\bibinfo {year} {2017})},\ \Eprint
  {https://arxiv.org/abs/1610.09713} {arXiv:1610.09713 [gr-qc]} \BibitemShut
  {NoStop}%
\bibitem [{\citenamefont {Hofmann}\ \emph {et~al.}(2016)\citenamefont
  {Hofmann}, \citenamefont {Barausse},\ and\ \citenamefont
  {Rezzolla}}]{Hofmann:2016yih}%
  \BibitemOpen
  \bibfield  {author} {\bibinfo {author} {\bibfnamefont {F.}~\bibnamefont
  {Hofmann}}, \bibinfo {author} {\bibfnamefont {E.}~\bibnamefont {Barausse}},\
  and\ \bibinfo {author} {\bibfnamefont {L.}~\bibnamefont {Rezzolla}},\ }\href
  {https://doi.org/10.3847/2041-8205/825/2/L19} {\bibfield  {journal} {\bibinfo
   {journal} {Astrophys. J.}\ }\textbf {\bibinfo {volume} {825}},\ \bibinfo
  {pages} {L19} (\bibinfo {year} {2016})},\ \Eprint
  {https://arxiv.org/abs/1605.01938} {arXiv:1605.01938 [gr-qc]} \BibitemShut
  {NoStop}%
\bibitem [{\citenamefont {Jim\'enez-Forteza}\ \emph {et~al.}(2017)\citenamefont
  {Jim\'enez-Forteza}, \citenamefont {Keitel}, \citenamefont {Husa},
  \citenamefont {Hannam}, \citenamefont {Khan},\ and\ \citenamefont
  {P{\"u}rrer}}]{Jimenez-Forteza:2016oae}%
  \BibitemOpen
  \bibfield  {author} {\bibinfo {author} {\bibfnamefont {X.}~\bibnamefont
  {Jim\'enez-Forteza}}, \bibinfo {author} {\bibfnamefont {D.}~\bibnamefont
  {Keitel}}, \bibinfo {author} {\bibfnamefont {S.}~\bibnamefont {Husa}},
  \bibinfo {author} {\bibfnamefont {M.}~\bibnamefont {Hannam}}, \bibinfo
  {author} {\bibfnamefont {S.}~\bibnamefont {Khan}},\ and\ \bibinfo {author}
  {\bibfnamefont {M.}~\bibnamefont {P{\"u}rrer}},\ }\href
  {https://doi.org/10.1103/PhysRevD.95.064024} {\bibfield  {journal} {\bibinfo
  {journal} {Phys. Rev.}\ }\textbf {\bibinfo {volume} {D95}},\ \bibinfo {pages}
  {064024} (\bibinfo {year} {2017})},\ \Eprint
  {https://arxiv.org/abs/1611.00332} {arXiv:1611.00332 [gr-qc]} \BibitemShut
  {NoStop}%
\bibitem [{\citenamefont {{Johnson-McDaniel}}\ \emph
  {et~al.}(2016)\citenamefont {{Johnson-McDaniel}}, \citenamefont {{Gupta}},
  \citenamefont {{Keitel}}, \citenamefont {{Ajith}}, \citenamefont
  {{Birnholtz}}, \citenamefont {{Ohme}},\ and\ \citenamefont
  {{Husa}}}]{NathanDCC}%
  \BibitemOpen
  \bibfield  {author} {\bibinfo {author} {\bibfnamefont {N.~K.}\ \bibnamefont
  {{Johnson-McDaniel}}}, \bibinfo {author} {\bibfnamefont {A.}~\bibnamefont
  {{Gupta}}}, \bibinfo {author} {\bibfnamefont {D.}~\bibnamefont {{Keitel}}},
  \bibinfo {author} {\bibfnamefont {P.}~\bibnamefont {{Ajith}}}, \bibinfo
  {author} {\bibfnamefont {O.}~\bibnamefont {{Birnholtz}}}, \bibinfo {author}
  {\bibfnamefont {F.}~\bibnamefont {{Ohme}}},\ and\ \bibinfo {author}
  {\bibfnamefont {S.}~\bibnamefont {{Husa}}},\ }\href
  {https://dcc.ligo.org/LIGO-T1600168-v5/public} {\bibfield  {journal}
  {\bibinfo  {journal} {LIGO DCC}\ ,\ \bibinfo {eid} {LIGO-T1600168-v6}}
  (\bibinfo {year} {2016})}\BibitemShut {NoStop}%
\bibitem [{\citenamefont {Varma}\ \emph {et~al.}(2019)\citenamefont {Varma},
  \citenamefont {Field}, \citenamefont {Scheel}, \citenamefont {Blackman},
  \citenamefont {Gerosa}, \citenamefont {Stein}, \citenamefont {Kidder},\ and\
  \citenamefont {Pfeiffer}}]{Varma:2019csw}%
  \BibitemOpen
  \bibfield  {author} {\bibinfo {author} {\bibfnamefont {V.}~\bibnamefont
  {Varma}}, \bibinfo {author} {\bibfnamefont {S.~E.}\ \bibnamefont {Field}},
  \bibinfo {author} {\bibfnamefont {M.~A.}\ \bibnamefont {Scheel}}, \bibinfo
  {author} {\bibfnamefont {J.}~\bibnamefont {Blackman}}, \bibinfo {author}
  {\bibfnamefont {D.}~\bibnamefont {Gerosa}}, \bibinfo {author} {\bibfnamefont
  {L.~C.}\ \bibnamefont {Stein}}, \bibinfo {author} {\bibfnamefont {L.~E.}\
  \bibnamefont {Kidder}},\ and\ \bibinfo {author} {\bibfnamefont {H.~P.}\
  \bibnamefont {Pfeiffer}},\ }\href
  {https://doi.org/10.1103/PhysRevResearch.1.033015} {\bibfield  {journal}
  {\bibinfo  {journal} {Phys. Rev. Research.}\ }\textbf {\bibinfo {volume}
  {1}},\ \bibinfo {pages} {033015} (\bibinfo {year} {2019})},\ \Eprint
  {https://arxiv.org/abs/1905.09300} {arXiv:1905.09300 [gr-qc]} \BibitemShut
  {NoStop}%
\bibitem [{\citenamefont {Boschini}\ \emph {et~al.}(2023)\citenamefont
  {Boschini} \emph {et~al.}}]{Boschini:2023ryi}%
  \BibitemOpen
  \bibfield  {author} {\bibinfo {author} {\bibfnamefont {M.}~\bibnamefont
  {Boschini}} \emph {et~al.},\ }\href
  {https://doi.org/10.1103/PhysRevD.108.084015} {\bibfield  {journal} {\bibinfo
   {journal} {Phys. Rev. D}\ }\textbf {\bibinfo {volume} {108}},\ \bibinfo
  {pages} {084015} (\bibinfo {year} {2023})},\ \Eprint
  {https://arxiv.org/abs/2307.03435} {arXiv:2307.03435 [gr-qc]} \BibitemShut
  {NoStop}%
\bibitem [{\citenamefont {Ferguson}\ \emph {et~al.}(2019)\citenamefont
  {Ferguson}, \citenamefont {Ghonge}, \citenamefont {Clark}, \citenamefont
  {Calderon~Bustillo}, \citenamefont {Laguna}, \citenamefont {Shoemaker},\ and\
  \citenamefont {Calderon~Bustillo}}]{Ferguson:2019slp}%
  \BibitemOpen
  \bibfield  {author} {\bibinfo {author} {\bibfnamefont {D.}~\bibnamefont
  {Ferguson}}, \bibinfo {author} {\bibfnamefont {S.}~\bibnamefont {Ghonge}},
  \bibinfo {author} {\bibfnamefont {J.~A.}\ \bibnamefont {Clark}}, \bibinfo
  {author} {\bibfnamefont {J.}~\bibnamefont {Calderon~Bustillo}}, \bibinfo
  {author} {\bibfnamefont {P.}~\bibnamefont {Laguna}}, \bibinfo {author}
  {\bibfnamefont {D.}~\bibnamefont {Shoemaker}},\ and\ \bibinfo {author}
  {\bibfnamefont {J.}~\bibnamefont {Calderon~Bustillo}},\ }\href
  {https://doi.org/10.1103/PhysRevLett.123.151101} {\bibfield  {journal}
  {\bibinfo  {journal} {Phys. Rev. Lett.}\ }\textbf {\bibinfo {volume} {123}},\
  \bibinfo {pages} {151101} (\bibinfo {year} {2019})},\ \Eprint
  {https://arxiv.org/abs/1905.03756} {arXiv:1905.03756 [gr-qc]} \BibitemShut
  {NoStop}%
\bibitem [{\citenamefont {Nagar}\ \emph {et~al.}(2018)\citenamefont {Nagar}
  \emph {et~al.}}]{Nagar:2018zoe}%
  \BibitemOpen
  \bibfield  {author} {\bibinfo {author} {\bibfnamefont {A.}~\bibnamefont
  {Nagar}} \emph {et~al.},\ }\href {https://doi.org/10.1103/PhysRevD.98.104052}
  {\bibfield  {journal} {\bibinfo  {journal} {Phys. Rev.}\ }\textbf {\bibinfo
  {volume} {D98}},\ \bibinfo {pages} {104052} (\bibinfo {year} {2018})},\
  \Eprint {https://arxiv.org/abs/1806.01772} {arXiv:1806.01772 [gr-qc]}
  \BibitemShut {NoStop}%
\bibitem [{\citenamefont {Nagar}\ \emph {et~al.}(2020)\citenamefont {Nagar},
  \citenamefont {Pratten}, \citenamefont {Riemenschneider},\ and\ \citenamefont
  {Gamba}}]{Nagar:2019wds}%
  \BibitemOpen
  \bibfield  {author} {\bibinfo {author} {\bibfnamefont {A.}~\bibnamefont
  {Nagar}}, \bibinfo {author} {\bibfnamefont {G.}~\bibnamefont {Pratten}},
  \bibinfo {author} {\bibfnamefont {G.}~\bibnamefont {Riemenschneider}},\ and\
  \bibinfo {author} {\bibfnamefont {R.}~\bibnamefont {Gamba}},\ }\href
  {https://doi.org/10.1103/PhysRevD.101.024041} {\bibfield  {journal} {\bibinfo
   {journal} {Phys. Rev. D}\ }\textbf {\bibinfo {volume} {101}},\ \bibinfo
  {pages} {024041} (\bibinfo {year} {2020})},\ \Eprint
  {https://arxiv.org/abs/1904.09550} {arXiv:1904.09550 [gr-qc]} \BibitemShut
  {NoStop}%
\bibitem [{\citenamefont {Carullo}\ \emph {et~al.}(2019)\citenamefont
  {Carullo}, \citenamefont {Riemenschneider}, \citenamefont {Tsang},
  \citenamefont {Nagar},\ and\ \citenamefont {Del~Pozzo}}]{Carullo:2018gah}%
  \BibitemOpen
  \bibfield  {author} {\bibinfo {author} {\bibfnamefont {G.}~\bibnamefont
  {Carullo}}, \bibinfo {author} {\bibfnamefont {G.}~\bibnamefont
  {Riemenschneider}}, \bibinfo {author} {\bibfnamefont {K.~W.}\ \bibnamefont
  {Tsang}}, \bibinfo {author} {\bibfnamefont {A.}~\bibnamefont {Nagar}},\ and\
  \bibinfo {author} {\bibfnamefont {W.}~\bibnamefont {Del~Pozzo}},\ }\href
  {https://doi.org/10.1088/1361-6382/ab185e} {\bibfield  {journal} {\bibinfo
  {journal} {Class. Quant. Grav.}\ }\textbf {\bibinfo {volume} {36}},\ \bibinfo
  {pages} {105009} (\bibinfo {year} {2019})},\ \Eprint
  {https://arxiv.org/abs/1811.08744} {arXiv:1811.08744 [gr-qc]} \BibitemShut
  {NoStop}%
\bibitem [{\citenamefont {Bernuzzi}\ \emph {et~al.}(2014)\citenamefont
  {Bernuzzi}, \citenamefont {Nagar}, \citenamefont {Balmelli}, \citenamefont
  {Dietrich},\ and\ \citenamefont {Ujevic}}]{Bernuzzi:2014kca}%
  \BibitemOpen
  \bibfield  {author} {\bibinfo {author} {\bibfnamefont {S.}~\bibnamefont
  {Bernuzzi}}, \bibinfo {author} {\bibfnamefont {A.}~\bibnamefont {Nagar}},
  \bibinfo {author} {\bibfnamefont {S.}~\bibnamefont {Balmelli}}, \bibinfo
  {author} {\bibfnamefont {T.}~\bibnamefont {Dietrich}},\ and\ \bibinfo
  {author} {\bibfnamefont {M.}~\bibnamefont {Ujevic}},\ }\href
  {https://doi.org/10.1103/PhysRevLett.112.201101} {\bibfield  {journal}
  {\bibinfo  {journal} {Phys.Rev.Lett.}\ }\textbf {\bibinfo {volume} {112}},\
  \bibinfo {pages} {201101} (\bibinfo {year} {2014})},\ \Eprint
  {https://arxiv.org/abs/1402.6244} {arXiv:1402.6244 [gr-qc]} \BibitemShut
  {NoStop}%
\bibitem [{\citenamefont {Healy}\ and\ \citenamefont
  {Lousto}(2022)}]{Healy:2022wdn}%
  \BibitemOpen
  \bibfield  {author} {\bibinfo {author} {\bibfnamefont {J.}~\bibnamefont
  {Healy}}\ and\ \bibinfo {author} {\bibfnamefont {C.~O.}\ \bibnamefont
  {Lousto}},\ }\href {https://doi.org/10.1103/PhysRevD.105.124010} {\bibfield
  {journal} {\bibinfo  {journal} {Phys. Rev. D}\ }\textbf {\bibinfo {volume}
  {105}},\ \bibinfo {pages} {124010} (\bibinfo {year} {2022})},\ \Eprint
  {https://arxiv.org/abs/2202.00018} {arXiv:2202.00018 [gr-qc]} \BibitemShut
  {NoStop}%
\bibitem [{\citenamefont {Chu}\ \emph {et~al.}(2009)\citenamefont {Chu},
  \citenamefont {Pfeiffer},\ and\ \citenamefont {Scheel}}]{Chu:2009md}%
  \BibitemOpen
  \bibfield  {author} {\bibinfo {author} {\bibfnamefont {T.}~\bibnamefont
  {Chu}}, \bibinfo {author} {\bibfnamefont {H.~P.}\ \bibnamefont {Pfeiffer}},\
  and\ \bibinfo {author} {\bibfnamefont {M.~A.}\ \bibnamefont {Scheel}},\
  }\href {https://doi.org/10.1103/PhysRevD.80.124051} {\bibfield  {journal}
  {\bibinfo  {journal} {Phys. Rev.}\ }\textbf {\bibinfo {volume} {D80}},\
  \bibinfo {pages} {124051} (\bibinfo {year} {2009})},\ \Eprint
  {https://arxiv.org/abs/0909.1313} {arXiv:0909.1313 [gr-qc]} \BibitemShut
  {NoStop}%
\bibitem [{\citenamefont {Lovelace}\ \emph {et~al.}(2011)\citenamefont
  {Lovelace}, \citenamefont {Scheel},\ and\ \citenamefont
  {Szilagyi}}]{Lovelace:2010ne}%
  \BibitemOpen
  \bibfield  {author} {\bibinfo {author} {\bibfnamefont {G.}~\bibnamefont
  {Lovelace}}, \bibinfo {author} {\bibfnamefont {M.}~\bibnamefont {Scheel}},\
  and\ \bibinfo {author} {\bibfnamefont {B.}~\bibnamefont {Szilagyi}},\ }\href
  {https://doi.org/10.1103/PhysRevD.83.024010} {\bibfield  {journal} {\bibinfo
  {journal} {Phys.Rev.}\ }\textbf {\bibinfo {volume} {D83}},\ \bibinfo {pages}
  {024010} (\bibinfo {year} {2011})},\ \Eprint
  {https://arxiv.org/abs/1010.2777} {arXiv:1010.2777 [gr-qc]} \BibitemShut
  {NoStop}%
\bibitem [{\citenamefont {Lovelace}\ \emph {et~al.}(2012)\citenamefont
  {Lovelace}, \citenamefont {Boyle}, \citenamefont {Scheel},\ and\
  \citenamefont {Szilagyi}}]{Lovelace:2011nu}%
  \BibitemOpen
  \bibfield  {author} {\bibinfo {author} {\bibfnamefont {G.}~\bibnamefont
  {Lovelace}}, \bibinfo {author} {\bibfnamefont {M.}~\bibnamefont {Boyle}},
  \bibinfo {author} {\bibfnamefont {M.~A.}\ \bibnamefont {Scheel}},\ and\
  \bibinfo {author} {\bibfnamefont {B.}~\bibnamefont {Szilagyi}},\ }\href
  {https://doi.org/10.1088/0264-9381/29/4/045003} {\bibfield  {journal}
  {\bibinfo  {journal} {Class. Quant. Grav.}\ }\textbf {\bibinfo {volume}
  {29}},\ \bibinfo {pages} {045003} (\bibinfo {year} {2012})},\ \Eprint
  {https://arxiv.org/abs/1110.2229} {arXiv:1110.2229 [gr-qc]} \BibitemShut
  {NoStop}%
\bibitem [{\citenamefont {Buchman}\ \emph {et~al.}(2012)\citenamefont
  {Buchman}, \citenamefont {Pfeiffer}, \citenamefont {Scheel},\ and\
  \citenamefont {Szilagyi}}]{Buchman:2012dw}%
  \BibitemOpen
  \bibfield  {author} {\bibinfo {author} {\bibfnamefont {L.~T.}\ \bibnamefont
  {Buchman}}, \bibinfo {author} {\bibfnamefont {H.~P.}\ \bibnamefont
  {Pfeiffer}}, \bibinfo {author} {\bibfnamefont {M.~A.}\ \bibnamefont
  {Scheel}},\ and\ \bibinfo {author} {\bibfnamefont {B.}~\bibnamefont
  {Szilagyi}},\ }\href {https://doi.org/10.1103/PhysRevD.86.084033} {\bibfield
  {journal} {\bibinfo  {journal} {Phys. Rev.}\ }\textbf {\bibinfo {volume}
  {D86}},\ \bibinfo {pages} {084033} (\bibinfo {year} {2012})},\ \Eprint
  {https://arxiv.org/abs/1206.3015} {arXiv:1206.3015 [gr-qc]} \BibitemShut
  {NoStop}%
\bibitem [{\citenamefont {Hemberger}\ \emph {et~al.}(2013)\citenamefont
  {Hemberger}, \citenamefont {Lovelace}, \citenamefont {Loredo}, \citenamefont
  {Kidder}, \citenamefont {Scheel}, \citenamefont {Szilágyi}, \citenamefont
  {Taylor},\ and\ \citenamefont {Teukolsky}}]{Hemberger:2013hsa}%
  \BibitemOpen
  \bibfield  {author} {\bibinfo {author} {\bibfnamefont {D.~A.}\ \bibnamefont
  {Hemberger}}, \bibinfo {author} {\bibfnamefont {G.}~\bibnamefont {Lovelace}},
  \bibinfo {author} {\bibfnamefont {T.~J.}\ \bibnamefont {Loredo}}, \bibinfo
  {author} {\bibfnamefont {L.~E.}\ \bibnamefont {Kidder}}, \bibinfo {author}
  {\bibfnamefont {M.~A.}\ \bibnamefont {Scheel}}, \bibinfo {author}
  {\bibfnamefont {B.}~\bibnamefont {Szilágyi}}, \bibinfo {author}
  {\bibfnamefont {N.~W.}\ \bibnamefont {Taylor}},\ and\ \bibinfo {author}
  {\bibfnamefont {S.~A.}\ \bibnamefont {Teukolsky}},\ }\href
  {https://doi.org/10.1103/PhysRevD.88.064014} {\bibfield  {journal} {\bibinfo
  {journal} {Phys. Rev.}\ }\textbf {\bibinfo {volume} {D88}},\ \bibinfo {pages}
  {064014} (\bibinfo {year} {2013})},\ \Eprint
  {https://arxiv.org/abs/1305.5991} {arXiv:1305.5991 [gr-qc]} \BibitemShut
  {NoStop}%
\bibitem [{\citenamefont {Scheel}\ \emph {et~al.}(2015)\citenamefont {Scheel},
  \citenamefont {Giesler}, \citenamefont {Hemberger}, \citenamefont {Lovelace},
  \citenamefont {Kuper}, \citenamefont {Boyle}, \citenamefont {Szil\'agyi},\
  and\ \citenamefont {Kidder}}]{Scheel:2014ina}%
  \BibitemOpen
  \bibfield  {author} {\bibinfo {author} {\bibfnamefont {M.~A.}\ \bibnamefont
  {Scheel}}, \bibinfo {author} {\bibfnamefont {M.}~\bibnamefont {Giesler}},
  \bibinfo {author} {\bibfnamefont {D.~A.}\ \bibnamefont {Hemberger}}, \bibinfo
  {author} {\bibfnamefont {G.}~\bibnamefont {Lovelace}}, \bibinfo {author}
  {\bibfnamefont {K.}~\bibnamefont {Kuper}}, \bibinfo {author} {\bibfnamefont
  {M.}~\bibnamefont {Boyle}}, \bibinfo {author} {\bibfnamefont
  {B.}~\bibnamefont {Szil\'agyi}},\ and\ \bibinfo {author} {\bibfnamefont
  {L.~E.}\ \bibnamefont {Kidder}},\ }\href
  {https://doi.org/10.1088/0264-9381/32/10/105009} {\bibfield  {journal}
  {\bibinfo  {journal} {Class. Quant. Grav.}\ }\textbf {\bibinfo {volume}
  {32}},\ \bibinfo {pages} {105009} (\bibinfo {year} {2015})},\ \Eprint
  {https://arxiv.org/abs/1412.1803} {arXiv:1412.1803 [gr-qc]} \BibitemShut
  {NoStop}%
\bibitem [{\citenamefont {Blackman}\ \emph {et~al.}(2015)\citenamefont
  {Blackman}, \citenamefont {Field}, \citenamefont {Galley}, \citenamefont
  {Szilágyi}, \citenamefont {Scheel}, \citenamefont {Tiglio},\ and\
  \citenamefont {Hemberger}}]{Blackman:2015pia}%
  \BibitemOpen
  \bibfield  {author} {\bibinfo {author} {\bibfnamefont {J.}~\bibnamefont
  {Blackman}}, \bibinfo {author} {\bibfnamefont {S.~E.}\ \bibnamefont {Field}},
  \bibinfo {author} {\bibfnamefont {C.~R.}\ \bibnamefont {Galley}}, \bibinfo
  {author} {\bibfnamefont {B.}~\bibnamefont {Szilágyi}}, \bibinfo {author}
  {\bibfnamefont {M.~A.}\ \bibnamefont {Scheel}}, \bibinfo {author}
  {\bibfnamefont {M.}~\bibnamefont {Tiglio}},\ and\ \bibinfo {author}
  {\bibfnamefont {D.~A.}\ \bibnamefont {Hemberger}},\ }\href
  {https://doi.org/10.1103/PhysRevLett.115.121102} {\bibfield  {journal}
  {\bibinfo  {journal} {Phys. Rev. Lett.}\ }\textbf {\bibinfo {volume} {115}},\
  \bibinfo {pages} {121102} (\bibinfo {year} {2015})},\ \Eprint
  {https://arxiv.org/abs/1502.07758} {arXiv:1502.07758 [gr-qc]} \BibitemShut
  {NoStop}%
\bibitem [{\citenamefont {Lovelace}\ \emph {et~al.}(2015)\citenamefont
  {Lovelace} \emph {et~al.}}]{Lovelace:2014twa}%
  \BibitemOpen
  \bibfield  {author} {\bibinfo {author} {\bibfnamefont {G.}~\bibnamefont
  {Lovelace}} \emph {et~al.},\ }\href
  {https://doi.org/10.1088/0264-9381/32/6/065007} {\bibfield  {journal}
  {\bibinfo  {journal} {Class. Quant. Grav.}\ }\textbf {\bibinfo {volume}
  {32}},\ \bibinfo {pages} {065007} (\bibinfo {year} {2015})},\ \Eprint
  {https://arxiv.org/abs/1411.7297} {arXiv:1411.7297 [gr-qc]} \BibitemShut
  {NoStop}%
\bibitem [{\citenamefont {Mroue}\ \emph {et~al.}(2013)\citenamefont {Mroue},
  \citenamefont {Scheel}, \citenamefont {Szilagyi}, \citenamefont {Pfeiffer},
  \citenamefont {Boyle} \emph {et~al.}}]{Mroue:2013xna}%
  \BibitemOpen
  \bibfield  {author} {\bibinfo {author} {\bibfnamefont {A.~H.}\ \bibnamefont
  {Mroue}}, \bibinfo {author} {\bibfnamefont {M.~A.}\ \bibnamefont {Scheel}},
  \bibinfo {author} {\bibfnamefont {B.}~\bibnamefont {Szilagyi}}, \bibinfo
  {author} {\bibfnamefont {H.~P.}\ \bibnamefont {Pfeiffer}}, \bibinfo {author}
  {\bibfnamefont {M.}~\bibnamefont {Boyle}}, \emph {et~al.},\ }\href
  {https://doi.org/10.1103/PhysRevLett.111.241104} {\bibfield  {journal}
  {\bibinfo  {journal} {Phys.Rev.Lett.}\ }\textbf {\bibinfo {volume} {111}},\
  \bibinfo {pages} {241104} (\bibinfo {year} {2013})},\ \Eprint
  {https://arxiv.org/abs/1304.6077} {arXiv:1304.6077 [gr-qc]} \BibitemShut
  {NoStop}%
\bibitem [{\citenamefont {Kumar}\ \emph {et~al.}(2015)\citenamefont {Kumar},
  \citenamefont {Barkett}, \citenamefont {Bhagwat}, \citenamefont {Afshari},
  \citenamefont {Brown}, \citenamefont {Lovelace}, \citenamefont {Scheel},\
  and\ \citenamefont {Szilágyi}}]{Kumar:2015tha}%
  \BibitemOpen
  \bibfield  {author} {\bibinfo {author} {\bibfnamefont {P.}~\bibnamefont
  {Kumar}}, \bibinfo {author} {\bibfnamefont {K.}~\bibnamefont {Barkett}},
  \bibinfo {author} {\bibfnamefont {S.}~\bibnamefont {Bhagwat}}, \bibinfo
  {author} {\bibfnamefont {N.}~\bibnamefont {Afshari}}, \bibinfo {author}
  {\bibfnamefont {D.~A.}\ \bibnamefont {Brown}}, \bibinfo {author}
  {\bibfnamefont {G.}~\bibnamefont {Lovelace}}, \bibinfo {author}
  {\bibfnamefont {M.~A.}\ \bibnamefont {Scheel}},\ and\ \bibinfo {author}
  {\bibfnamefont {B.}~\bibnamefont {Szilágyi}},\ }\href
  {https://doi.org/10.1103/PhysRevD.92.102001} {\bibfield  {journal} {\bibinfo
  {journal} {Phys. Rev.}\ }\textbf {\bibinfo {volume} {D92}},\ \bibinfo {pages}
  {102001} (\bibinfo {year} {2015})},\ \Eprint
  {https://arxiv.org/abs/1507.00103} {arXiv:1507.00103 [gr-qc]} \BibitemShut
  {NoStop}%
\bibitem [{\citenamefont {Chu}\ \emph {et~al.}(2016)\citenamefont {Chu},
  \citenamefont {Fong}, \citenamefont {Kumar}, \citenamefont {Pfeiffer},
  \citenamefont {Boyle}, \citenamefont {Hemberger}, \citenamefont {Kidder},
  \citenamefont {Scheel},\ and\ \citenamefont {Szilagyi}}]{Chu:2015kft}%
  \BibitemOpen
  \bibfield  {author} {\bibinfo {author} {\bibfnamefont {T.}~\bibnamefont
  {Chu}}, \bibinfo {author} {\bibfnamefont {H.}~\bibnamefont {Fong}}, \bibinfo
  {author} {\bibfnamefont {P.}~\bibnamefont {Kumar}}, \bibinfo {author}
  {\bibfnamefont {H.~P.}\ \bibnamefont {Pfeiffer}}, \bibinfo {author}
  {\bibfnamefont {M.}~\bibnamefont {Boyle}}, \bibinfo {author} {\bibfnamefont
  {D.~A.}\ \bibnamefont {Hemberger}}, \bibinfo {author} {\bibfnamefont {L.~E.}\
  \bibnamefont {Kidder}}, \bibinfo {author} {\bibfnamefont {M.~A.}\
  \bibnamefont {Scheel}},\ and\ \bibinfo {author} {\bibfnamefont
  {B.}~\bibnamefont {Szilagyi}},\ }\href
  {https://doi.org/10.1088/0264-9381/33/16/165001} {\bibfield  {journal}
  {\bibinfo  {journal} {Class. Quant. Grav.}\ }\textbf {\bibinfo {volume}
  {33}},\ \bibinfo {pages} {165001} (\bibinfo {year} {2016})},\ \Eprint
  {https://arxiv.org/abs/1512.06800} {arXiv:1512.06800 [gr-qc]} \BibitemShut
  {NoStop}%
\bibitem [{\citenamefont {Boyle}\ \emph {et~al.}(2019)\citenamefont {Boyle}
  \emph {et~al.}}]{Boyle:2019kee}%
  \BibitemOpen
  \bibfield  {author} {\bibinfo {author} {\bibfnamefont {M.}~\bibnamefont
  {Boyle}} \emph {et~al.},\ }\href {https://doi.org/10.1088/1361-6382/ab34e2}
  {\bibfield  {journal} {\bibinfo  {journal} {Class. Quant. Grav.}\ }\textbf
  {\bibinfo {volume} {36}},\ \bibinfo {pages} {195006} (\bibinfo {year}
  {2019})},\ \Eprint {https://arxiv.org/abs/1904.04831} {arXiv:1904.04831
  [gr-qc]} \BibitemShut {NoStop}%
\bibitem [{SXS()}]{SXS:catalog}%
  \BibitemOpen
  \href@noop {} {\bibinfo {title} {{SXS Gravitational Waveform Database}}},\
  \bibinfo {howpublished}
  {\url{https://data.black-holes.org/waveforms/index.html}}\BibitemShut
  {NoStop}%
\bibitem [{\citenamefont {L{\"o}ffler}\ \emph {et~al.}(2012)\citenamefont
  {L{\"o}ffler} \emph {et~al.}}]{Loffler:2011ay}%
  \BibitemOpen
  \bibfield  {author} {\bibinfo {author} {\bibfnamefont {F.}~\bibnamefont
  {L{\"o}ffler}} \emph {et~al.},\ }\href
  {https://doi.org/10.1088/0264-9381/29/11/115001} {\bibfield  {journal}
  {\bibinfo  {journal} {Class. Quant. Grav.}\ }\textbf {\bibinfo {volume}
  {29}},\ \bibinfo {pages} {115001} (\bibinfo {year} {2012})},\ \Eprint
  {https://arxiv.org/abs/1111.3344} {arXiv:1111.3344 [gr-qc]} \BibitemShut
  {NoStop}%
\bibitem [{\citenamefont {Brandt}\ and\ \citenamefont
  {Br{\"u}gmann}(1997)}]{Brandt:1997tf}%
  \BibitemOpen
  \bibfield  {author} {\bibinfo {author} {\bibfnamefont {S.}~\bibnamefont
  {Brandt}}\ and\ \bibinfo {author} {\bibfnamefont {B.}~\bibnamefont
  {Br{\"u}gmann}},\ }\href {https://doi.org/10.1103/PhysRevLett.78.3606}
  {\bibfield  {journal} {\bibinfo  {journal} {Phys. Rev. Lett.}\ }\textbf
  {\bibinfo {volume} {78}},\ \bibinfo {pages} {3606} (\bibinfo {year}
  {1997})},\ \Eprint {https://arxiv.org/abs/gr-qc/9703066}
  {arXiv:gr-qc/9703066} \BibitemShut {NoStop}%
\bibitem [{\citenamefont {Ansorg}\ \emph {et~al.}(2004)\citenamefont {Ansorg},
  \citenamefont {Br{\"u}gmann},\ and\ \citenamefont {Tichy}}]{Ansorg:2004ds}%
  \BibitemOpen
  \bibfield  {author} {\bibinfo {author} {\bibfnamefont {M.}~\bibnamefont
  {Ansorg}}, \bibinfo {author} {\bibfnamefont {B.}~\bibnamefont
  {Br{\"u}gmann}},\ and\ \bibinfo {author} {\bibfnamefont {W.}~\bibnamefont
  {Tichy}},\ }\href {https://doi.org/10.1103/PhysRevD.70.064011} {\bibfield
  {journal} {\bibinfo  {journal} {Phys. Rev.}\ }\textbf {\bibinfo {volume}
  {D70}},\ \bibinfo {pages} {064011} (\bibinfo {year} {2004})},\ \Eprint
  {https://arxiv.org/abs/gr-qc/0404056} {arXiv:gr-qc/0404056} \BibitemShut
  {NoStop}%
\bibitem [{\citenamefont {Baumgarte}\ and\ \citenamefont
  {Shapiro}(1999)}]{Baumgarte:1998te}%
  \BibitemOpen
  \bibfield  {author} {\bibinfo {author} {\bibfnamefont {T.~W.}\ \bibnamefont
  {Baumgarte}}\ and\ \bibinfo {author} {\bibfnamefont {S.~L.}\ \bibnamefont
  {Shapiro}},\ }\href {https://doi.org/10.1103/PhysRevD.59.024007} {\bibfield
  {journal} {\bibinfo  {journal} {Phys. Rev.}\ }\textbf {\bibinfo {volume}
  {D59}},\ \bibinfo {pages} {024007} (\bibinfo {year} {1999})},\ \Eprint
  {https://arxiv.org/abs/gr-qc/9810065} {arXiv:gr-qc/9810065} \BibitemShut
  {NoStop}%
\bibitem [{\citenamefont {Shibata}\ and\ \citenamefont
  {Nakamura}(1995)}]{Shibata:1995we}%
  \BibitemOpen
  \bibfield  {author} {\bibinfo {author} {\bibfnamefont {M.}~\bibnamefont
  {Shibata}}\ and\ \bibinfo {author} {\bibfnamefont {T.}~\bibnamefont
  {Nakamura}},\ }\href {https://doi.org/10.1103/PhysRevD.52.5428} {\bibfield
  {journal} {\bibinfo  {journal} {Phys. Rev.}\ }\textbf {\bibinfo {volume}
  {D52}},\ \bibinfo {pages} {5428} (\bibinfo {year} {1995})}\BibitemShut
  {NoStop}%
\bibitem [{\citenamefont {Regge}\ and\ \citenamefont
  {Wheeler}(1957)}]{Regge:1957td}%
  \BibitemOpen
  \bibfield  {author} {\bibinfo {author} {\bibfnamefont {T.}~\bibnamefont
  {Regge}}\ and\ \bibinfo {author} {\bibfnamefont {J.~A.}\ \bibnamefont
  {Wheeler}},\ }\href@noop {} {\bibfield  {journal} {\bibinfo  {journal} {Phys.
  Rev.}\ }\textbf {\bibinfo {volume} {108}},\ \bibinfo {pages} {1063} (\bibinfo
  {year} {1957})}\BibitemShut {NoStop}%
\bibitem [{\citenamefont {Zerilli}(1970)}]{Zerilli:1970se}%
  \BibitemOpen
  \bibfield  {author} {\bibinfo {author} {\bibfnamefont {F.~J.}\ \bibnamefont
  {Zerilli}},\ }\href {https://doi.org/10.1103/PhysRevLett.24.737} {\bibfield
  {journal} {\bibinfo  {journal} {Phys. Rev. Lett.}\ }\textbf {\bibinfo
  {volume} {24}},\ \bibinfo {pages} {737} (\bibinfo {year} {1970})}\BibitemShut
  {NoStop}%
\bibitem [{\citenamefont {Nagar}\ and\ \citenamefont
  {Rezzolla}(2005)}]{Nagar:2005ea}%
  \BibitemOpen
  \bibfield  {author} {\bibinfo {author} {\bibfnamefont {A.}~\bibnamefont
  {Nagar}}\ and\ \bibinfo {author} {\bibfnamefont {L.}~\bibnamefont
  {Rezzolla}},\ }\href {https://doi.org/10.1088/0264-9381/22/16/R01} {\bibfield
   {journal} {\bibinfo  {journal} {Class. Quant. Grav.}\ }\textbf {\bibinfo
  {volume} {22}},\ \bibinfo {pages} {R167} (\bibinfo {year} {2005})},\ \Eprint
  {https://arxiv.org/abs/gr-qc/0502064} {arXiv:gr-qc/0502064} \BibitemShut
  {NoStop}%
\bibitem [{\citenamefont {Martel}\ and\ \citenamefont
  {Poisson}(2005)}]{Martel:2005ir}%
  \BibitemOpen
  \bibfield  {author} {\bibinfo {author} {\bibfnamefont {K.}~\bibnamefont
  {Martel}}\ and\ \bibinfo {author} {\bibfnamefont {E.}~\bibnamefont
  {Poisson}},\ }\href {https://doi.org/10.1103/PhysRevD.71.104003} {\bibfield
  {journal} {\bibinfo  {journal} {Physical Review D (Particles, Fields,
  Gravitation, and Cosmology)}\ }\textbf {\bibinfo {volume} {71}},\ \bibinfo
  {eid} {104003} (\bibinfo {year} {2005})}\BibitemShut {NoStop}%
\bibitem [{\citenamefont {Bernuzzi}\ and\ \citenamefont
  {Nagar}(2010)}]{Bernuzzi:2010ty}%
  \BibitemOpen
  \bibfield  {author} {\bibinfo {author} {\bibfnamefont {S.}~\bibnamefont
  {Bernuzzi}}\ and\ \bibinfo {author} {\bibfnamefont {A.}~\bibnamefont
  {Nagar}},\ }\href {https://doi.org/10.1103/PhysRevD.81.084056} {\bibfield
  {journal} {\bibinfo  {journal} {Phys. Rev.}\ }\textbf {\bibinfo {volume}
  {D81}},\ \bibinfo {pages} {084056} (\bibinfo {year} {2010})},\ \Eprint
  {https://arxiv.org/abs/1003.0597} {arXiv:1003.0597 [gr-qc]} \BibitemShut
  {NoStop}%
\bibitem [{\citenamefont {Bernuzzi}\ \emph {et~al.}(2011)\citenamefont
  {Bernuzzi}, \citenamefont {Nagar},\ and\ \citenamefont
  {Zenginoglu}}]{Bernuzzi:2011aj}%
  \BibitemOpen
  \bibfield  {author} {\bibinfo {author} {\bibfnamefont {S.}~\bibnamefont
  {Bernuzzi}}, \bibinfo {author} {\bibfnamefont {A.}~\bibnamefont {Nagar}},\
  and\ \bibinfo {author} {\bibfnamefont {A.}~\bibnamefont {Zenginoglu}},\
  }\href {https://doi.org/10.1103/PhysRevD.84.084026} {\bibfield  {journal}
  {\bibinfo  {journal} {Phys.Rev.}\ }\textbf {\bibinfo {volume} {D84}},\
  \bibinfo {pages} {084026} (\bibinfo {year} {2011})},\ \Eprint
  {https://arxiv.org/abs/1107.5402} {arXiv:1107.5402 [gr-qc]} \BibitemShut
  {NoStop}%
\bibitem [{\citenamefont {Bernuzzi}\ \emph {et~al.}(2012)\citenamefont
  {Bernuzzi}, \citenamefont {Nagar},\ and\ \citenamefont
  {Zenginoglu}}]{Bernuzzi:2012ku}%
  \BibitemOpen
  \bibfield  {author} {\bibinfo {author} {\bibfnamefont {S.}~\bibnamefont
  {Bernuzzi}}, \bibinfo {author} {\bibfnamefont {A.}~\bibnamefont {Nagar}},\
  and\ \bibinfo {author} {\bibfnamefont {A.}~\bibnamefont {Zenginoglu}},\
  }\href {https://doi.org/10.1103/PhysRevD.86.104038} {\bibfield  {journal}
  {\bibinfo  {journal} {Phys.Rev.}\ }\textbf {\bibinfo {volume} {D86}},\
  \bibinfo {pages} {104038} (\bibinfo {year} {2012})},\ \Eprint
  {https://arxiv.org/abs/1207.0769} {arXiv:1207.0769 [gr-qc]} \BibitemShut
  {NoStop}%
\bibitem [{\citenamefont {Damour}\ and\ \citenamefont
  {Nagar}(2014)}]{Damour:2014yha}%
  \BibitemOpen
  \bibfield  {author} {\bibinfo {author} {\bibfnamefont {T.}~\bibnamefont
  {Damour}}\ and\ \bibinfo {author} {\bibfnamefont {A.}~\bibnamefont {Nagar}},\
  }\href {https://doi.org/10.1103/PhysRevD.90.024054} {\bibfield  {journal}
  {\bibinfo  {journal} {Phys.Rev.}\ }\textbf {\bibinfo {volume} {D90}},\
  \bibinfo {pages} {024054} (\bibinfo {year} {2014})},\ \Eprint
  {https://arxiv.org/abs/1406.0401} {arXiv:1406.0401 [gr-qc]} \BibitemShut
  {NoStop}%
\bibitem [{\citenamefont {London}(2020)}]{London:2018gaq}%
  \BibitemOpen
  \bibfield  {author} {\bibinfo {author} {\bibfnamefont {L.~T.}\ \bibnamefont
  {London}},\ }\href {https://doi.org/10.1103/PhysRevD.102.084052} {\bibfield
  {journal} {\bibinfo  {journal} {Phys. Rev. D}\ }\textbf {\bibinfo {volume}
  {102}},\ \bibinfo {pages} {084052} (\bibinfo {year} {2020})},\ \Eprint
  {https://arxiv.org/abs/1801.08208} {arXiv:1801.08208 [gr-qc]} \BibitemShut
  {NoStop}%
\bibitem [{\citenamefont {Waskom}\ \emph {et~al.}(2022)\citenamefont {Waskom}
  \emph {et~al.}}]{core}%
  \BibitemOpen
  \bibfield  {author} {\bibinfo {author} {\bibfnamefont {M.}~\bibnamefont
  {Waskom}} \emph {et~al.},\ }\href {https://git.tpi.uni-jena.de/core}
  {\bibinfo {title} {core-watpy: v0.1.1 (may 2023)}} (\bibinfo {year}
  {2022})\BibitemShut {NoStop}%
\bibitem [{\citenamefont {Collette}(2013)}]{hdf5}%
  \BibitemOpen
  \bibfield  {author} {\bibinfo {author} {\bibfnamefont {A.}~\bibnamefont
  {Collette}},\ }\href@noop {} {\emph {\bibinfo {title} {Python and HDF5}}}\
  (\bibinfo  {publisher} {O'Reilly},\ \bibinfo {year} {2013})\BibitemShut
  {NoStop}%
\bibitem [{\citenamefont {Pezoa}\ \emph {et~al.}(2016)\citenamefont {Pezoa},
  \citenamefont {Reutter}, \citenamefont {Suarez}, \citenamefont {Ugarte},\
  and\ \citenamefont {Vrgo{\v{c}}}}]{json}%
  \BibitemOpen
  \bibfield  {author} {\bibinfo {author} {\bibfnamefont {F.}~\bibnamefont
  {Pezoa}}, \bibinfo {author} {\bibfnamefont {J.~L.}\ \bibnamefont {Reutter}},
  \bibinfo {author} {\bibfnamefont {F.}~\bibnamefont {Suarez}}, \bibinfo
  {author} {\bibfnamefont {M.}~\bibnamefont {Ugarte}},\ and\ \bibinfo {author}
  {\bibfnamefont {D.}~\bibnamefont {Vrgo{\v{c}}}},\ }\bibfield  {booktitle}
  {\emph {\bibinfo {booktitle} {25th International Conference on World Wide
  Web}},\ }\href@noop {} {\ ,\ \bibinfo {pages} {263} (\bibinfo {year}
  {2016})}\BibitemShut {NoStop}%
\bibitem [{\citenamefont {{LIGO Scientific Collaboration}}(2018)}]{lalsuite}%
  \BibitemOpen
  \bibfield  {author} {\bibinfo {author} {\bibnamefont {{LIGO Scientific
  Collaboration}}},\ }\href {https://doi.org/10.7935/GT1W-FZ16} {\bibinfo
  {title} {{LIGO} {A}lgorithm {L}ibrary - {LALS}uite}},\ \bibinfo
  {howpublished} {free software (GPL)} (\bibinfo {year} {2018})\BibitemShut
  {NoStop}%
\bibitem [{\citenamefont {Hunter}(2007)}]{matplotlib}%
  \BibitemOpen
  \bibfield  {author} {\bibinfo {author} {\bibfnamefont {J.~D.}\ \bibnamefont
  {Hunter}},\ }\href {https://doi.org/10.1109/MCSE.2007.55} {\bibfield
  {journal} {\bibinfo  {journal} {Comput. Sci. Eng.}\ }\textbf {\bibinfo
  {volume} {9}},\ \bibinfo {pages} {90} (\bibinfo {year} {2007})}\BibitemShut
  {NoStop}%
\bibitem [{\citenamefont {Harris}\ \emph {et~al.}(2020)\citenamefont {Harris}
  \emph {et~al.}}]{numpy}%
  \BibitemOpen
  \bibfield  {author} {\bibinfo {author} {\bibfnamefont {C.~R.}\ \bibnamefont
  {Harris}} \emph {et~al.},\ }\href {https://doi.org/10.1038/s41586-020-2649-2}
  {\bibfield  {journal} {\bibinfo  {journal} {Nature (London)}\ }\textbf
  {\bibinfo {volume} {585}},\ \bibinfo {pages} {357} (\bibinfo {year}
  {2020})}\BibitemShut {NoStop}%
\bibitem [{\citenamefont {{W}es {M}c{K}inney}(2010)}]{pandas}%
  \BibitemOpen
  \bibfield  {author} {\bibinfo {author} {\bibnamefont {{W}es {M}c{K}inney}},\
  }\bibfield  {booktitle} {\emph {\bibinfo {booktitle} {{P}roceedings of the
  9th {P}ython in {S}cience {C}onference}},\ }\href
  {https://doi.org/10.25080/Majora-92bf1922-00a} {\ ,\ \bibinfo {pages} {56 }
  (\bibinfo {year} {2010})}\BibitemShut {NoStop}%
\bibitem [{\citenamefont {pandas~development team}(2020)}]{pandas_zenodo}%
  \BibitemOpen
  \bibfield  {author} {\bibinfo {author} {\bibfnamefont {T.}~\bibnamefont
  {pandas~development team}},\ }\href {https://doi.org/10.5281/zenodo.3509134}
  {\bibinfo {title} {pandas-dev/pandas: Pandas}} (\bibinfo {year}
  {2020})\BibitemShut {NoStop}%
\bibitem [{\citenamefont {Carullo}\ \emph {et~al.}(2023)\citenamefont
  {Carullo}, \citenamefont {Del~Pozzo},\ and\ \citenamefont
  {Veitch}}]{pyRing_2p3p0}%
  \BibitemOpen
  \bibfield  {author} {\bibinfo {author} {\bibfnamefont {G.}~\bibnamefont
  {Carullo}}, \bibinfo {author} {\bibfnamefont {W.}~\bibnamefont {Del~Pozzo}},\
  and\ \bibinfo {author} {\bibfnamefont {J.}~\bibnamefont {Veitch}},\ }\href
  {https://doi.org/10.5281/zenodo.8165508} {\bibinfo {title} {pyring}}
  (\bibinfo {year} {2023})\BibitemShut {NoStop}%
\bibitem [{\citenamefont {Virtanen}\ \emph {et~al.}(2020)\citenamefont
  {Virtanen}, \citenamefont {Gommers}, \citenamefont {Oliphant}, \citenamefont
  {Haberland}, \citenamefont {Reddy}, \citenamefont {Cournapeau}, \citenamefont
  {Burovski}, \citenamefont {Peterson}, \citenamefont {{Weckesser}},
  \citenamefont {{Bright}}, \citenamefont {{van der Walt}}, \citenamefont
  {{Brett}}, \citenamefont {{Wilson}}, \citenamefont {{Jarrod Millman}},
  \citenamefont {{Mayorov}}, \citenamefont {{Nelson}}, \citenamefont {{Jones}},
  \citenamefont {{Kern}}, \citenamefont {{Larson}}, \citenamefont {{Carey}},
  \citenamefont {{Polat}}, \citenamefont {{Feng}}, \citenamefont {{Moore}},
  \citenamefont {{Vand erPlas}}, \citenamefont {{Laxalde}}, \citenamefont
  {{Perktold}}, \citenamefont {{Cimrman}}, \citenamefont {{Henriksen}},
  \citenamefont {{Quintero}}, \citenamefont {{Harris}}, \citenamefont
  {{Archibald}}, \citenamefont {{Ribeiro}}, \citenamefont {{Pedregosa}},
  \citenamefont {{van Mulbregt}},\ and\ \citenamefont
  {{Contributors}}}]{scipy}%
  \BibitemOpen
  \bibfield  {author} {\bibinfo {author} {\bibfnamefont {P.}~\bibnamefont
  {Virtanen}}, \bibinfo {author} {\bibfnamefont {R.}~\bibnamefont {Gommers}},
  \bibinfo {author} {\bibfnamefont {T.~E.}\ \bibnamefont {Oliphant}}, \bibinfo
  {author} {\bibfnamefont {M.}~\bibnamefont {Haberland}}, \bibinfo {author}
  {\bibfnamefont {T.}~\bibnamefont {Reddy}}, \bibinfo {author} {\bibfnamefont
  {D.}~\bibnamefont {Cournapeau}}, \bibinfo {author} {\bibfnamefont
  {E.}~\bibnamefont {Burovski}}, \bibinfo {author} {\bibfnamefont
  {P.}~\bibnamefont {Peterson}}, \bibinfo {author} {\bibfnamefont
  {W.}~\bibnamefont {{Weckesser}}}, \bibinfo {author} {\bibfnamefont
  {J.}~\bibnamefont {{Bright}}}, \bibinfo {author} {\bibfnamefont {S.~J.}\
  \bibnamefont {{van der Walt}}}, \bibinfo {author} {\bibfnamefont
  {M.}~\bibnamefont {{Brett}}}, \bibinfo {author} {\bibfnamefont
  {J.}~\bibnamefont {{Wilson}}}, \bibinfo {author} {\bibfnamefont
  {K.}~\bibnamefont {{Jarrod Millman}}}, \bibinfo {author} {\bibfnamefont
  {N.}~\bibnamefont {{Mayorov}}}, \bibinfo {author} {\bibfnamefont {A.~R.~J.}\
  \bibnamefont {{Nelson}}}, \bibinfo {author} {\bibfnamefont {E.}~\bibnamefont
  {{Jones}}}, \bibinfo {author} {\bibfnamefont {R.}~\bibnamefont {{Kern}}},
  \bibinfo {author} {\bibfnamefont {E.}~\bibnamefont {{Larson}}}, \bibinfo
  {author} {\bibfnamefont {C.}~\bibnamefont {{Carey}}}, \bibinfo {author}
  {\bibfnamefont {l.}~\bibnamefont {{Polat}}}, \bibinfo {author} {\bibfnamefont
  {Y.}~\bibnamefont {{Feng}}}, \bibinfo {author} {\bibfnamefont {E.~W.}\
  \bibnamefont {{Moore}}}, \bibinfo {author} {\bibfnamefont {J.}~\bibnamefont
  {{Vand erPlas}}}, \bibinfo {author} {\bibfnamefont {D.}~\bibnamefont
  {{Laxalde}}}, \bibinfo {author} {\bibfnamefont {J.}~\bibnamefont
  {{Perktold}}}, \bibinfo {author} {\bibfnamefont {R.}~\bibnamefont
  {{Cimrman}}}, \bibinfo {author} {\bibfnamefont {I.}~\bibnamefont
  {{Henriksen}}}, \bibinfo {author} {\bibfnamefont {E.~A.}\ \bibnamefont
  {{Quintero}}}, \bibinfo {author} {\bibfnamefont {C.~R.}\ \bibnamefont
  {{Harris}}}, \bibinfo {author} {\bibfnamefont {A.~M.}\ \bibnamefont
  {{Archibald}}}, \bibinfo {author} {\bibfnamefont {A.~H.}\ \bibnamefont
  {{Ribeiro}}}, \bibinfo {author} {\bibfnamefont {F.}~\bibnamefont
  {{Pedregosa}}}, \bibinfo {author} {\bibfnamefont {P.}~\bibnamefont {{van
  Mulbregt}}},\ and\ \bibinfo {author} {\bibfnamefont {S.~.~.}\ \bibnamefont
  {{Contributors}}},\ }\href@noop {} {\bibfield  {journal} {\bibinfo  {journal}
  {Nature Methods}\ } (\bibinfo {year} {2020})}\BibitemShut {NoStop}%
\bibitem [{\citenamefont {Waskom}\ \emph {et~al.}(2021)\citenamefont {Waskom}
  \emph {et~al.}}]{seaborn}%
  \BibitemOpen
  \bibfield  {author} {\bibinfo {author} {\bibfnamefont {M.}~\bibnamefont
  {Waskom}} \emph {et~al.},\ }\href {https://doi.org/10.5281/zenodo.592845}
  {\bibinfo {title} {mwaskom/seaborn: v0.11.2 (august 2021)}} (\bibinfo {year}
  {2021})\BibitemShut {NoStop}%
\bibitem [{\citenamefont {Boyle}\ and\ \citenamefont
  {Scheel}(2023)}]{Boyle_The_sxs_package_2023}%
  \BibitemOpen
  \bibfield  {author} {\bibinfo {author} {\bibfnamefont {M.}~\bibnamefont
  {Boyle}}\ and\ \bibinfo {author} {\bibfnamefont {M.}~\bibnamefont {Scheel}},\
  }\href {https://doi.org/10.5281/zenodo.4034006} {\bibinfo {title} {{The sxs
  package}}} (\bibinfo {year} {2023})\BibitemShut {NoStop}%
\end{thebibliography}%

\end{document}